\documentclass[11pt]{article}
\usepackage[utf8]{inputenc}
\usepackage{xcolor}
\usepackage{hyperref}
\usepackage{amsmath}
\usepackage{amsfonts}
\usepackage{amssymb}
\usepackage{stmaryrd} 
\usepackage{physics}
\usepackage{graphicx}
\usepackage[margin=1in]{geometry}
\usepackage{adjustbox}
\usepackage{appendix}
\usepackage{cite}

\newcommand{\U}{$U_q(\mathfrak{sl}_n)$} 

\graphicspath{ {./diagrams/} }

\begin{document}

\thispagestyle{empty}

\begin{center}

\Large{$U_q(\mathfrak{sl}_n)$ web models and $\mathbb{Z}_n$ spin interfaces}

\vskip 1cm

\large{Augustin Lafay$^{1}$, Azat M.\ Gainutdinov$^{2,3}$, and Jesper Lykke Jacobsen$^{1,4,5,6}$}

\vspace{1.0cm}

{\sl\small $^1$ Laboratoire de Physique de l'\'Ecole Normale Sup\'erieure, ENS, Universit\'e PSL, \\
CNRS, Sorbonne Universit\'e, Universit\'e de Paris, F-75005 Paris, France\\}

{\sl\small $^2$
Institut Denis Poisson, CNRS, Universit\'e de Tours, Universit\'e d'Orl\'eans, \\Parc de Grammont, F-37200 Tours, France\\}

 {\sl\small $^3$  
National Research University Higher School of Economics,\\ Usacheva str., 6, Moscow, Russia \\}

{\sl\small $^4$ Sorbonne Universit\'e, \'Ecole Normale Sup\'erieure, CNRS, \\
Laboratoire de Physique (LPENS), F-75005 Paris, France\\}

{\sl\small $^5$ Universit\'e Paris Saclay, CNRS, CEA, Institut de Physique Th\'eorique, \\ F-91191 Gif-sur-Yvette, France\\}

{\sl\small $^6$ Institut des Hautes \'Etudes Scientifiques, Universit\'e Paris Saclay, CNRS, \\
Le Bois-Marie, 35 route de Chartres, F-91440 Bures-sur-Yvette, France\\}
 
\end{center}

\begin{abstract}

This is the first in a series of papers devoted to generalisations of statistical loop models.
We define a lattice model of \U\ webs on the honeycomb lattice, for $n \ge 2$. It is a statistical model of closed, cubic graphs
with certain non-local Boltzmann weights that can be computed from spider relations. For $n=2$, 
  the model has no branchings and reduces to the well-known O($N$) loop model 
introduced by Nienhuis \cite{Nienhuis}. 
In the general case, 
we show that the web model possesses a particular point, at $q=e^{i \pi/(n+1)}$, where the partition function is
proportional to that of a $\mathbb{Z}_n$-symmetric chiral spin model on the dual lattice. Moreover, under this equivalence,
the graphs given by the configurations of the web model are in bijection with the domain walls of the spin model.
For $n=2$, this equivalence reduces to the well-known relation between the Ising and O($1$) models. We define as well an open \U\ web model
on a simply connected domain with a boundary, and discuss in particular the role of defects on the boundary.

\end{abstract}

\newpage 

\section{Introduction}

The study of two-dimensional lattice models of loops is an extremely fruitful research topic at the intersection between
statistical and condensed matter physics, mathematical physics, and various branches of mathematics. Loop models have physical applications in 
lattice polymers \cite{DS-NPB87}, level lines of random surfaces~\cite{KH95}, domain walls in spin systems \cite{GamsaCardy07},
and equipotential lines in quantum models of electron gases \cite{GLR99}, to mention but a few examples.
Their study calls for an array of mathematical techniques, including integrability \cite{Nienhuis,Baxter86,WNS92,IntReview}, knot theory~\cite{Kauf87},
representation theory of cellular algebras~\cite{AlgReview}, and category theory \cite{Wang,GL1,GS16}.
In the so-called continuum limit, where the mesh of the lattice is taken to zero, and under certain conditions,
the loops form conformally invariant ensembles of random curves \cite{LoopReview}.
The continuum limit of loop models makes rich contact with conformal field theory (CFT) \cite{CFTbook},
and---on the mathematical side---probabilistic approaches such as conformal loop ensembles (CLE) \cite{CLE}
and Schramm-Loewner evolution (SLE) \cite{SLE}.

Generally speaking, the loop models that we
have in mind are obtained as follows. First we fix some regular two-dimensional lattice, consisting of nodes and links. A configuration
of the loop model is a drawing, on top of the links of the lattice, of a set of self-avoiding and mutually avoiding closed loops, subject
to specific rules of avoidance at the nodes. A link covered by one segment of a loop is referred to as a {\em bond}. The statistical weight
of a loop configuration then consists of a local and a non-local part. The local part gives simply a certain weight per bond (in the terminology
of lattice polymers this is the {\em monomer fugacity}) and specific weights for the possible arrangements at each node. Importantly, the non-local part 
attributes a weight $N$ per loop, regardless of its size. It is precisely this non-local loop weights that is crucial for particular physical applications---for
instance, the limit $N \to 0$ produces lattice polymers---and provides the interest of loop models within algebra and topology.

\begin{figure}
\begin{align*}
     \vcenter{\hbox{\includegraphics[scale=0.2]{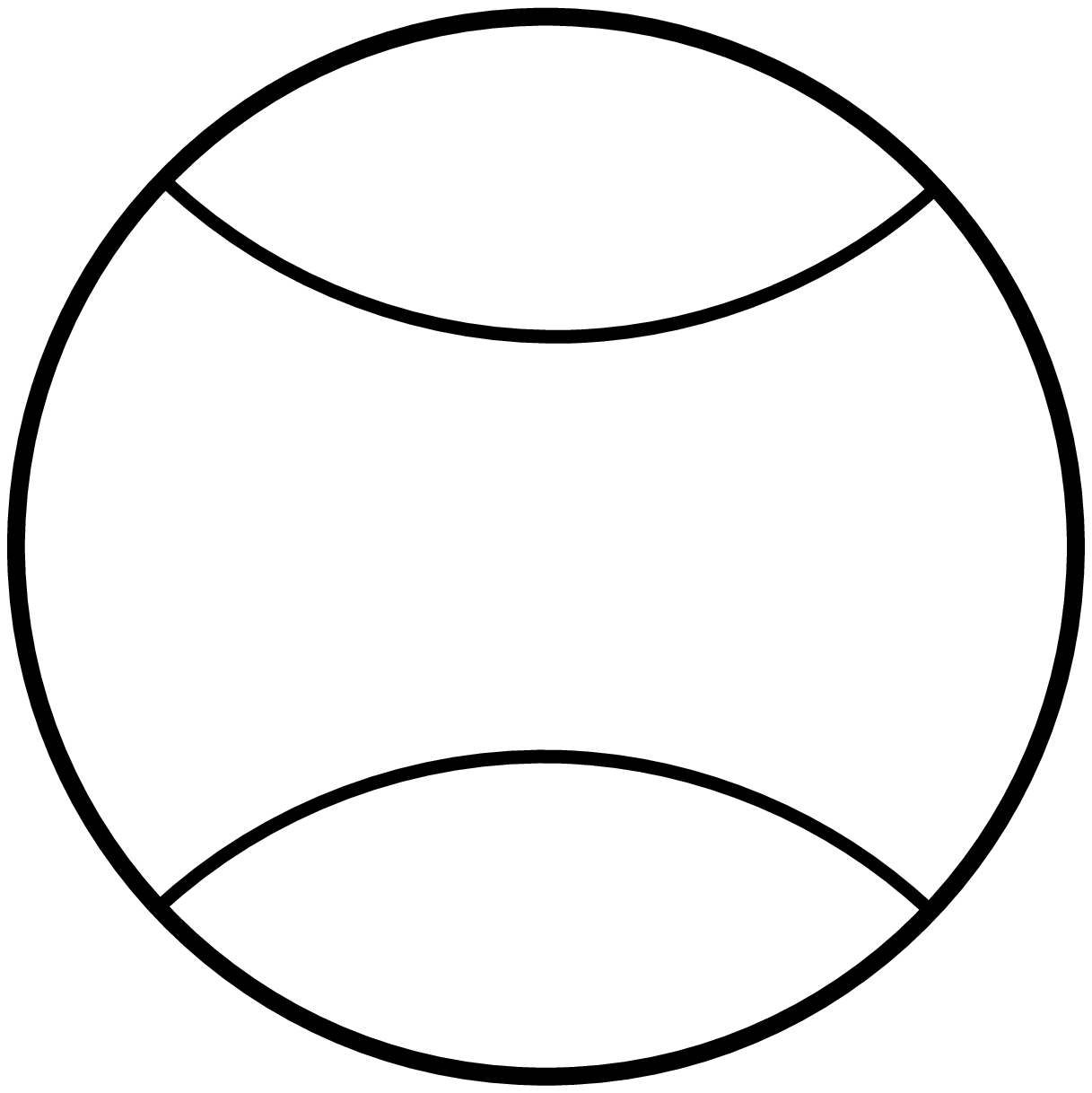}}}\qquad 
     \vcenter{\hbox{\includegraphics[scale=0.1]{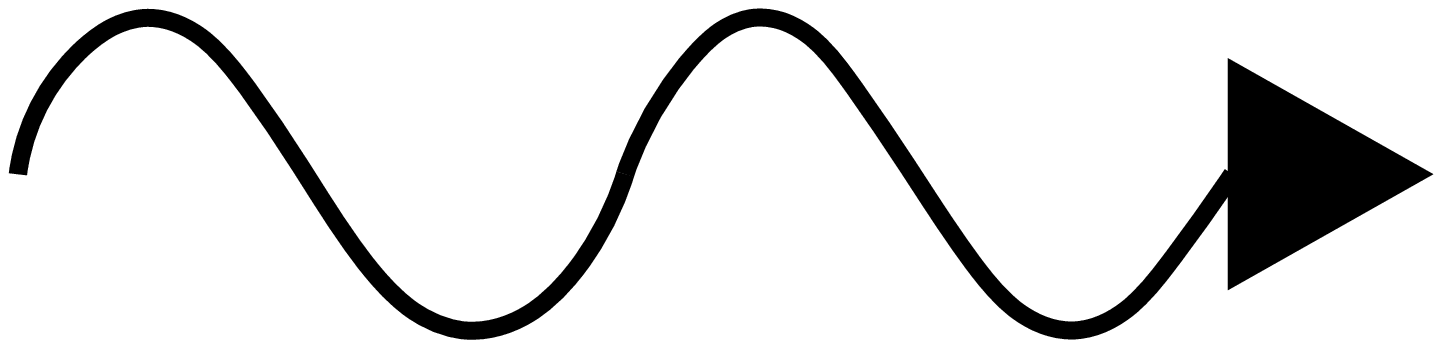}}} \qquad \vcenter{\hbox{\includegraphics[scale=0.2]{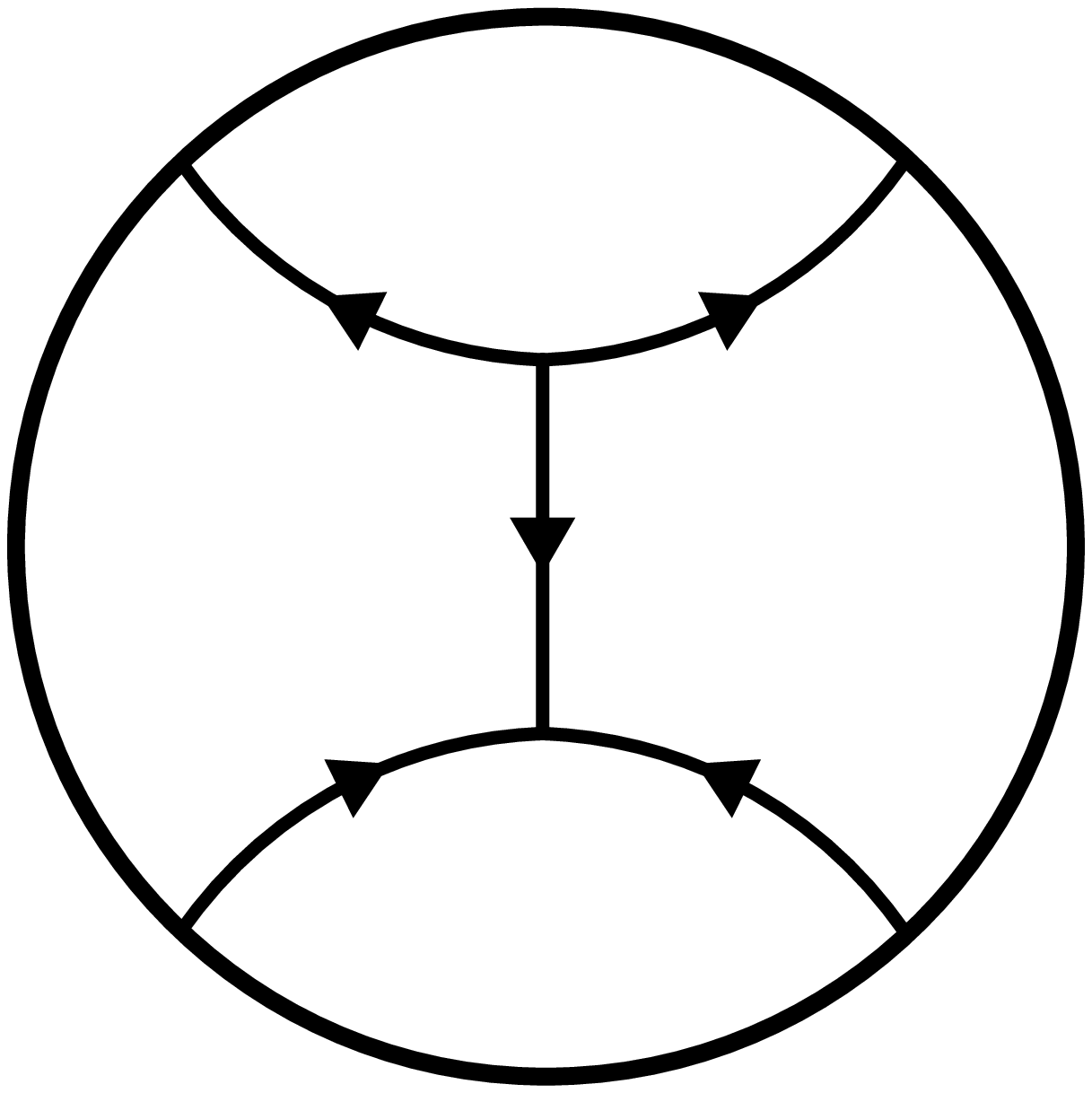}}}
\end{align*}
\caption{Generalisation of the loop models to $U_q(\mathfrak{sl}_3)$ web models where the local connectivity (on the left) is replaced by a bifurcation configuration (on the right).}
\label{fig:spider}
\end{figure}

Recent years have seen the emergence of interest in more complex models of random geometry, involving extended structures which---unlike loops---can
undergo branchings and bifurcations. We shall refer to such structures as {\em webs}. They are relevant for the physical description of domain walls
in spin systems \cite{Dubail10,Dubail10b,PiccoSantachiara11}, where the spin can take more than
two values, or for certain network models motivated by
topological phases~\cite{Kitaev03,LevinWen05,Fendley08}.
In parallel, the mathematical literature has seen the definition of algebraic structures
whose geometrical representation takes the form of bifurcating objects
 \cite{kuperberg1991quantum, Kuperberg_1996, MOY, kim2003graphical, JK05, Mor07,  Wu,  Cautis_2014} also known as ``spiders". To be precise, the algebra underlying the description of loops is the celebrated
Temperley-Lieb algebra~\cite{TL71} and its 
close cousins
(with higher-spin representations \cite{BirmanWenzl,Murakami}, and versions with dilution \cite{WNS92,Grimm}, with colours \cite{GrimmPearce,GrimmMartin} or the fully-packed versions \cite{KdGN96,JK98,KJ98}). The simplest example of
an algebraic construction accounting for bifurcations is the so-called Kuperberg spider~\cite{Kuperberg_1996}, see an example on the right of Figure~\ref{fig:spider}.
The study of such structures seems however to have been driven
by rather formal motivations, such as  classification of invariants of the quantum group $U_q(\mathfrak{sl}_n)$ with $n \ge 3$ in its tensor-product representations~\cite{Kuperberg_1996, kim2003graphical, Cautis_2014}, 
 and for construction of isotopy invariants of links and knots  generalising the Jones polynomials~\cite{MOY,JK05, Wu}. The usual loop
case, described by the Temperley-Lieb algebra, is related to $U_q(\mathfrak{sl}_2)$ where the Temperley-Lieb diagrams classify the invariants in a certain tensor power of the fundamental representation of $U_q(\mathfrak{sl}_2)$.

\smallskip
The purpose of this paper is to define and initiate the study of a simple physical model of dilute webs enjoying $U_q(\mathfrak{sl}_n)$ symmetry, for any $n \ge 2$. The quantum group invariance property is manifested locally at the bifurcation points, via the spider operators and the relations that they satisfy.
We  formulate our statistical model in the simplest possible setting, namely on a hexagonal lattice, so that for $n=2$ it coincides with
the well-known and popular $O(N)$ model of loops due to Nienhuis \cite{Nienhuis}. When $n \ge 3$ our dilute web models involve bifurcations as well. 
A simple example of such a generalisation for $n=3$ is illustrated in Figure~\ref{fig:spider} where we have a trivalent graph generalising the ``local" loop configuration.
In this case, the mentioned  $U_q(\mathfrak{sl}_3)$ symmetry
means that the (lower) bifurcation point corresponds to the projection from the product of two fundamental representations onto the anti-fundamental one of $U_q(\mathfrak{sl}_3)$; the anti-fundamental nature of the latter is presented graphically by the arrow pointing at the opposite direction, namely downward. Note that in contrast with the $\mathfrak{sl}_2$ or loop case we have no contribution of vacuum states here, and this is why a bifurcation appears. 

In Section~\ref{sec:Kuperberg}, we begin by introducing the web model on a cylinder for the first non-trivial case $n=3$, which is related to the Kuperberg spider shown in Figure~\ref{fig:spider},
and establish a bijection with interfaces in the $\mathbb{Z}_3$ spin model   in Section~\ref{sec:Z3-interfaces}.
Then, we present in Section~\ref{sec:Cautis} the further generalisation to $U_q(\mathfrak{sl}_n)$ web models valid for arbitrary $n \ge 2$. 
In the definition of the models we need only to know local relations or moves for the spider diagrams (that represent $U_q(\mathfrak{sl}_n)$ invariant operators) and fortunately a complete set of such moves was extensively studied by mathematicians, so we do not even need to develop any $U_q(\mathfrak{sl}_n)$ representation theory.
In this paper, we follow the diagrammatical formalism of Cautis-Kamnitzer-Morrison  for $\mathfrak{sl}_n$ calculus from~\cite{Cautis_2014}.
In the literature there are a few other related, though not necessarily equivalent, formalisms~\cite{MOY, kim2003graphical, JK05}, and we choose the one given in~\cite{Cautis_2014} mainly for practical reasons, to make physical properties (like CPT invariance) of our models evident.
Moreover, the formalism of Cautis-Kamnitzer-Morrison becomes quite crucial when we study open variants of our models with boundary defects.

In Section~\ref{sec:spin},
we make explicit the relevance of these
 $U_q(\mathfrak{sl}_n)$
 web models for the
description of interfaces in a wide class of $\mathbb{Z}_n$ symmetric spin systems, including the Potts model,
generalising a well-known correspondence between
loops and Ising domain walls or percolation hulls.
Finally, in Section~\ref{sec:open} we extend the treatment of the \U\ web model to a simply connected domain, where we allow for 
defects or ``spider legs" to reside on the boundary of the domain. 
Our conclusions and directions for further work are presented in Section~\ref{sec:conclusion}.

\smallskip
Throughout the discussion will be kept as accessible and non-technical as possible. In particular,
the knowledge about  $U_q(\mathfrak{sl}_n)$ quantum groups, their representation theory and spiders is not needed to understand this paper, and we shall only occasionally
provide a few remarks about these connections. The algebraic origin of the diagrammatic objects will however turn out essential in order to
define a vertex-model formulation of the web models with purely {\em local} weights, in the same way as was done for the usual loop models.
Establishing such a local formulation for the $U_q(\mathfrak{sl}_n)$ web models will form the object of a subsequent paper \cite{loc}.
We also expect the models to exhibit critical behaviour and leave for future work the description of their continuum limits in terms of CFT.

\newcommand\qbin[2]{\left[\begin{matrix} #1 \\ #2 \end{matrix} \right]_{q}}
\newcommand\qbinom[2]{\left[\begin{matrix} #1 \\ #2 \end{matrix} \right]}

\medskip

\textbf{Notations.}\;
Let $q \in \mathbb{C}$ be an arbitrary non-zero
 complex number (but $q \neq \pm 1$). The statistical weights of the web models will be defined in terms of so-called
$q$-numbers $[k]_q$, with $k \in \mathbb{N}$, defined by
\begin{equation}
    [k]_q =\frac{q^k-q^{-k}}{q-q^{-1}} \,.
\end{equation}
Note that the $q$-numbers reduce to the ordinary integers, $[k]_q \to k$, in the limit $q \to 1$.
We shall also need the corresponding $q$-factorial and $q$-binomial coefficients:
\begin{equation*}
    [k]_q!=\prod_{1\leq i\leq k} [i]_q \;, \qquad
    \qbin{n}{k}=\frac{[n]_q!}{[k]_q![n-k]_q!} \,,
\end{equation*}
with the convention $[0]_q! = 1$ and $\footnotesize\qbin{n}{0}=1$.

\section{Kuperberg web model}
\label{sec:Kuperberg}

In order to make the general definition and properties of web models clearer, we begin by exposing the simplest case after the $O(N)$ loop model ($n=2$), namely, the
web model based on the invariant theory of $U_q(\mathfrak{sl}_3)$. This model as well as its generalisations to any $n \ge 3$ will eventually be defined in
a unified way in Section~\ref{sec:Cautis}, by using the webs of 
Cautis-Kamnitzer-Morrison~\cite{Cautis_2014}. However, to keep the discussion as simple as possible,
we first give in the present section an equivalent formulation of the $n=3$ case using the webs of the Kuperberg spider \cite{Kuperberg_1996},
which may be
more familiar.
 The two models will be shown to be equivalent in Section~\ref{sec:Cautis}. We begin with the introduction of 
the Kuperberg web model and then describe its relation with $\mathbb{Z}_3$ spin models.

\subsection{Kuperberg web model}

Our model is defined on an underlying finite hexagonal lattice $\mathbb{H}$, with boundary conditions such that the lattice is planar (e.g., free or periodic in one direction).
Using the terminology of the introduction, $\mathbb{H}$ consists of {\em nodes} and {\em links}, each node being adjacent to three links
(we say that $\mathbb{H}$ is {\em trivalent}).

A configuration $c$ of the Kuperberg web model is a certain subset of links in $\mathbb{H}$---represented
by drawing {\em bonds} on top of the links in the subset---subject to a number of constraints that we now describe. First, $c$ must form a {\em closed} planar
graph, meaning that a node can be adjacent to $0, 2$ or $3$ bonds, but never $1$. A node adjacent to $3$ bonds (a 3-valent node) is called a {\em vertex} of $c$. A path of
consecutive bonds, going either from one vertex to a distinct vertex via a succession of 2-valent nodes, or forming a closed path of 2-valent nodes,
is called an {\em edge} of $c$. An edge of the latter type
(closed path) is also called a {\em loop}. Second, we give an orientation
to each bond, and impose that all bonds along one edge be consistently oriented. Each edge has the orientation inherited from its constituent bonds.
Third, we demand each vertex of $c$ to be either a {\em source} (i.e., adjacent to 3 outgoing edges) or a {\em sink} (i.e., adjacent to 3 ingoing edges).

With these definitions, any configuration $c$ can be drawn as a graph consisting of vertices and oriented edges.
We call such a graph a {\em web}.
By construction, the graph $c$ is closed and planar. It can be considered an {\em abstract} graph,
in the sense that it can be drawn without reference to the underlying lattice $\mathbb{H}$, but it is nevertheless still embedded in the plane.
Moreover, $c$ is trivalent and {\em bipartite}\/: each oriented edge (except for those edges that form loops) goes from a source to a sink, so the set of sources and the set of sinks provide
a bipartition of $c$. Notice finally that $c$ is not necessarily connected; indeed, it may have several connected components.

\begin{figure}
\begin{center}
    \includegraphics[scale=0.3]{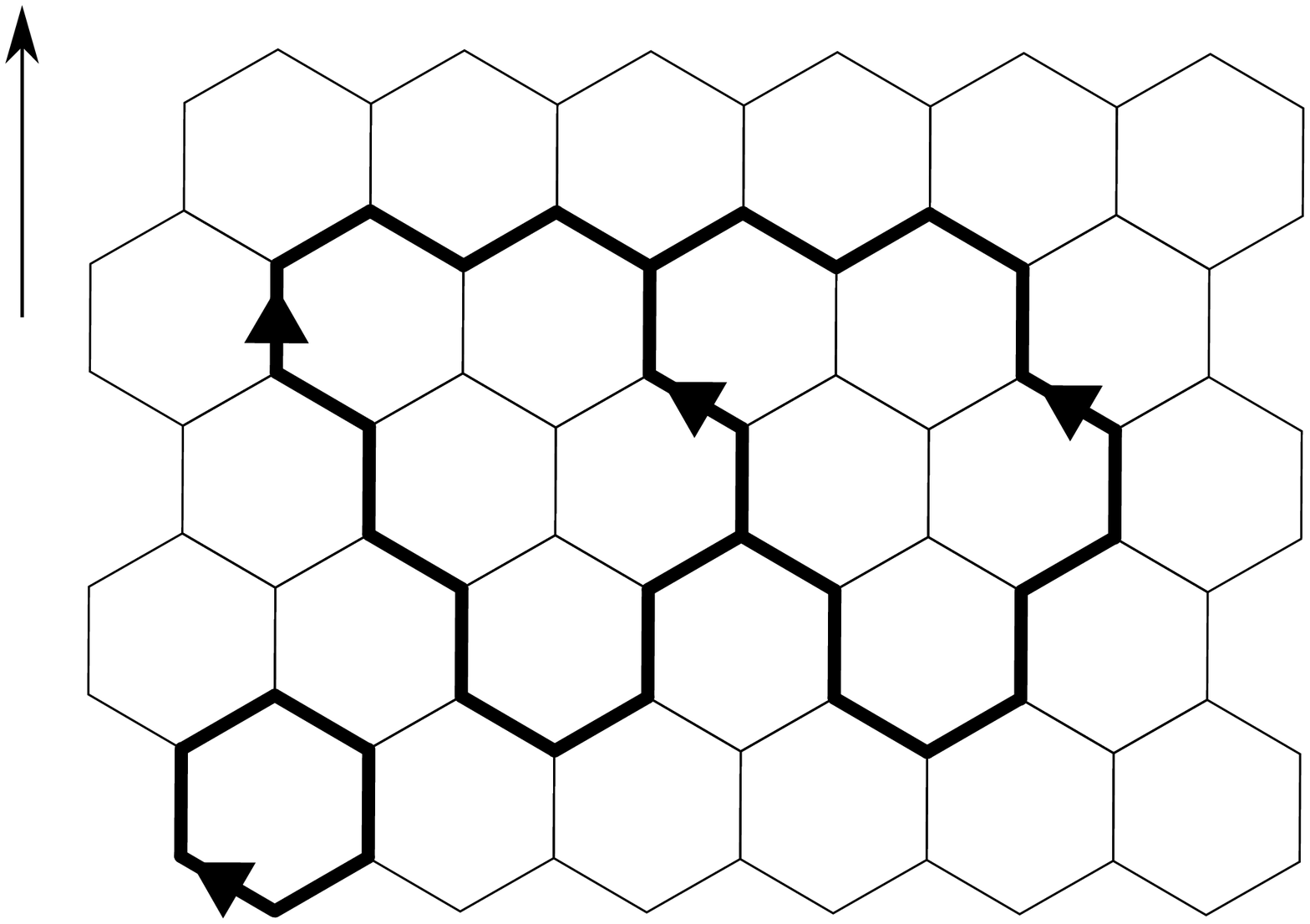} \qquad \qquad \includegraphics[scale=0.3]{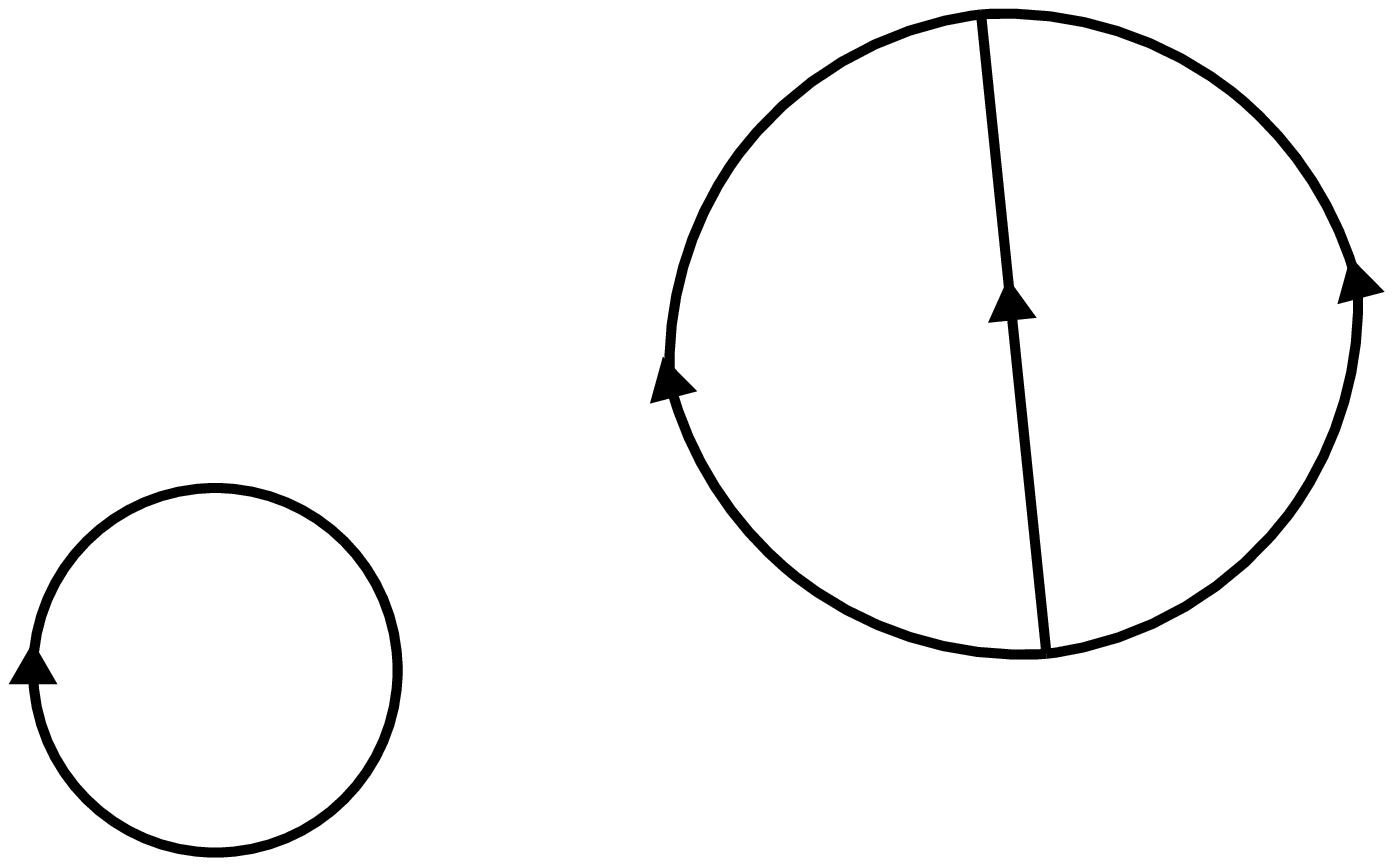}
\end{center}
    \caption{Left panel: A configuration on $\mathbb{H}$ of weight $x_1^{22}x_2^{13}yz[2]_q[3]_q^2$. The arrow is parallel to the axis of the cylinder. The left and right sides of the drawing are identified by periodic boundary conditions. Right panel: The same configuration drawn as a web.}
    \label{fig:config}
\end{figure}

For definiteness, we consider here the case of periodic boundary conditions%
\footnote{Open boundary conditions will be treated in Section \ref{sec:open}.}
and accordingly embed $\mathbb{H}$ in a cylinder,
such that one third of its links are parallel to the axis of the cylinder (see Figure \ref{fig:config}). 

We fix an orientation of this axis and define an oriented bond to be {\em upward} (resp.\ {\em downward}) if it has a positive (resp.\ negative) projection
onto the chosen orientation.

\medskip

We now describe how to weigh a given configuration $c$. Firstly, webs are given a non-local weight~$w_{\rm K}(c)$. This weight is computed according to the following rules, taken from \cite{Kuperberg_1996}: 
\begin{subequations}\label{eq:Kup-rels}
\label{3rules}
\begin{align}
    \vcenter{\hbox{\includegraphics[scale=0.2]{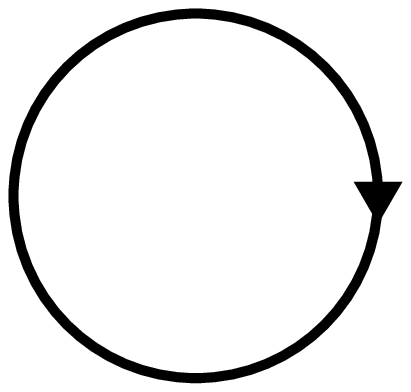}}}&\quad=\quad[3]_q\label{eq:Kup-rule1}\\[5pt]
    \vcenter{\hbox{\includegraphics[scale=0.2]{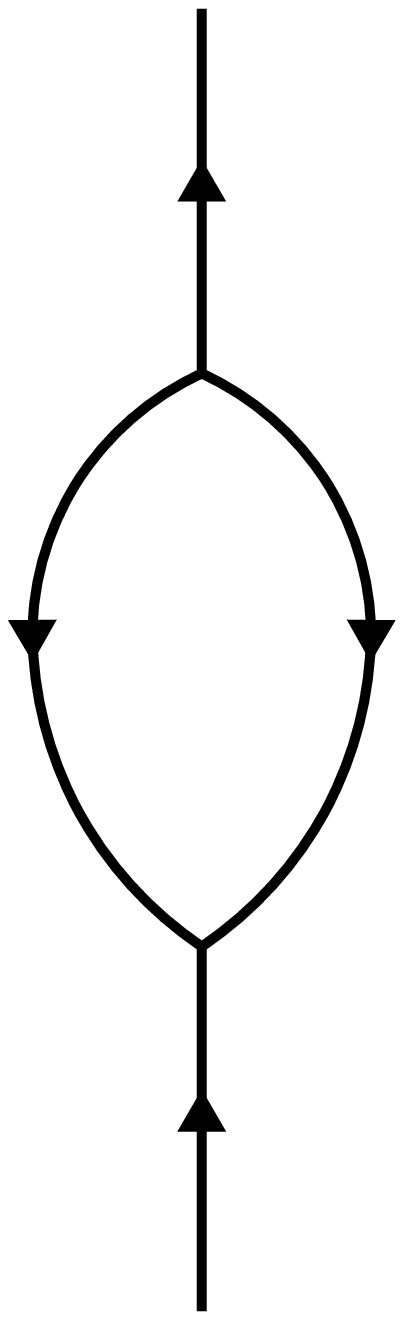}}}&\quad=\quad[2]_q\;\vcenter{\hbox{\includegraphics[scale=0.2]{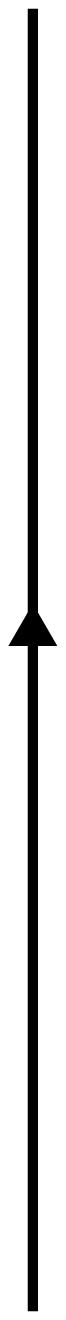}}}\\[5pt]
    \vcenter{\hbox{\includegraphics[scale=0.2]{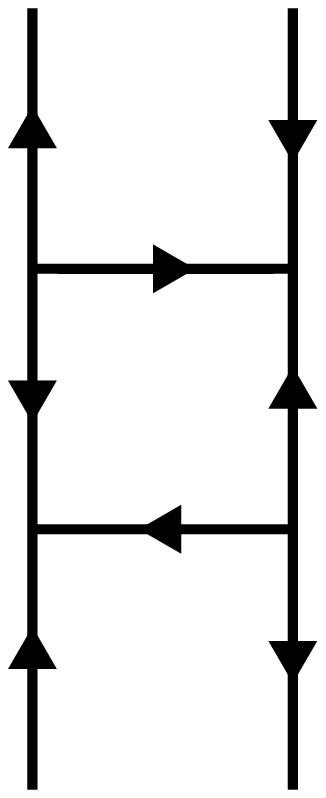}}}&\quad=\quad\vcenter{\hbox{\includegraphics[scale=0.2]{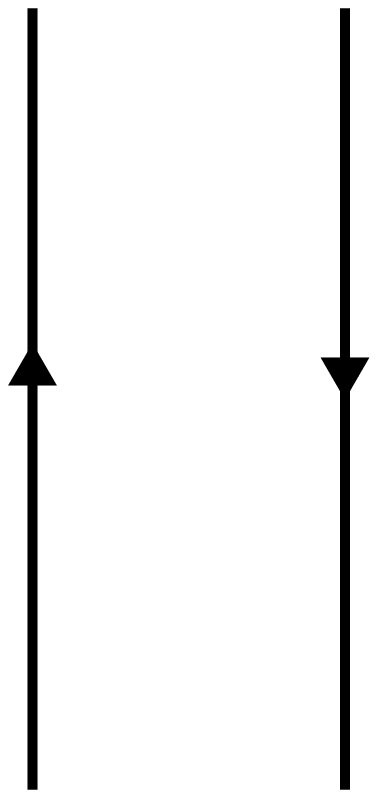}}}\quad +\quad\vcenter{\hbox{\includegraphics[scale=0.2]{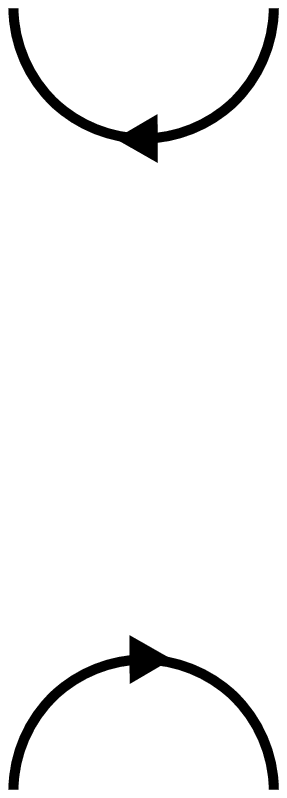}}}\label{kupsquare}
\end{align}
\end{subequations}
The first rule replaces a loop by the weight $[3]_q$. The second rule replaces every subgraph of the shape of a two-sided polygon (or digon) by an
edge with the weight $[2]_q$. Finally, the third rule replaces every subgraph of the shape of a four-sided polygon (or ``square'') by a formal sum of
subgraphs given in the right-hand side of (\ref{kupsquare}). Note that the first rule gives the same weight $[3]_q$ to any loop, regardless of its
orientation (clockwise or anticlockwise), and likewise the following two rules also hold true with all the arrow directions reversed.

The fact that webs are bipartite implies of course
that only even-sided polygons can occur. It is a non-trivial topological fact that any connected component of a web that is not a loop contains
at least one two-sided or one four-sided polygon. To see this, suppose that it is not the case, i.e, consider a non-empty web that contains neither
a loop, nor a two-sided or a four-sided polygon. Denote by $V$, $E$, and $F$, respectively, the number of vertices, edges and faces in the web.
As the graph in trivalent, one has $2E=3V$ by the hand-shake lemma, so from the Euler relation $F-E+V=F-\frac{1}{2}V=2$.
Now, by assumption, each of the faces is bounded by at least $6$ vertices, and since each vertex is surrounded by $3$ faces,
one has $3V\geq 6F$ implying $F-\frac{1}{2}V=2\leq 0$, a 
contradiction.%
\footnote{We note that the argument will not work in other geometries, like a torus, but we do not treat them here.}
Of course, a web can contain six-sided or more complicated polygons but from the above argument it necessarily contains a two-sided or a four-sided polygon attached to them, so that using the second and third rules in~\eqref{3rules}, the polygons reduce in size.
Therefore, proceeding recursively, any web 
is reduced to a collection of loops, which will each be replaced by the weight~$[3]_q$
by~\eqref{eq:Kup-rule1}. In other words, the set of three rules will replace any web $c$ by a corresponding non-local weight $w_{\rm K}(c)$.

It was shown in \cite{kuperberg1991quantum} that the set of three rules (\ref{3rules}) is well defined, in the sense that applying them in any order to $c$ will lead to the same
weight $w_{\rm K}(c)$. Therefore, $w_{\rm K}(c)$ is {\em the} weight of the web $c$. These rules were  in fact shown to be the only ones, up to a {\em rescaling} of vertices, to be well-defined for the class of graphs considered. We will use and precisely define the concept of vertex rescaling in the next subsection.

As an example of the application of the rules \eqref{3rules}, it is easy to see that the non-local weight of the web $c$ shown in Figure~\ref{fig:config} is $w_{\rm K}(c) = [2]_q[3]_q^2$.

\medskip
 
Secondly, we multiply this non-local weight by local fugacities for vertices and bonds. Each upward (resp.\ downward) bond is given the
fugacity $x_1$ (resp.\ $x_2$). A vertex is given the fugacity $y$ (resp.\ $z$) if it is a sink (resp.\ source).
Observe that, because the webs are closed, there are as many sinks as sources. Hence the partition function depends only on the product $yz$.

\medskip

Summarising, the total weight of a configuration $c$ is given by the product of local fugacities times the non-local weight $w_{\rm K}(c)$ obtained by the reduction procedure. The partition function thus reads:
\begin{align} \label{Z_K}
    Z_{\rm K} =\sum_{c\in K} x_1^{N_1}x_2^{N_2}(yz)^{N_V}w_{\rm K}(c) \,,
\end{align}
where $N_1$ (resp.\ $N_2$) is the number of upward (resp.\ downward) bonds, $N_V$ is the number of sink/source pairs of vertices, 
and 
 $K$ denotes the set of  Kuperberg webs on the finite hexagonal lattice $\mathbb{H}$.
The model is discretely rotationally invariant when $x_1=x_2$.

\subsection{Relation with $\mathbb{Z}_3$ spin models}
\label{sec:Z3-interfaces}

We will now show that the Kuperberg web model is equivalent, at a special point, to a chiral $\mathbb{Z}_3$ spin model in the most general case. More precisely, we give a map from spin configurations to web configurations such that webs correspond to interfaces between spin clusters. We will show that, under this map, the two partition functions agree up to an overall factor.

We first formulate the spin model in terms of a low-temperature expansion. The spins $\sigma_i$ take three different values,
$\sigma_i \in \mathbb{Z}_3 := \{0,1,2\}$, and are defined on the nodes $i$ of a triangular lattice~$\mathbb{T}$.
This triangular lattice is the dual of the hexagonal lattice $\mathbb{H}$ considered above, i.e., $\mathbb{T} = \mathbb{H}^*$.
A link $(ij)$ of $\mathbb{T}$ is 
specified by the two nodes, $i$ and $j$, on which it is incident.
 Viewing $\mathbb{T}$ along the oriented
axis defined in Figure~\ref{fig:config} determines uniquely whether $j$ is to the right or left of $i$ (since, by construction of the dual
lattice, the segment $[ij]$  cannot be parallel to the axis).
The weight describing the (chiral) interaction along any link $(ij)$ of $\mathbb{T}$ is then defined to be $x_{\sigma_j - \sigma_i}$
if $j$ is to the right of $i$. We recall that all differences between spins are computed within $\mathbb{Z}_3$,
i.e., modulo $3$. This defines three interaction parameters, $x_0$, $x_1$ and $x_2$.
Clearly the interactions are invariant by a global $\mathbb{Z}_3$ action, that is, upon shifting all spins by the same amount.

\begin{figure}
\begin{center}
    \includegraphics[scale=0.3]{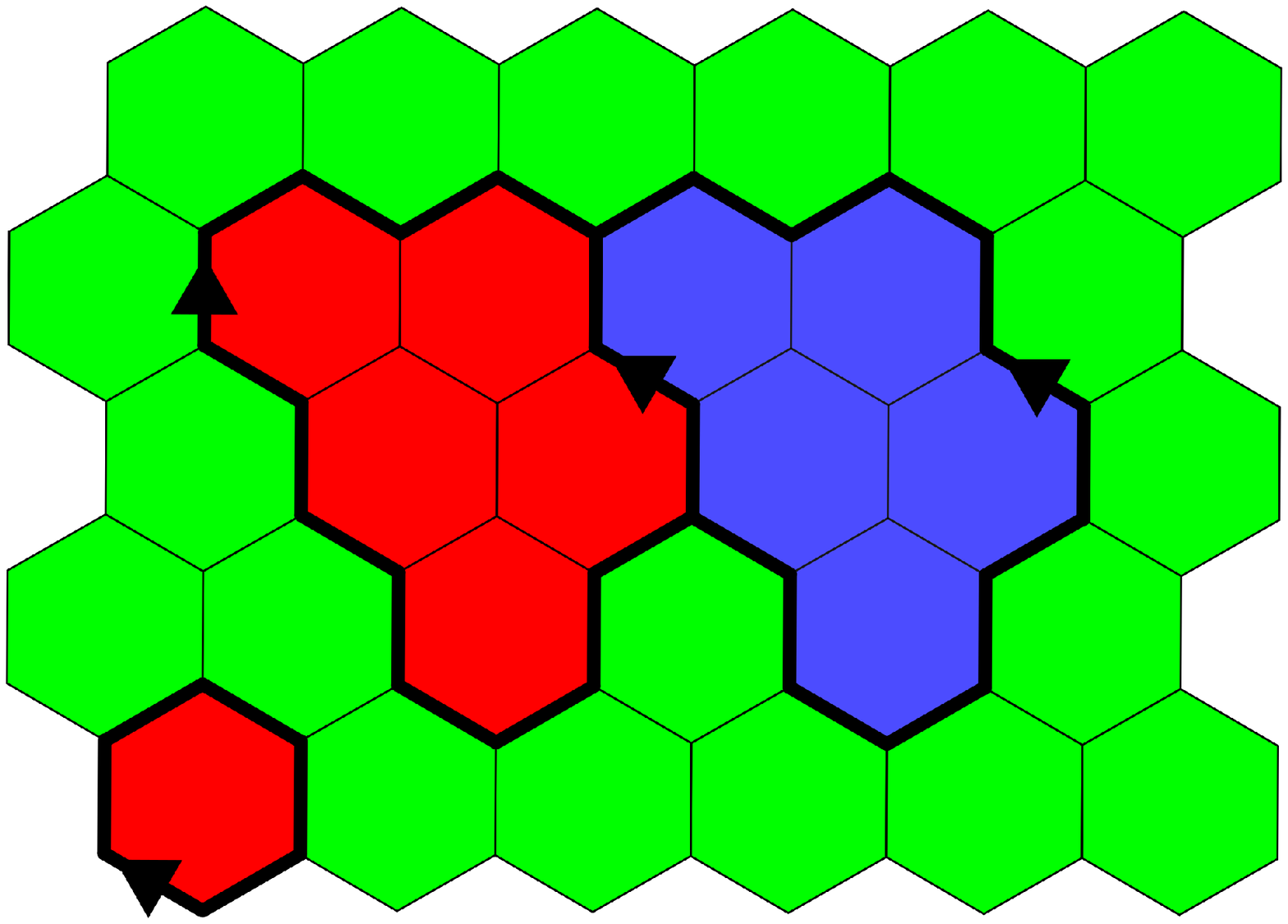} \quad \includegraphics[scale=0.3]{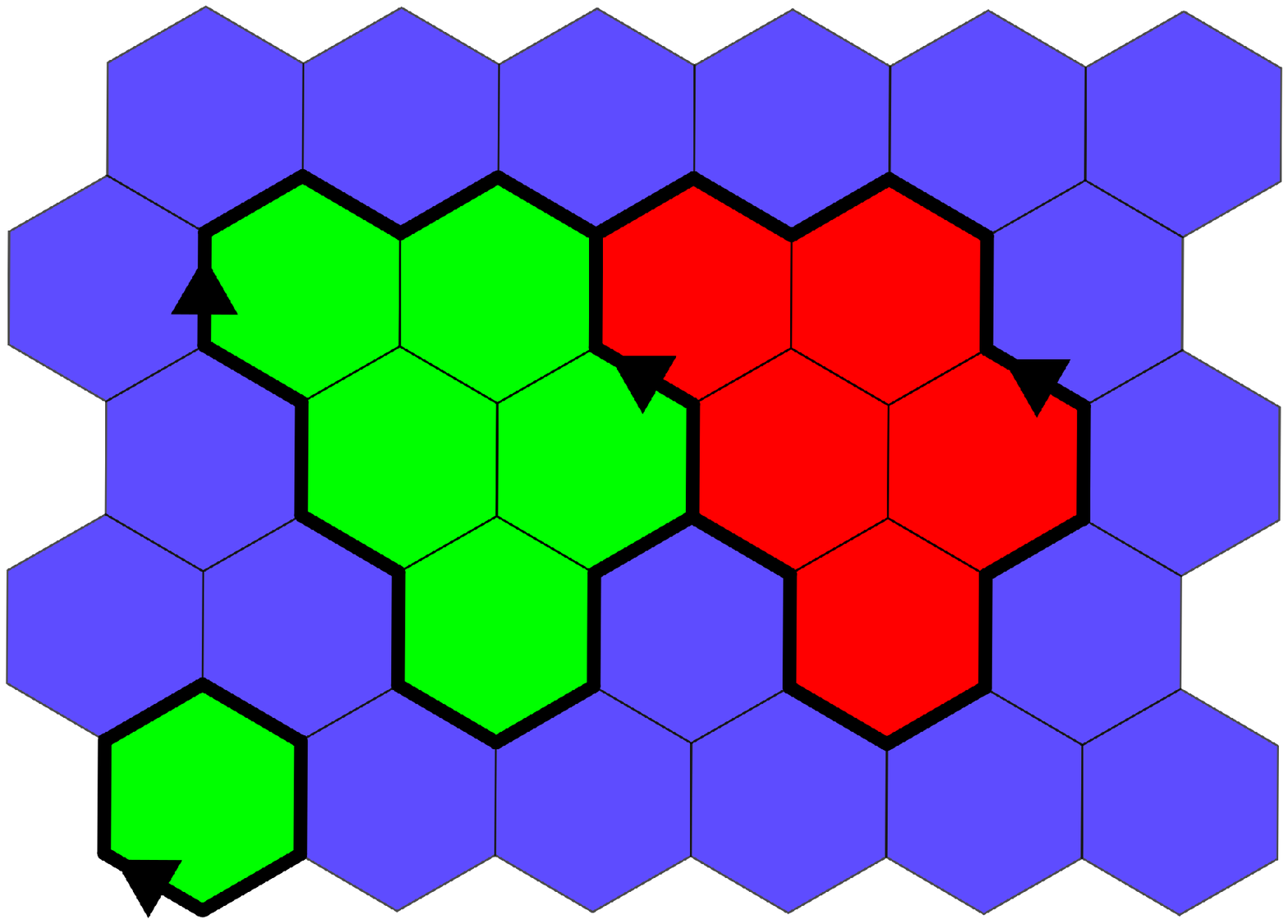}
    \caption{Two spin configurations, with the colours $\{{\rm red},{\rm blue},{\rm green}\}$ representing the spin values $\mathbb{Z}_3 := \{0,1,2\}$.
    These configurations are related by a global $\mathbb{Z}_3$ shift and hence produce the same Kuperberg web.}\label{fig:configspin}
\end{center}
\end{figure}

If we normalise the interactions by setting $x_0 = 1$, this spin model associates a non-trivial weight, $x_1$ or $x_2$, to each
piece of domain wall between unequal spins. More precisely, a piece of domain wall is a link of $\mathbb{H}$ that is dual to a
link $(ij)$ of $\mathbb{T}$ satisfying $\sigma_i \neq \sigma_j$. We call such a piece of domain wall a {\em bond} on $\mathbb{H}$.
We can further orient the bonds by the following rule. When~$j$ is to the right of $i$, we give the bond the upward (resp.\ downward) orientation
if $\sigma_j - \sigma_i = 1$ (resp.\ $\sigma_j - \sigma_i = 2$).  
With these rules, the vertices (i.e., nodes of the lattice adjacent to $3$ bonds) appear at a junction of three spin clusters and we observe that they are either sources or sinks according to how the spins around them are arranged. Indeed, the three spin colours  follow each other in cyclic order, $0 \to 1 \to 2 \to 0$, as one turns around a sink (resp.\ source) vertex
in the anticlockwise (resp.\ clockwise) direction.
It is thus clear that a configuration of the spin model produces a configuration of oriented bonds on $\mathbb{H}$
which is precisely a Kuperberg web. Notice however that, since the rules are defined in terms of spin differences, two spin configurations that are related
by a global $\mathbb{Z}_3$ action give rise to the same web (see Figure \ref{fig:configspin}). Conversely, a web configuration specifies
all spin differences in $\mathbb{Z}_3$ and so gives rise to a spin configuration, provided that we fix the value of one reference spin.
In other words, we have given a bijection between web configurations and spin configurations modulo the global $\mathbb{Z}_3$ action.

As to the statistical weight of a given spin configuration, the local bond weights $x_1$ and $x_2$ can readily be seen to have the same meaning
as in our previous definition of the Kuperberg web model. This implies that the partition function of the spin model can be written
\begin{align}
  \label{Z_spin}
    Z_{\rm spin}=3\sum_{c\in K} x_1^{N_1}x_2^{N_2} \,,
\end{align}
where the sum is over 
the same set of  Kuperberg webs $c$  on the finite hexagonal lattice $\mathbb{H}$, and the overall factor of $3$ accounts for the global $\mathbb{Z}_3$ invariance in the bijection.
The quantities $N_1$ and $N_2$ have the same meaning as in~\eqref{Z_K}.

The difference between the web partition function $Z_{\rm K}$ in \eqref{Z_K} and the spin partition function $Z_{\rm spin}$ in \eqref{Z_spin} is clearly
that the latter assigns neither a local weight to the sink/source vertices, nor the non-local topological weight to each connected component of the web.
In other words, the Kuperberg web model is equivalent to the spin model (i.e., $Z_{\rm spin} = 3 Z_{\rm K}$) provided that one can impose the
following relation for all configurations:
\begin{align}
  \label{wKprime}
    w_{\rm K}'(c) := (yz)^{N_V}w_{\rm K}(c)=1 \,.
\end{align}
This equivalence is achieved at the special point
\begin{subequations}
\begin{align}
  q &= e^{i\frac{\pi}{4}} \,, \\
  yz &= 2^{-\frac{1}{2}} \,.
\end{align}
\end{subequations}
To see this, observe first that one can compute the partial weight $w_{\rm K}'(c)$ in the same way as $w_{\rm K}(c)$ is computed, but using certain deformed rules.
Indeed, one can {\em rescale} the vertices in Kuperberg rules in order to incorporate the $y$ and $z$ fugacities, by defining the following ``dressed''
vertices:
\begin{subequations} \label{dressed_vertices_K}
\begin{align}
    \vcenter{\hbox{\includegraphics[scale=0.2]{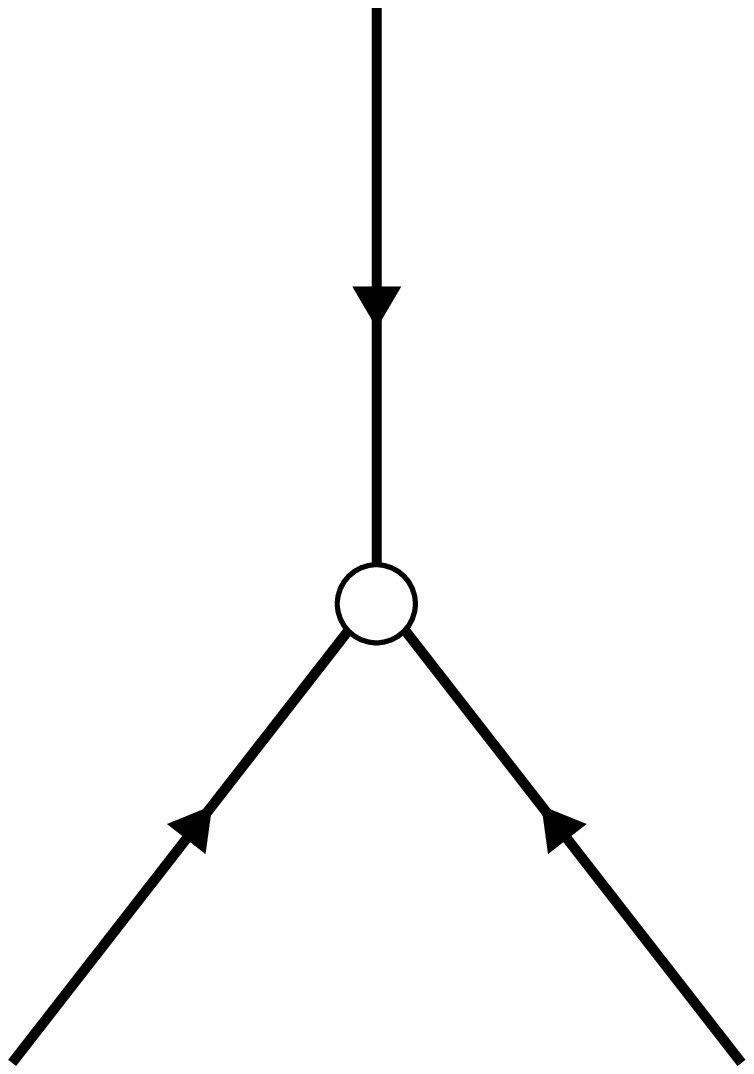}}}&\quad=\quad y\quad\vcenter{\hbox{\includegraphics[scale=0.2]{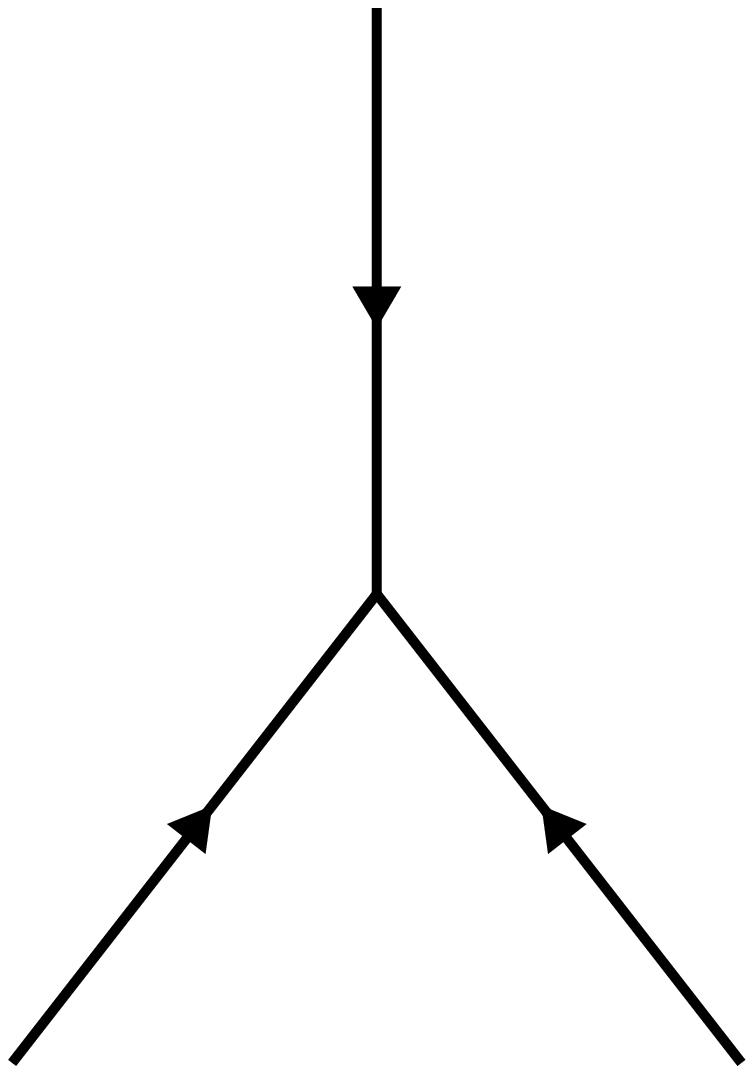}}}
\end{align}
and
\begin{align}
    \vcenter{\hbox{\includegraphics[scale=0.2]{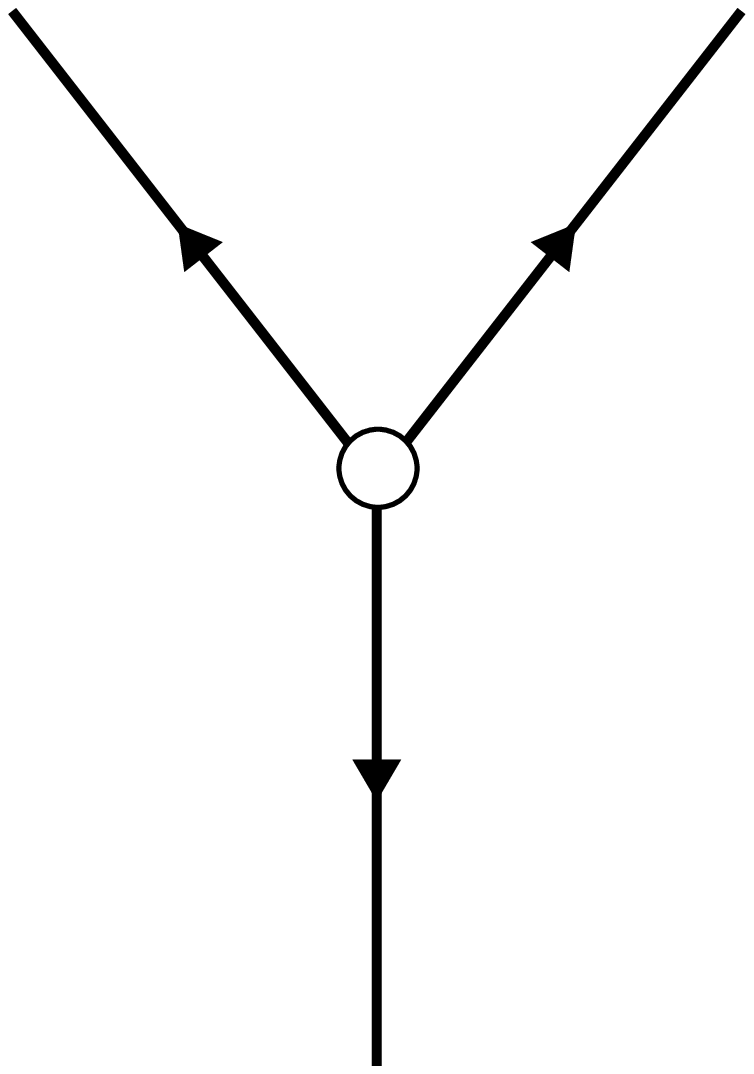}}}&\quad=\quad z\quad\vcenter{\hbox{\includegraphics[scale=0.2]{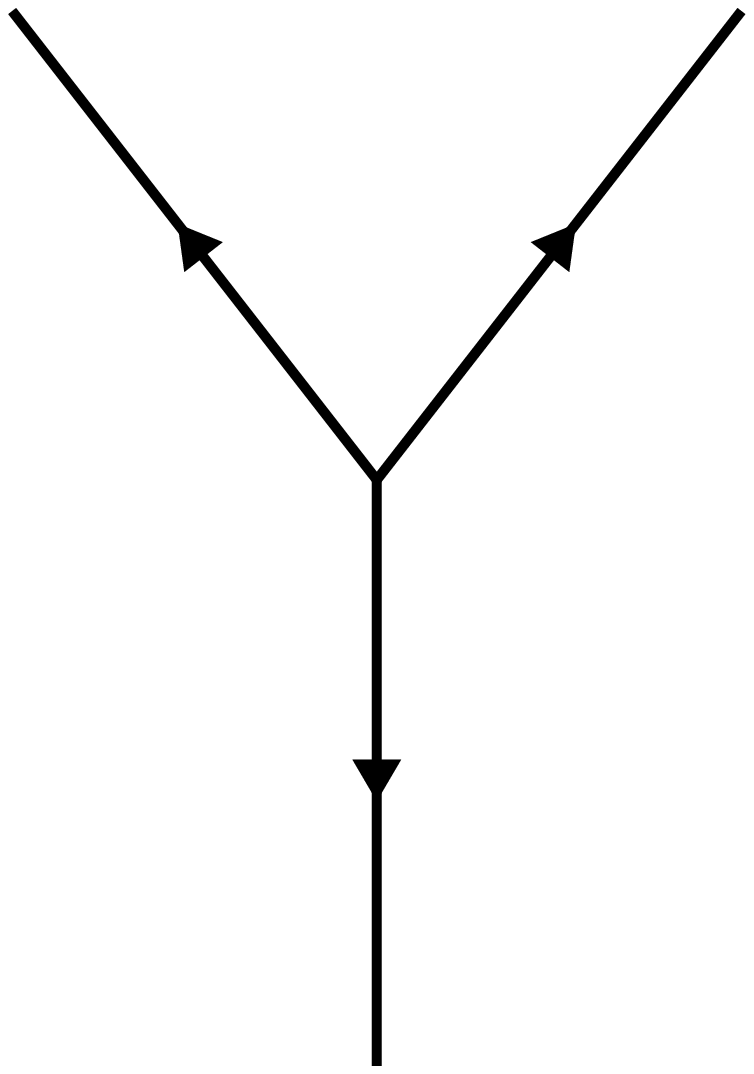}}}
\end{align}
\end{subequations} 
The computation of $w_{\rm K}'(c)$ is then made using the deformed relations 
\begin{subequations}\label{eq:deformed-rules-Kup}
\begin{align}
    \vcenter{\hbox{\includegraphics[scale=0.2]{diagrams/rel1kup.eps}}}&\quad=\quad[3]_q\\[5pt]
    \vcenter{\hbox{\includegraphics[scale=0.2]{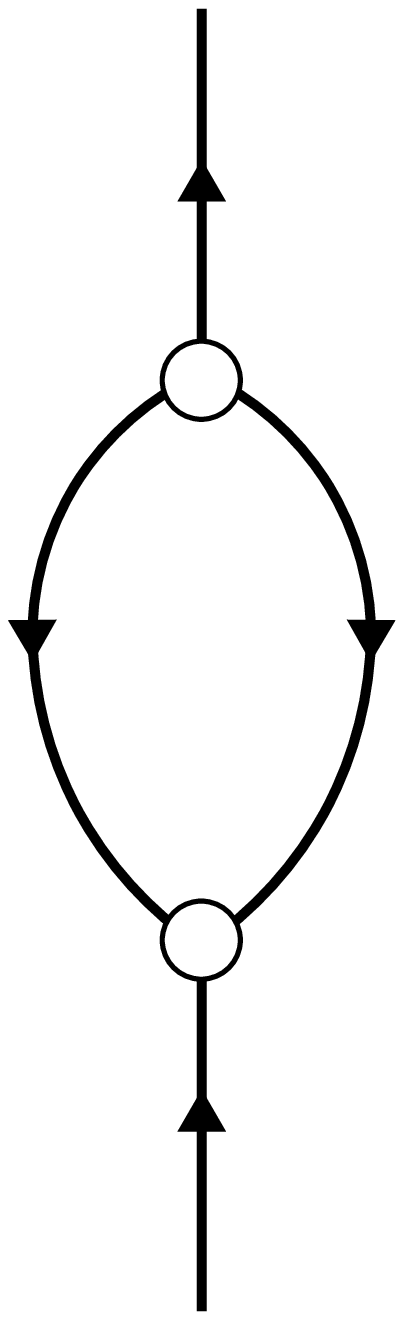}}}&\quad=\quad yz\,[2]_q\quad\vcenter{\hbox{\includegraphics[scale=0.2]{diagrams/rel2kup2.eps}}}\\[5pt]
    \vcenter{\hbox{\includegraphics[scale=0.2]{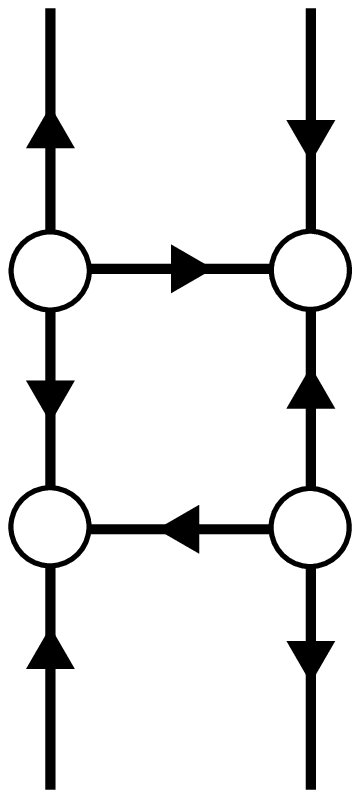}}}&\quad=\quad 
    (yz)^2
    \quad\vcenter{\hbox{\includegraphics[scale=0.2]{diagrams/rel3kup2.eps}}}\quad +\quad (yz)^2\quad\vcenter{\hbox{\includegraphics[scale=0.2]{diagrams/rel3kup3.eps}}}
\end{align}
\end{subequations}
We will then call \textit{dressed webs}, the webs that are made of the rescaled vertices~\eqref{dressed_vertices_K}. 
To evaluate the weight for closed dressed webs we use the deformed relations~\eqref{eq:deformed-rules-Kup}. By the construction, these relations  are well defined, i.e., the result does not depend on the way the dressed web is reduced, because the evaluation with the deformed rules is equivalent to the evaluation with the standard Kuperberg rules~\eqref{eq:Kup-rels} provided that we initially accounted for the vertex fugacities.
The dressed webs will turn out to provide a convenient notation in the following.

At $q=e^{i\frac{\pi}{4}}$, one has $[3]_q=1$ and $[2]_q=\sqrt{2}$. Moreover, when $yz=2^{-\frac{1}{2}}$ one obtains
\begin{subequations}
\label{specialrules}
\begin{align}
    \vcenter{\hbox{\includegraphics[scale=0.2]{diagrams/rel1kup.eps}}}&\quad=\quad 1\label{defrule1}\\[5pt] 
    \vcenter{\hbox{\includegraphics[scale=0.2]{diagrams/rel2kup1new.eps}}}&\quad=\quad\vcenter{\hbox{\includegraphics[scale=0.2]{diagrams/rel2kup2.eps}}}\label{defrule2}\\[5pt] 
    \vcenter{\hbox{\includegraphics[scale=0.2]{diagrams/rel3kup1new.eps}}}&\quad=\quad\frac{1}{2}\quad\vcenter{\hbox{\includegraphics[scale=0.2]{diagrams/rel3kup2.eps}}}\quad+\quad\frac{1}{2}\quad\vcenter{\hbox{\includegraphics[scale=0.2]{diagrams/rel3kup3.eps}}} \label{defrule3}
\end{align}
\end{subequations}

The important point in the relations \eqref{specialrules} is that for any of the three rules, the sum of prefactors of the graphs on the left-hand side is equal to the sum of prefactors of the graphs on the right-hand side. As a consequence, the partial weight $w_{\rm K}'(c)$ of any configuration is $1$, so that \eqref{wKprime} is satisfied.
It is not difficult to check this on an example:
\begin{align}
\label{reducexample}
    \vcenter{\hbox{\includegraphics[scale=0.15]{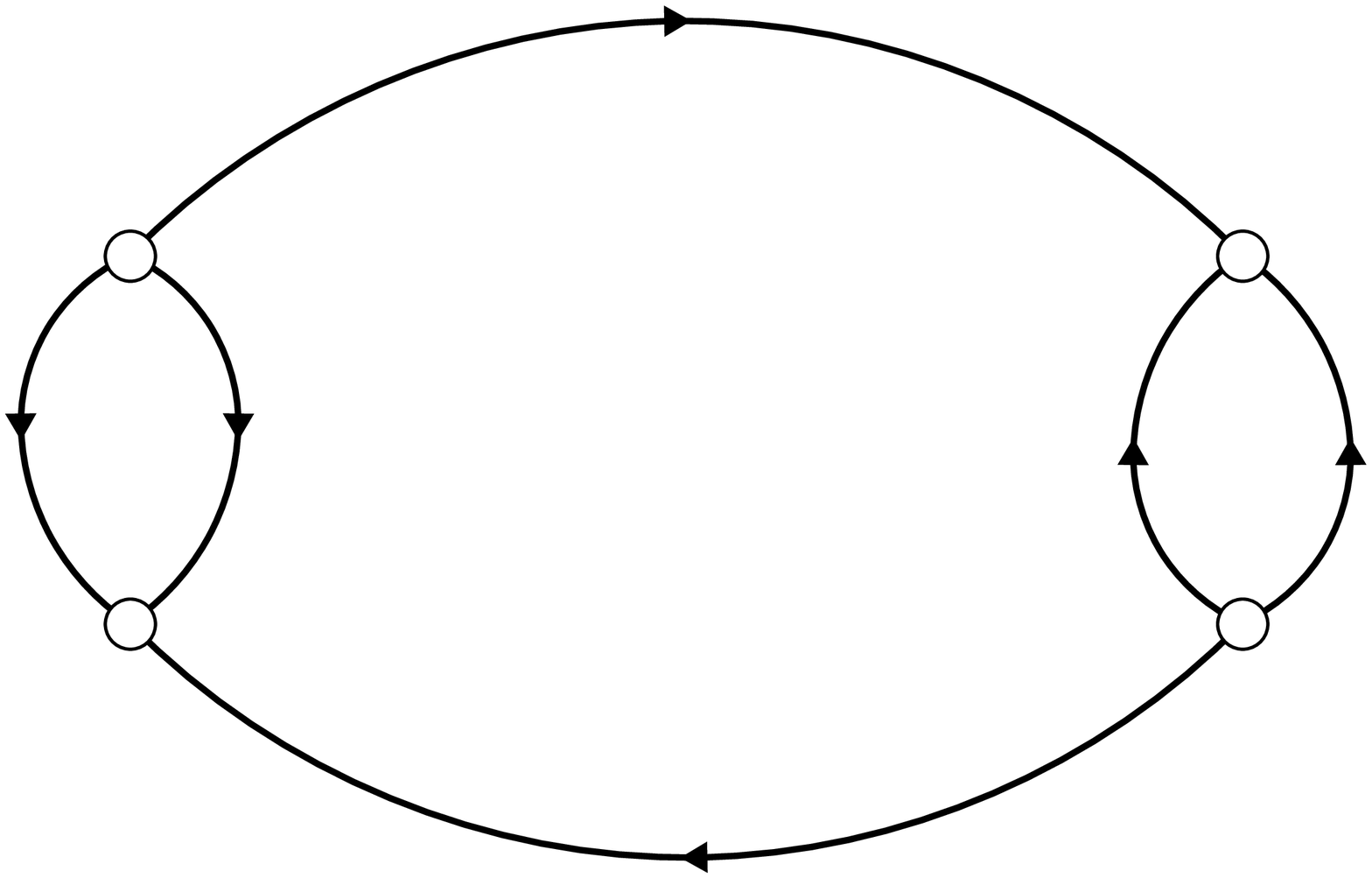}}}&\quad=\quad\frac{1}{2}\ \ \vcenter{\hbox{\includegraphics[scale=0.15]{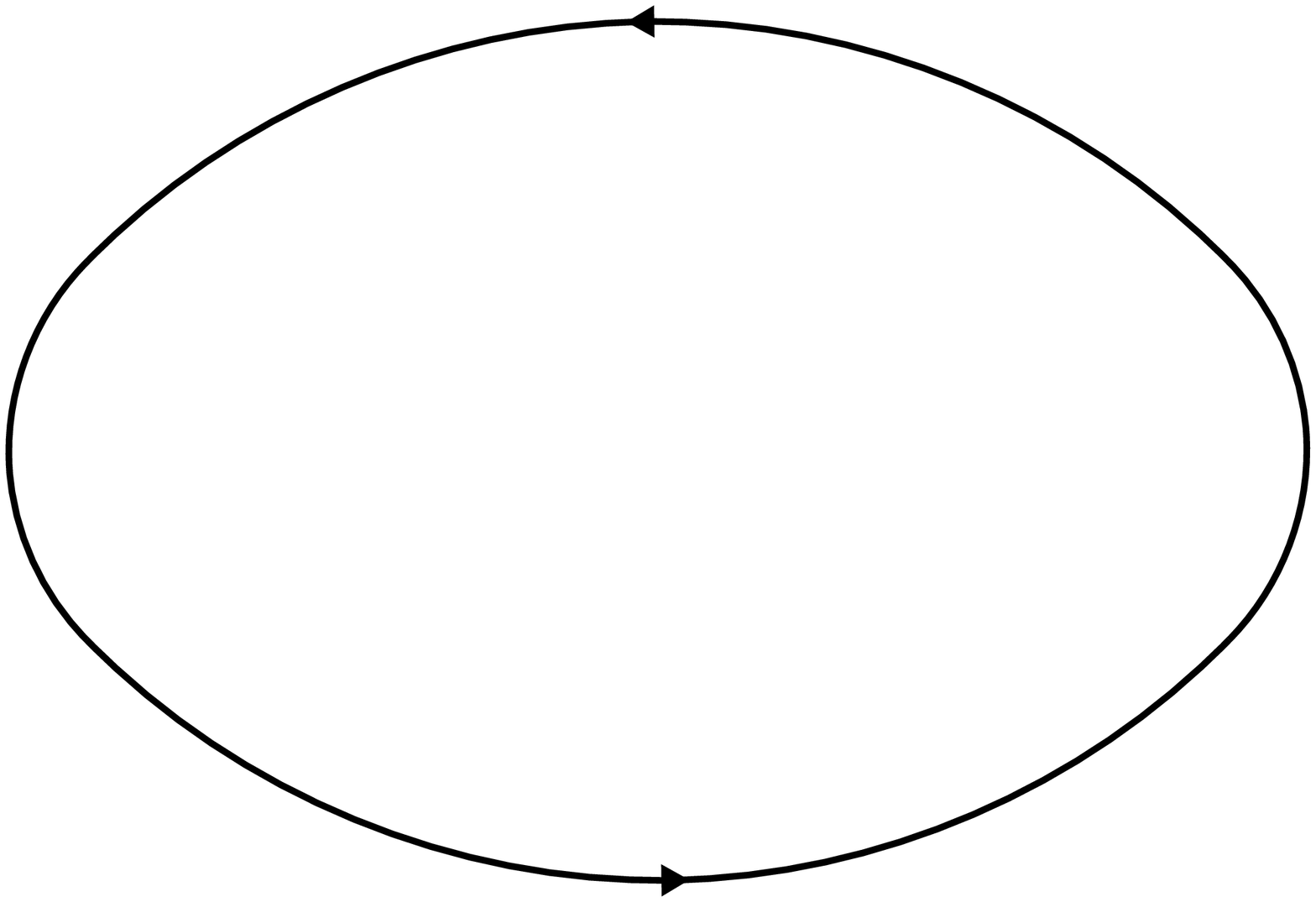}}}\quad +\quad \frac{1}{2}\ \ \vcenter{\hbox{\includegraphics[scale=0.2]{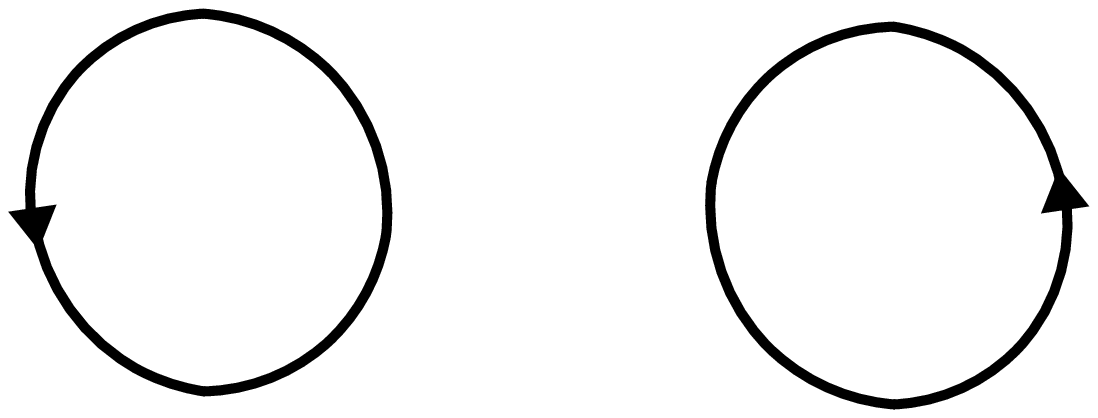}}} \quad =\quad 1 \,.
\end{align}
We have here used first the rule (\ref{defrule3}) and then (\ref{defrule1}). Alternatively, applying first (\ref{defrule2}) and then (\ref{defrule1}) results in the
same weight, as it should.

To show that $w_{\rm K}'(c) = 1$ for any $c$, consider the vector space generated by all dressed webs, including the empty one, that contribute during a reduction  process of $c$.
This is a finite-dimensional space because $c$ is a finite graph. In the example above, the space is of dimension $4$, with a basis given by the four dressed webs appearing successively in \eqref{reducexample}. Each stage appearing in the reduction process corresponds to a linear combination of dressed webs, that is,
a vector in this space. The sum of prefactors is the same on both sides of any of the reduction rules \eqref{specialrules}, implying that the sum of the components
in the basis of dressed webs remains constant upon applying any of these relations. As this sum is $1$ when we begin the reduction of a given dressed web,
it must also be $1$ when we end with the empty web. This completes the proof.

\medskip

We conclude this section with a few  brief remarks on  critical  points of the Kuperberg web model.
The $3$-state Potts model---equivalent to a $\mathbb{Z}_3$ clock model without chirality ($x_1=x_2$)---exhibits a second-order phase transition for a critical value $x_{\rm c} > 0$ of its interaction parameter, $x_1=x_2=x_{\rm c}$.%
\footnote{The model also has rich critical behaviour in the antiferromagnetic ($-1 < x < 0$) and unphysical ($x < -1$) parts of the parameter space.
This behaviour is only partially understood, and moreover depends crucially on the representation chosen (spins, loops, heights, etc.) \cite{Salas}.}
Hence, at the corresponding point, the Kuperberg web model is critical as well. This is analogous to the situation encountered in the loop case with the Ising model.
Thus we expect this critical point of the web model to be part of a whole critical submanifold of its parameter space, obtained by varying the deformation
parameter $q$ and presumably adjusting the vertex fugacities $y, z$ accordingly.
This submanifold cannot coincide with that of the critical $Q$-state Potts model (although they intersect at $Q=3$, as we have seen),
because the latter does not generically possess a $U_q(\mathfrak{sl}_3)$ symmetry.
Moreover, observe that for $x_1=x_2=1$, the $\mathbb{Z}_3$ spin model describes (equally weighted) three-colourings of the triangular lattice.
We can then ask whether the web model may lead to interesting critical observables for the latter.
Recall that  for $n=2$ the Ising model in the corresponding equally-weighted limit is related to percolation hulls.
Investigation of the phase content of the Kuperberg web model will be done numerically thanks to a local transfer matrix formulation in a forthcoming paper \cite{loc}.

\medskip

The following sections, though technically more demanding, are a direct generalisation of the above construction to the higher-rank cases.

\section{\U\ web models}
\label{sec:Cautis}

We now extend the definition of web models to the \U\ case. This requires a new setup, based on the spider of Cautis {\em et al.}~\cite{Cautis_2014}.

\subsection{Definition of the models}
\label{sec:def-of-models}

The stage is the same as before. The models are defined by weighting subgraphs of an underlying hexagonal lattice $\mathbb{H}$. This lattice is again embedded in a cylinder, such that one third of its links are parallel to the axis of the cylinder, which we orient as before. The subgraphs in question, which we shall still call
{\em webs}, have been defined in \cite{Cautis_2014} in connection with the representation theory of \U. As for $n=3$, webs are closed, oriented, planar
and trivalent graphs, but unlike the $n=3$ case they are not necessarily bipartite anymore. In addition, the edges now carry a label $i\in \llbracket 1,n-1 \rrbracket$,
which can be thought of as an integer flow.%
\footnote{We use the notation $\llbracket a,b \rrbracket$ for the set of integers $\{a,a+1,\ldots,b-1,b\}$.}
On the representation theoretical side, the label is indicating a fundamental representation of \U, while the orientation of the edge specifies the
presence of the fundamental representation itself or of its dual---the interested reader can find more details on this in \cite{Cautis_2014}.

\begin{figure}
\begin{center}
    \includegraphics[scale=0.25]{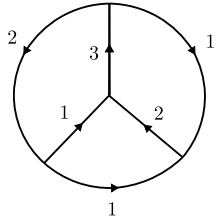}
\end{center}
    \caption{A simple web that is not bipartite, for $n\geq 4$.}
    \label{fig:simple-web}
\end{figure}

The graphs are generated by the following trivalent and bivalent vertices
\begin{center}
\includegraphics[scale=0.2]{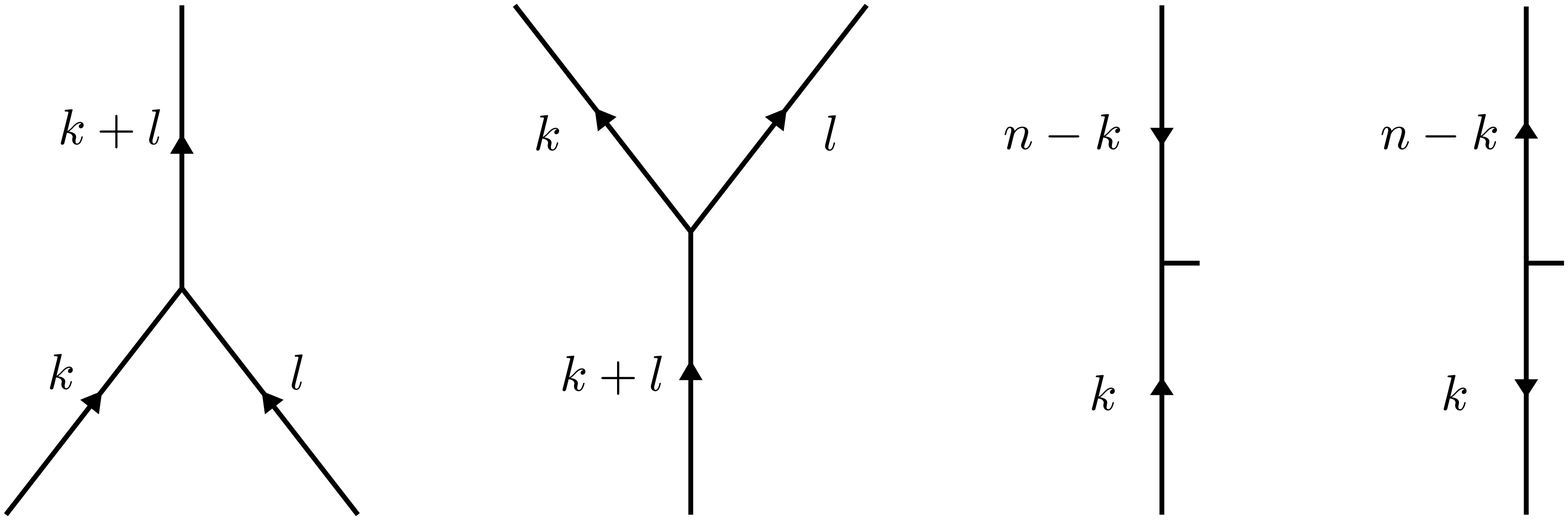}
\end{center}
where the last two bivalent vertices will be called {\em tags}. Note that trivalent vertices conserve the flow label strictly, whereas tags conserve it 
only modulo $n$. The tags come in two different variants, which can be considered sources and sinks of $n$ units of flow. Moreover, as we shall
see below, there is a distinction between tags placed on the left or right side of an edge.
Webs without tags will be referred to as {\em simple webs}. An example of a simple web is shown in Figure~\ref{fig:simple-web}; notice in particular
that this graph is not bipartite. 
As before, there exist well-defined reduction rules to assign a number to any web~\cite{Cautis_2014}:
\begin{subequations}
\label{spiderrules}
\begin{align}
    \vcenter{\hbox{\includegraphics[scale=0.2]{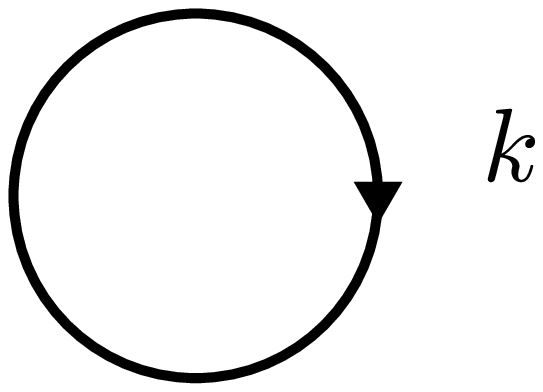}}}&\;=\;\qbinom{n}{k}_q\label{eq1}\\[5pt]
    \vcenter{\hbox{\includegraphics[scale=0.2]{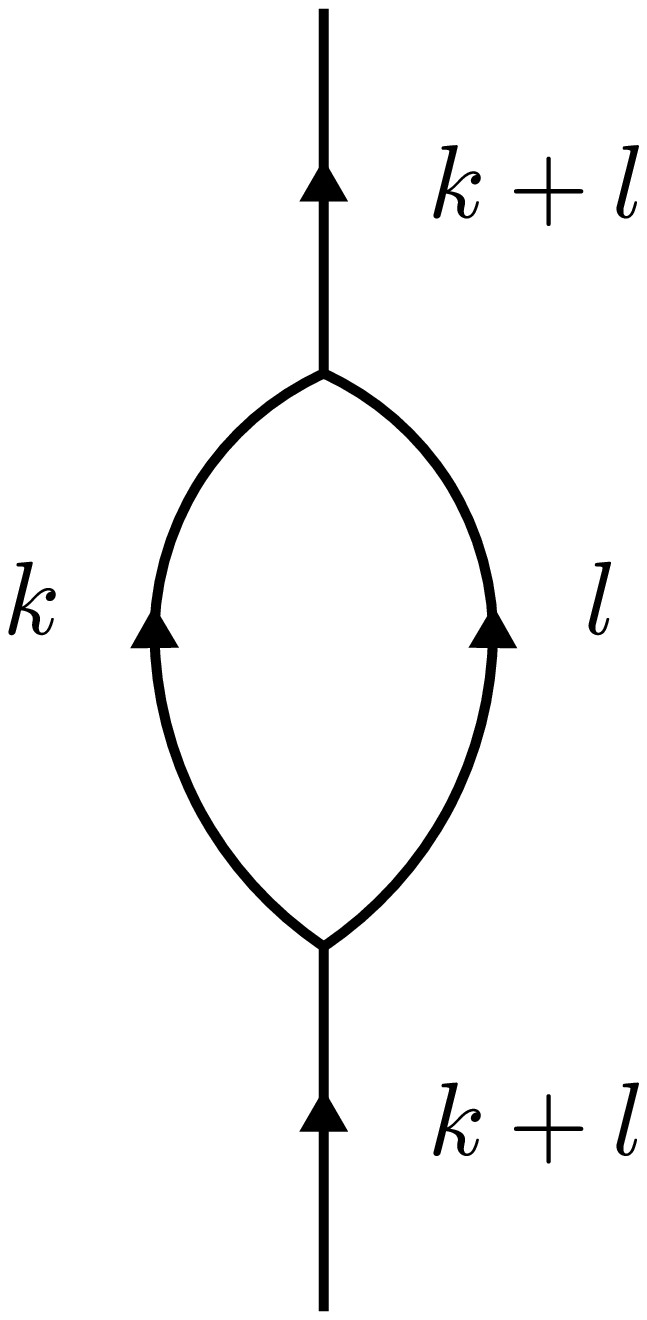}}}&\;=\;\qbinom{k+l}{k}_q\vcenter{\hbox{\includegraphics[scale=0.2]{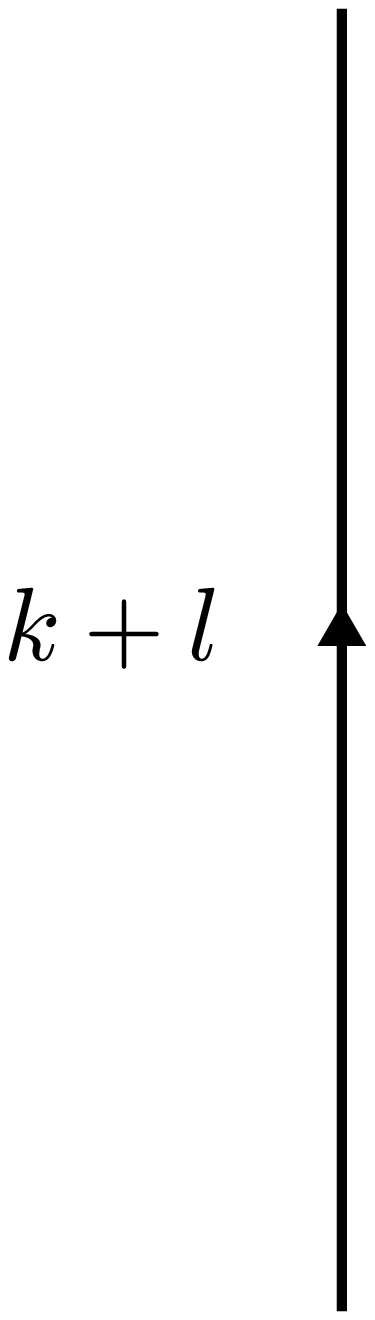}}}\label{eq2}\\[5pt]
    \vcenter{\hbox{\includegraphics[scale=0.2]{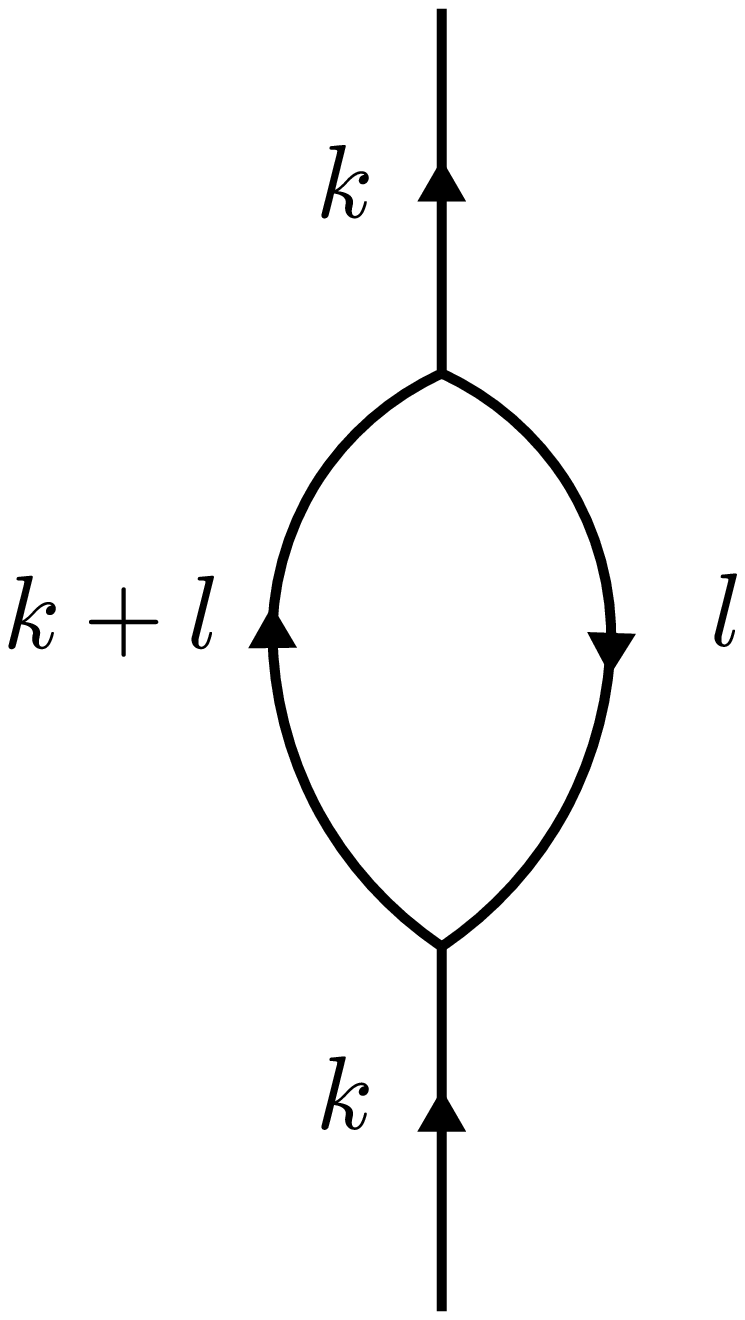}}}&\;=\;\qbinom{n-k}{l}_q\vcenter{\hbox{\includegraphics[scale=0.2]{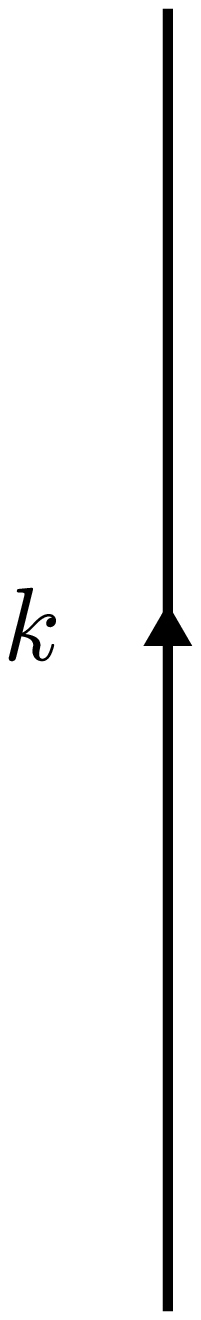}}}\label{eq3} 
\end{align}
including an associativity rule:
\begin{align}
    \vcenter{\hbox{\includegraphics[scale=0.2]{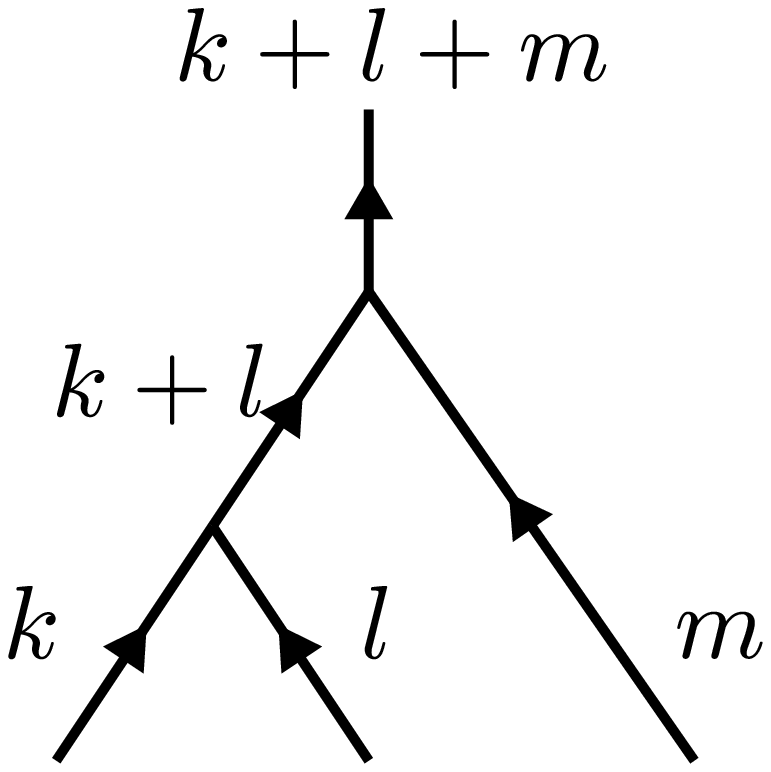}}}&\;\;=\;\;\vcenter{\hbox{\includegraphics[scale=0.2]{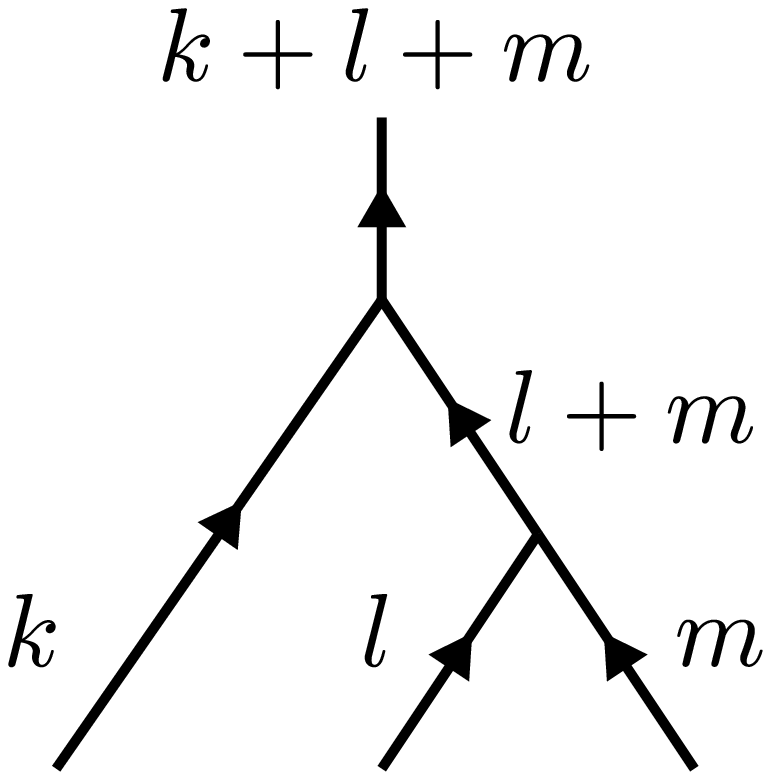}}}\label{eq4} 
\end{align}
a square rule:
\begin{align}
    \vcenter{\hbox{\includegraphics[scale=0.2]{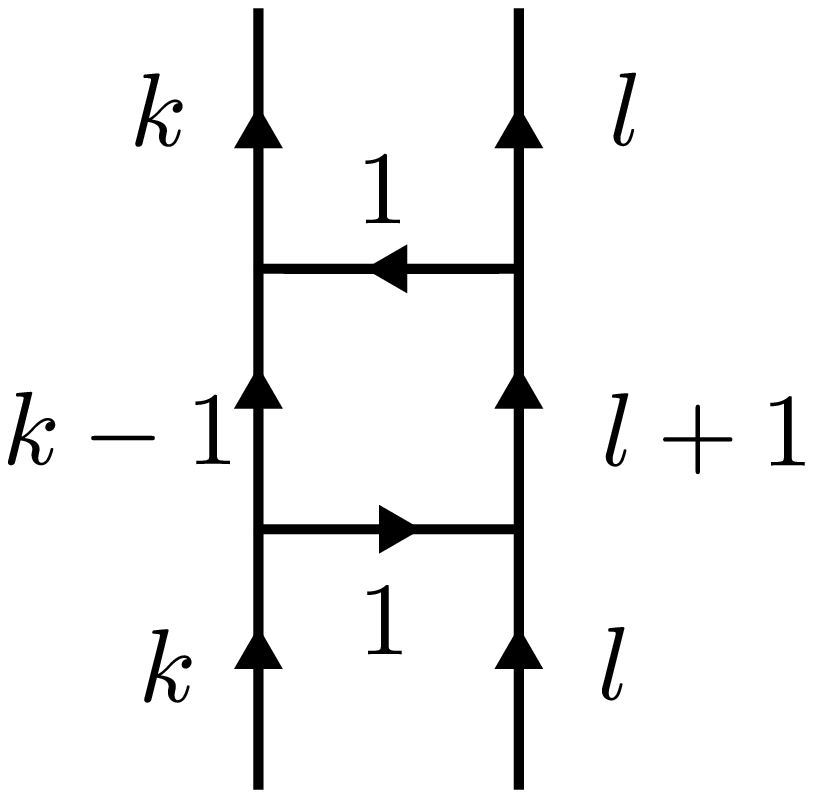}}}&\;\;=\;\;\vcenter{\hbox{\includegraphics[scale=0.2]{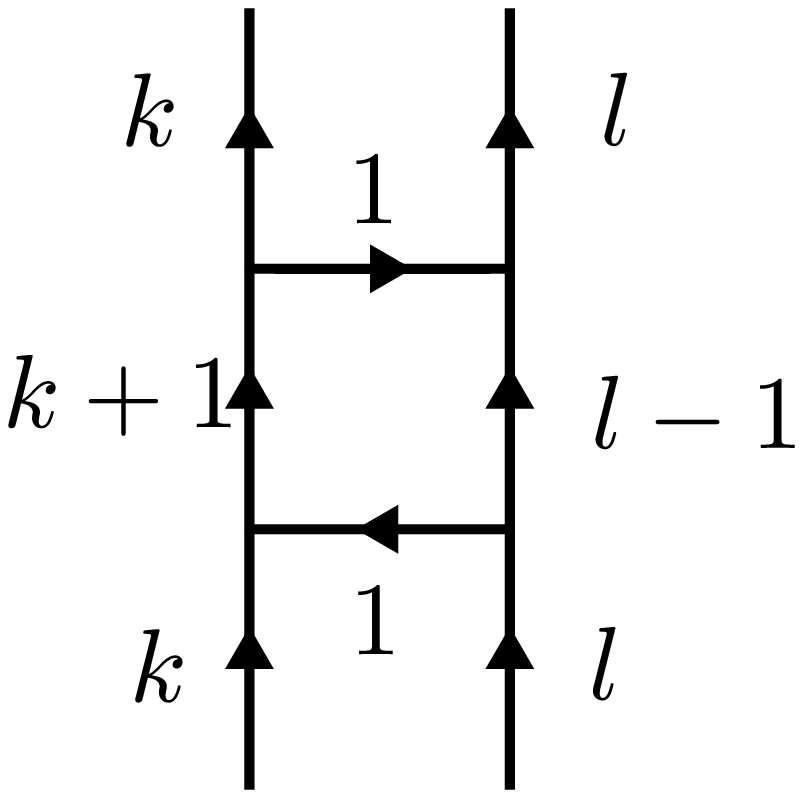}}}\;+\;[k-l]_q\;\vcenter{\hbox{\includegraphics[scale=0.2]{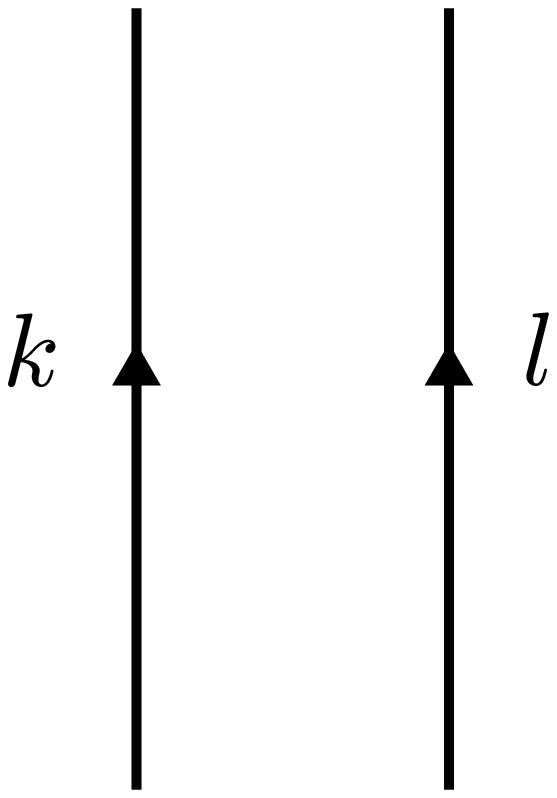}}}\label{eq9} 
\end{align}
and the tag rules:
\begin{align}
    \vcenter{\hbox{\includegraphics[scale=0.17]{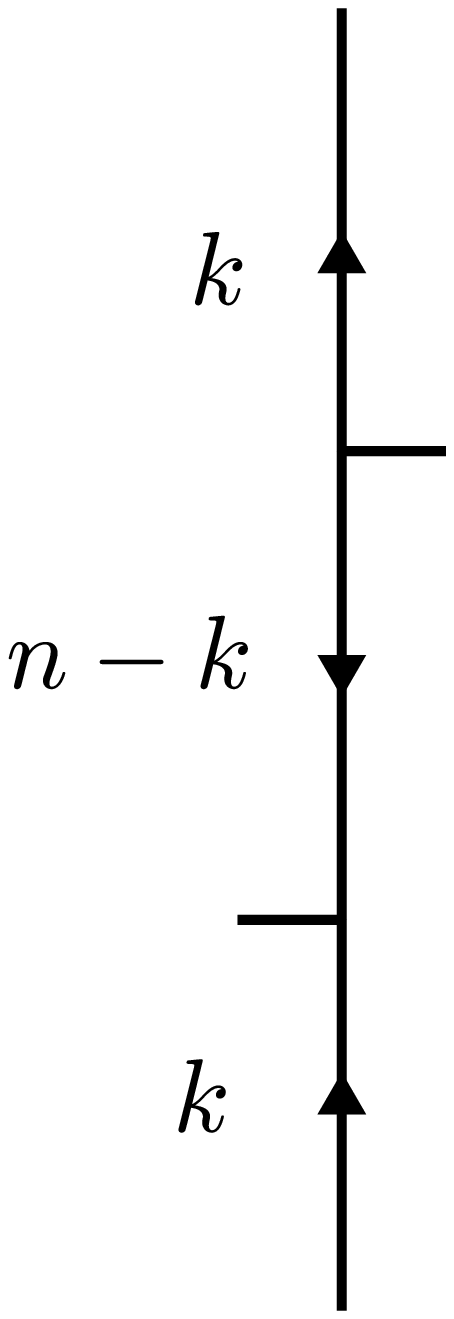}}}&\;\;=\;\;\vcenter{\hbox{\includegraphics[scale=0.17]{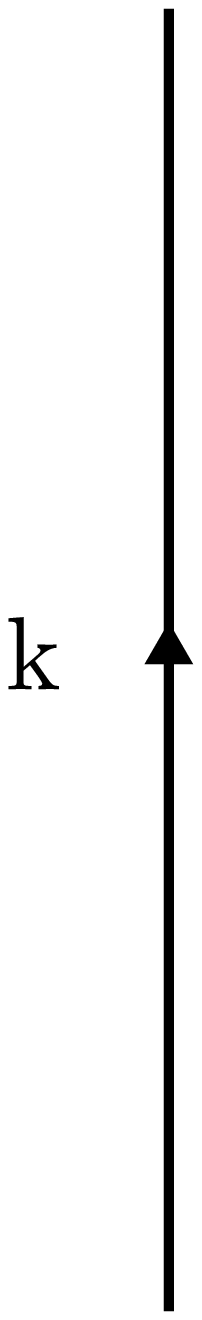}}}\label{eq6}\\[5pt]
    \vcenter{\hbox{\includegraphics[scale=0.2]{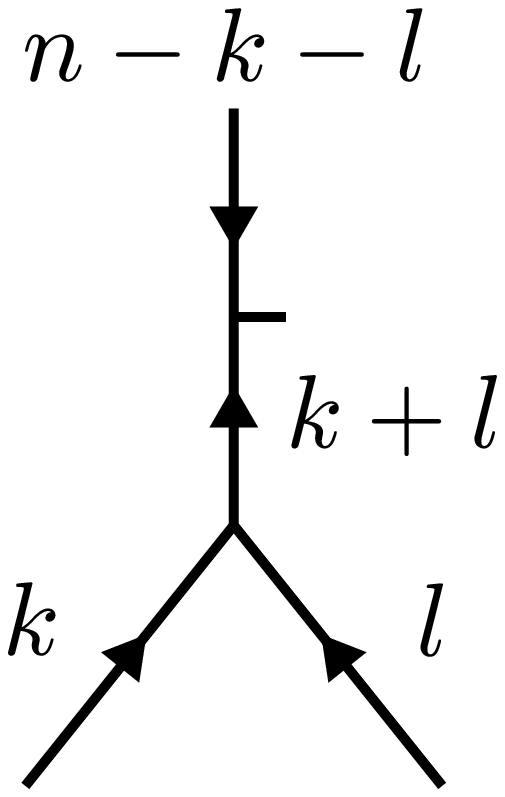}}}&\;\;=\;\;\vcenter{\hbox{\includegraphics[scale=0.2]{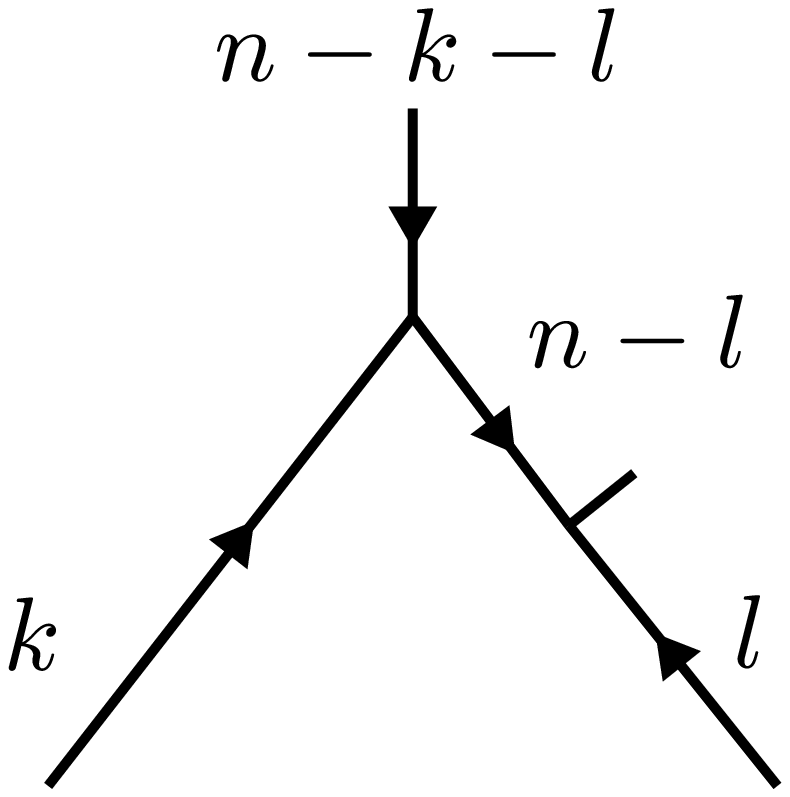}}}\label{eq7}\\[5pt]
    \vcenter{\hbox{\includegraphics[scale=0.2]{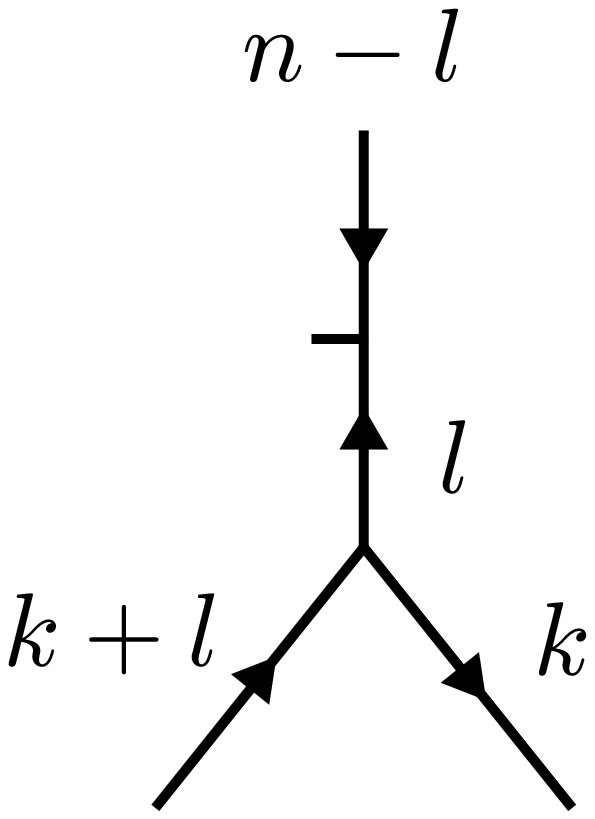}}}&\;\;=\;\;\vcenter{\hbox{\includegraphics[scale=0.2]{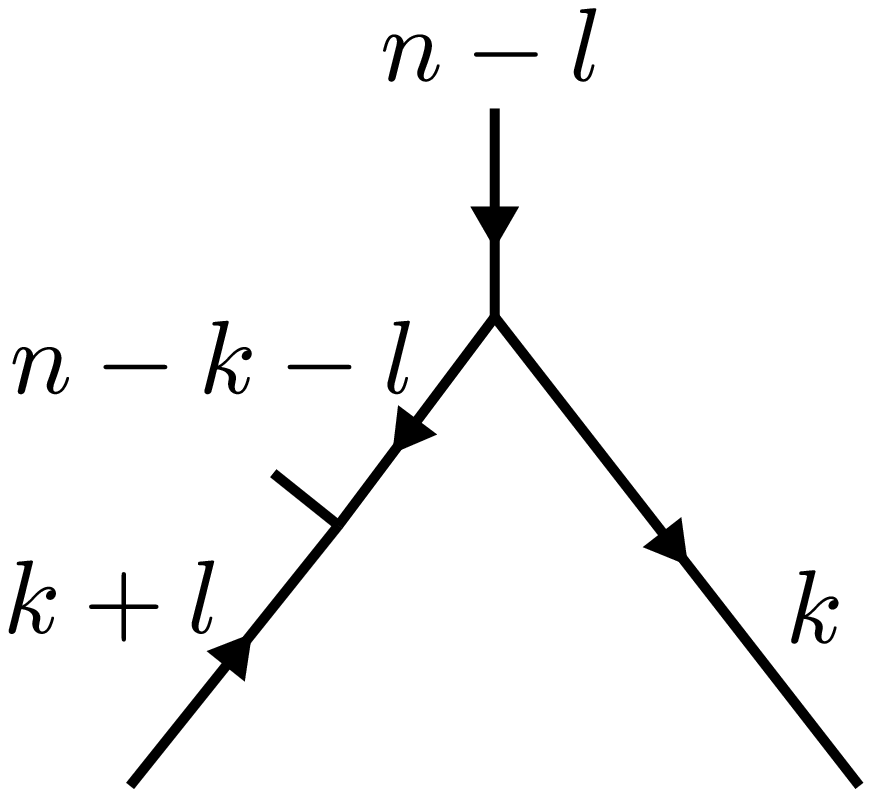}}}\label{eq8}\\[5pt]
     \vcenter{\hbox{\includegraphics[scale=0.17]{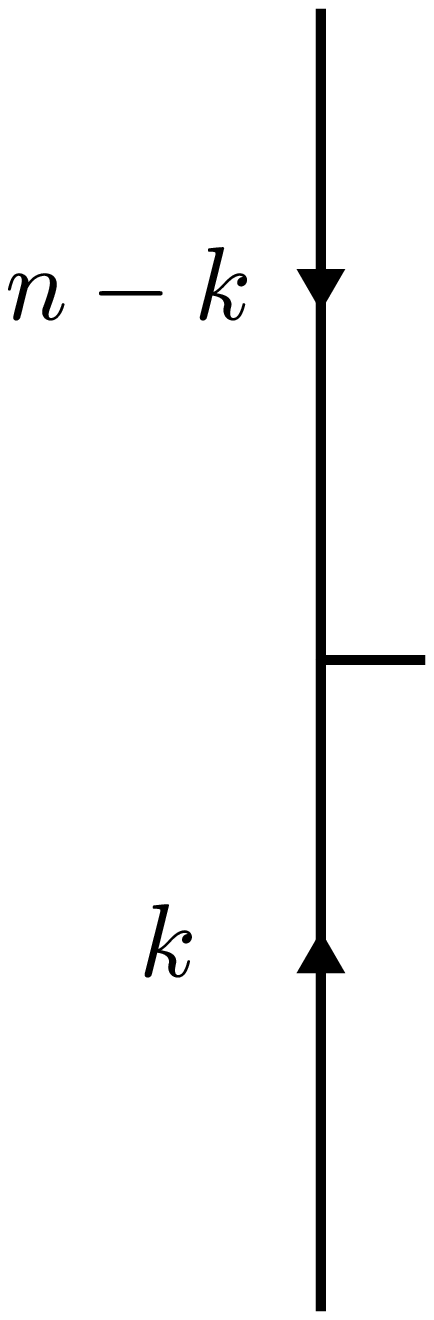}}}&\;\;=\;\;(-1)^{k(n-k)}\vcenter{\hbox{\includegraphics[scale=0.17]{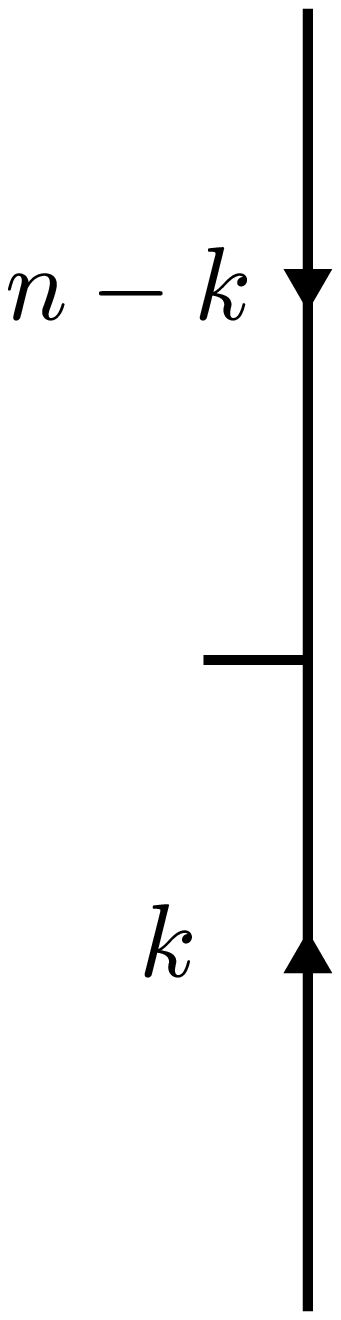}}}\label{eq5}
\end{align}
\end{subequations}
together with their mirrored and the arrow-reversed versions.
By arrow reversal, we mean the simultaneous change of all arrow orientations as well as the interchange of the tag orientations (i.e., the tags are flipped).
All edge labels take values in $\llbracket 1,n-1 \rrbracket$, except for \eqref{eq9} where edges in the interior ``square'' are allowed to take values
in $\llbracket 0,n \rrbracket$. To recover the usual range, $\llbracket 1,n-1 \rrbracket$, for all labels we delete any edge with label $0$ from the web, 
whereas an edge with label $n$ is replaced by a tag according to the rules
\begin{subequations}
\label{prescription}
\begin{align}
    \vcenter{\hbox{\includegraphics[scale=0.2]{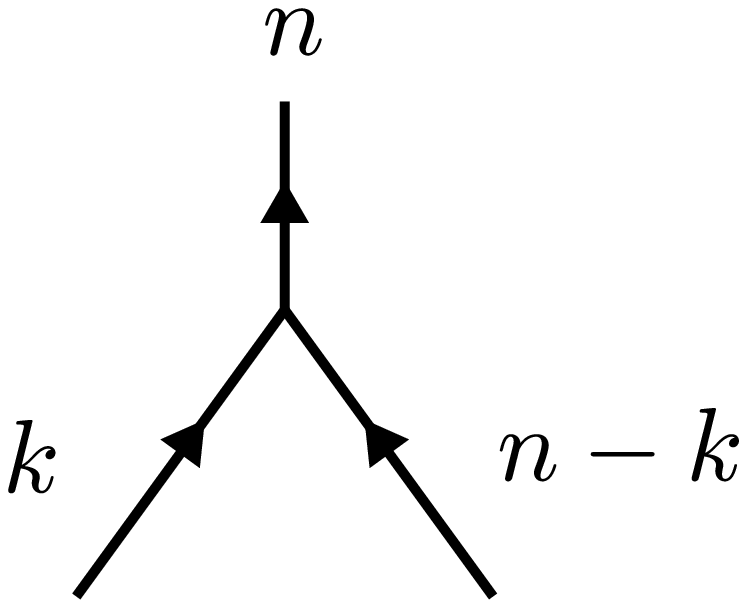}}}&\;=\;\vcenter{\hbox{\includegraphics[scale=0.2]{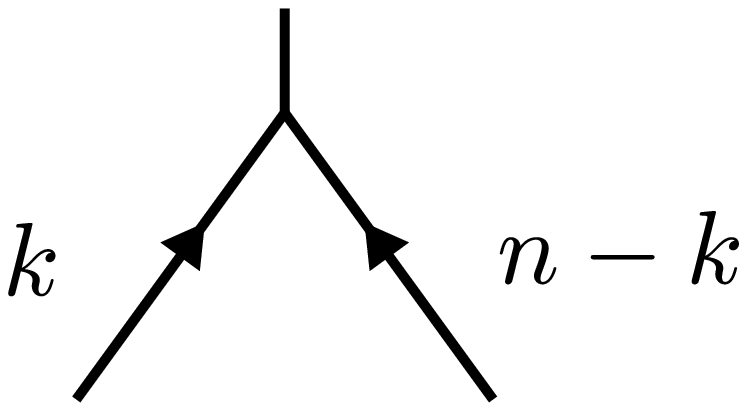}}}\\[5pt]
    \vcenter{\hbox{\includegraphics[scale=0.2]{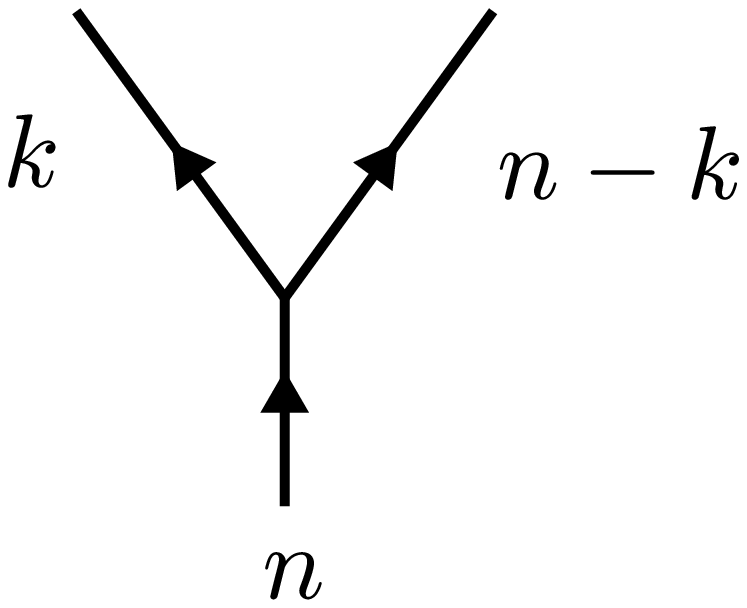}}}&\;=\;\vcenter{\hbox{\includegraphics[scale=0.2]{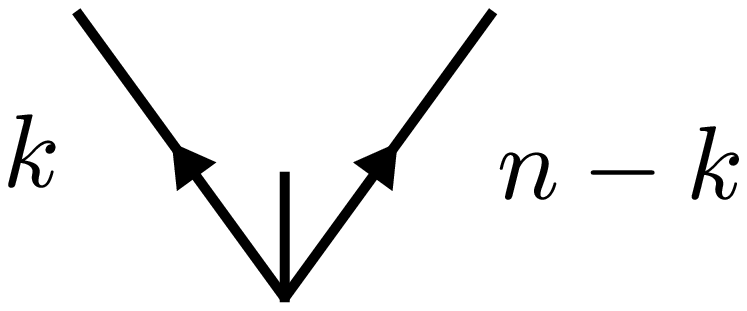}}}
\end{align}
\end{subequations}

We note that the set of rules~\eqref{spiderrules} is smaller than the set presented in \cite[Sec.\,2.2]{Cautis_2014}, but the two sets are nevertheless equivalent (as explained in the remarks on page 8 of \cite{Cautis_2014}). The extended set of rules in~\cite{Cautis_2014} might however be useful in actual computations. For example, the square diagram appearing on the left-hand side of~\eqref{eq9}, and with the horizontal edges labelled by some integer flows $r$ and $s$, is reduced\footnote{The reduction uses first the ``fusion of flows" rule~\cite[Eq.\,(2.9)]{Cautis_2014} that follows from~\eqref{eq2} and~\eqref{eq4}, and then a repeated application of~\eqref{eq9}.}  to a linear combination of squares with the flows $r-t$ and $s-t$ (for all allowed positive $t$) at the horizontal edges, and so on. However, a general practical algorithm for reducing any web is not know, as pointed out in~\cite{Cautis_2014}. It is a highly non-trivial algebraic result of the work~\cite{Cautis_2014} that the rules~\eqref{spiderrules}  are indeed enough to perform a complete reduction of any web.
In this paper, we do not need an explicit general algorithm but we will come back to the evaluation problem within a transfer-matrix formalism in the next paper.

\medskip

Our statistical model will be defined in terms of closed simple webs. The tag rules will be needed only in order to show that the  model is well defined.
In contrast to the Kuperberg web model (defined in Section \ref{sec:Kuperberg}), the allowed configurations are the {\em equivalence classes}
of embedded simple webs, under an equivalence relation that we now describe.

\begin{figure}
\begin{center}
    \includegraphics[scale=0.3]{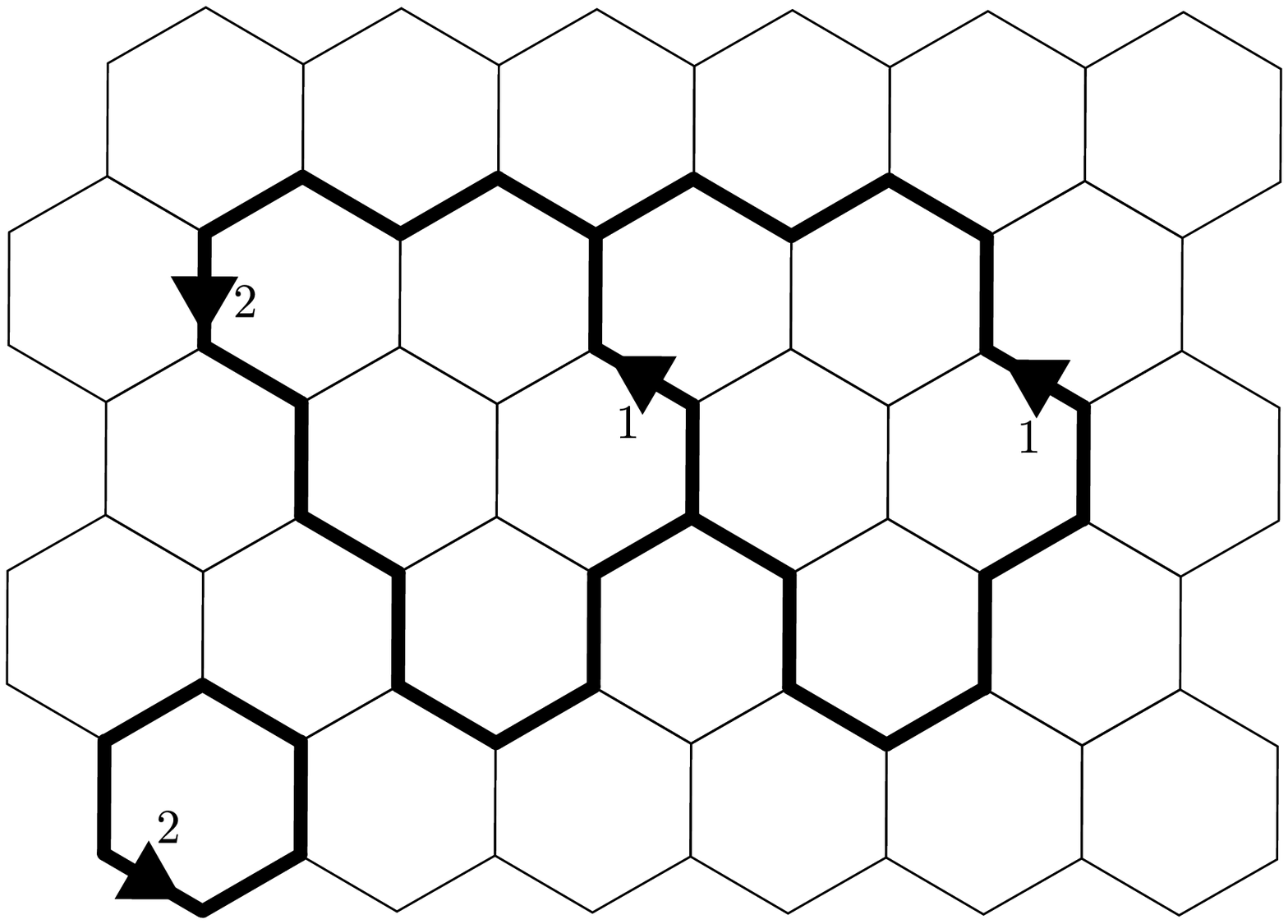} \quad \includegraphics[scale=0.3]{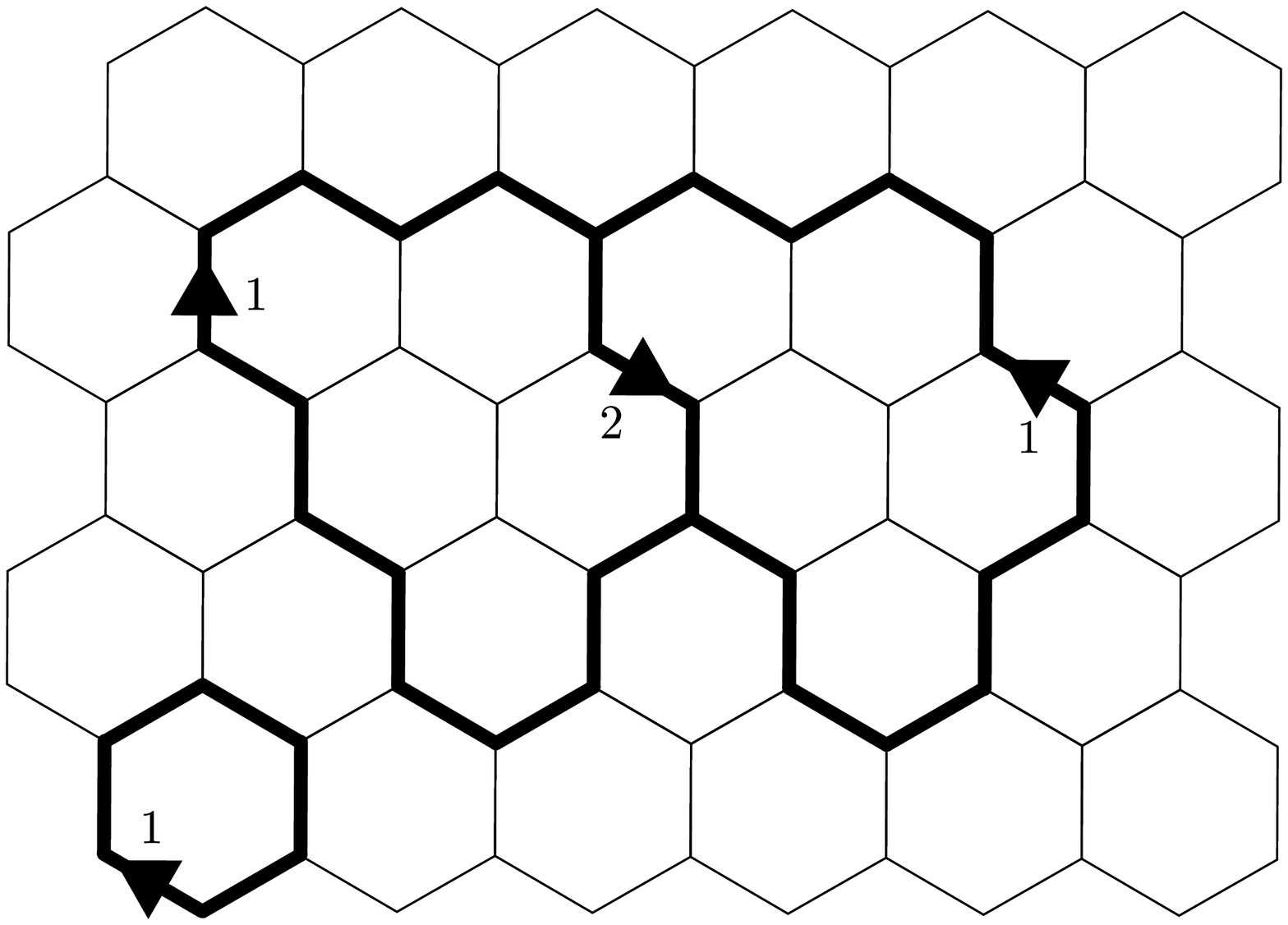}
        \end{center}
    \caption{Two equivalent simple webs (for the case $n=3$). Note that taking into account~\eqref{constraints} they have  equal weight $x_1^{11}x_{-1}^5x_2^8x_{-2}^{11}y_{1,1;2}y_{2;1,1}[2]_q[3]_q^2=x_1^{19}x_{-1}^{13}x_{-2}^3y_{1,1;2}y_{2;1,1}[2]_q[3]_q^2$.}
\label{webconf}
\end{figure}

We define the {\em CPT transformation} of an edge as the simultaneous reversal of its orientation and the replacement $k \mapsto n-k$ of
its flow label. In general, applying a CPT transformation to a single edge of a simple web will wreck the flow conservation at its adjacent
vertices, hence leading to a graph which is no longer a simple web. However, the simultaneous CPT transformations of two edges adjacent
to a common vertex $v$, one entering the vertex and the other exiting from it, will maintain the flow conservation at $v$. Define
a {\em transition cycle} of a simple web to be a closed path of consistently oriented edges (each edge entering a given vertex being followed
by another edge exiting from it, or vice versa). It follows that the simultaneous CPT transformation of each edge in a transition cycle
will transform the simple web into another simple web. We may now define the equivalence relation. Two simple webs, $w_1$ and $w_2$,
are equivalent (and we write $w_1 \sim w_2$) if they are related by the CPT transformation of a finite sequence of transition cycles.
Figure \ref{webconf} provides an example of two equivalent simple webs.

The CPT transformation is motivated by representation theoretical considerations. Indeed, the fundamental representation labelled by $k$
is isomorphic to the dual of the one labelled by $n-k$. The isomorphism and its inverse are unique up to a scalar factor and are given on the
spider side by the tags, up to a scalar. In a particle physics analogy, this means that we do not distinguish between a particle
flowing in some direction and its antiparticle flowing in the opposite direction. This analogy moreover justifies the name `CPT transformation'.

Let $[c]$ denote the equivalence class of the simple web $c$: we have $c_1,c_2 \in [c]$ if and only if $c_1 \sim c_2$.
Let ${\cal C}$ denote the set of all equivalence classes of simple webs on $\mathbb{H}$. Our web model is defined
over the configuration space ${\cal C}$, by assigning a statistical weight $w([c])$ to each $[c] \in {\cal C}$.

The weight $w([c])$ is the product of two factors, a local weight and a non-local weight. Let $w_s(c)$ denote the non-local weight assigned to
a simple web by the rules (\ref{spiderrules}). In Appendix~\ref{sec:app-equiv-weights}, we
invoke the tag rules and
 prove that
\begin{equation} \label{appAeq}
 c_1 \sim c_2 \Rightarrow w_s(c_1) = w_s(c_2) \,.
\end{equation}
We can therefore write $w_s([c])$ for the non-local part of the weight for the given  class $[c]$.

The local part of the weight consists of fugacities for vertices and bonds. To a bond covered by an edge labelled by $k$ flowing upward
(resp.\ downward) we assign the fugacity $x_k$ (resp.~$x_{-k}$). Due to the flow conservation, a vertex must be adjacent to either two
incoming edges and one outgoing edge, or to two outgoing edges and one incoming edge. In the former case, for a vertex with
edges labelled $k$ and $l$ flowing inward and an edge $k+l$ flowing outward, and the three of them being arranged in this order anticlockwise around the vertex,
 we assign the fugacity $y_{k,l;k+l}$. 
In the latter case, for a vertex with edges labelled $k$ and $l$ flowing outward and an edge $k+l$ flowing inward, arranged in this order clockwise around the vertex, we assign the fugacity $y_{k+l;k,l}$.\footnote{Note that in both cases, the labels of incoming (resp.\ outgoing) edges are written before (resp.\ after) the semicolon.} 
Moreover, we demand that the vertex fugacities be \textsl{rotationally invariant}, which means that two  vertices related to each other by a transformation under the full symmetry group of the lattice $\mathbb{H}$ get  the same fugacity. In other words, the vertex fugacity is independant of the embedding of a given web into $\mathbb{H}$. For the local part of the weight to be independent of the chosen representative $c \in [c]$ of the equivalence class $[c]$, we also impose the following constraints:
\begin{subequations}
\label{constraints}
\begin{align}
    &x_{-k}=x_{n-k}\quad &k \in \llbracket 1,n-1 \rrbracket \\
    &y_{k,l;m}=y_{l,n-m;n-k}=y_{n-m,k;n-l}\quad &m=k+l\text{ and }k,l,m \in \llbracket 1,n-1 \rrbracket \\
    &y_{m;k,l}=y_{n-k;l,n-m}=y_{n-l;n-m,k}\quad &m=k+l\text{ and }k,l,m \in \llbracket 1,n-1 \rrbracket
\end{align}
\end{subequations}

Note that here the rotational invariance is used implicitly: after we make a CPT transformation the vertex will have new labels, but will also be rotated.\footnote{We note that we however do not require rotational invariance for the bond fugacities.}

The total weight given to a configuration $[c] \in {\cal C}$ is then
\begin{equation} \label{total_weight}
    w([c])= \left(\prod_{k\in \llbracket -n+1,n-1 \rrbracket} x_k^{N_k}\right) \left(\prod_{\substack{k,l\in \llbracket 1,n-1 \rrbracket \\ k+l\in \llbracket 1,n-1 \rrbracket}} y_{k,l;k+l}^{N_{k,l;k+l}} \right) \left( \prod_{\substack{k,l\in \llbracket 1,n-1 \rrbracket \\ k+l\in \llbracket 1,n-1 \rrbracket}} y_{k+l;k,l}^{N_{k+l;k,l}} \right) w_s([c]) \,,
\end{equation}
where as usual $N_{\cdots}$ denotes the number of occurrences of the corresponding bond or vertex types.
We use the convenient notation $x_0=1$, which can be understood as normalising the weight by fixing the fugacity of an empty link.

We can finally write the partition functions defining the web models:
\begin{align}
    Z=\sum_{[c]\in \mathcal{C}} w([c]) \,.
 \label{Z_web}
\end{align}
Remark that the discrete rotational invariance of the underlying lattice $\mathbb{H}$ is recovered if we choose $x_k=x_{n-k}$ for any $k \in \llbracket 1,n-1 \rrbracket$.

\subsection{Equivalence between the $U_q(\mathfrak{sl}_3)$ web model and the Kuperberg web model}

The $U_q(\mathfrak{sl}_3)$ web model defined in the last subsection is equivalent to the Kuperberg web model of Section~\ref{sec:Kuperberg}.
This should not come as a surprise, since their building blocks, the webs, are just two ways of naming the same objects on the representation theoretical side.

To prove the equivalence, we shall proceed in two steps. First, we build a bijection $h$ between equivalence classes of $U_q(\mathfrak{sl}_3)$ simple webs
and Kuperberg webs. This induces a bijection between configurations of the two models. Second, we tune the parameters, so that each configuration
acquires the same weight in either model.

We first describe the mapping $h$. For a given simple web $c$, replace each edge carrying a label $2$ by an edge carrying label $1$ on its central part,
by introducing a pair of tags using \eqref{eq6}. Observe that the orientation of tags is immaterial, since $n=3$ being odd implies that $(-1)^{k(n-k)} = 1$
for any $k$ in \eqref{eq5}. After this transformation, the labels $2$ are now confined at the extremities of edges, next to the vertices.
Note that, again because $n$ is odd,
tags can be moved and oriented freely around a vertex, by means of relations \eqref{eq5} and \eqref{eq7}--\eqref{eq8}.
We can use this freedom to get only two different situations:
\begin{align}
   \vcenter{\hbox{\includegraphics[scale=0.2]{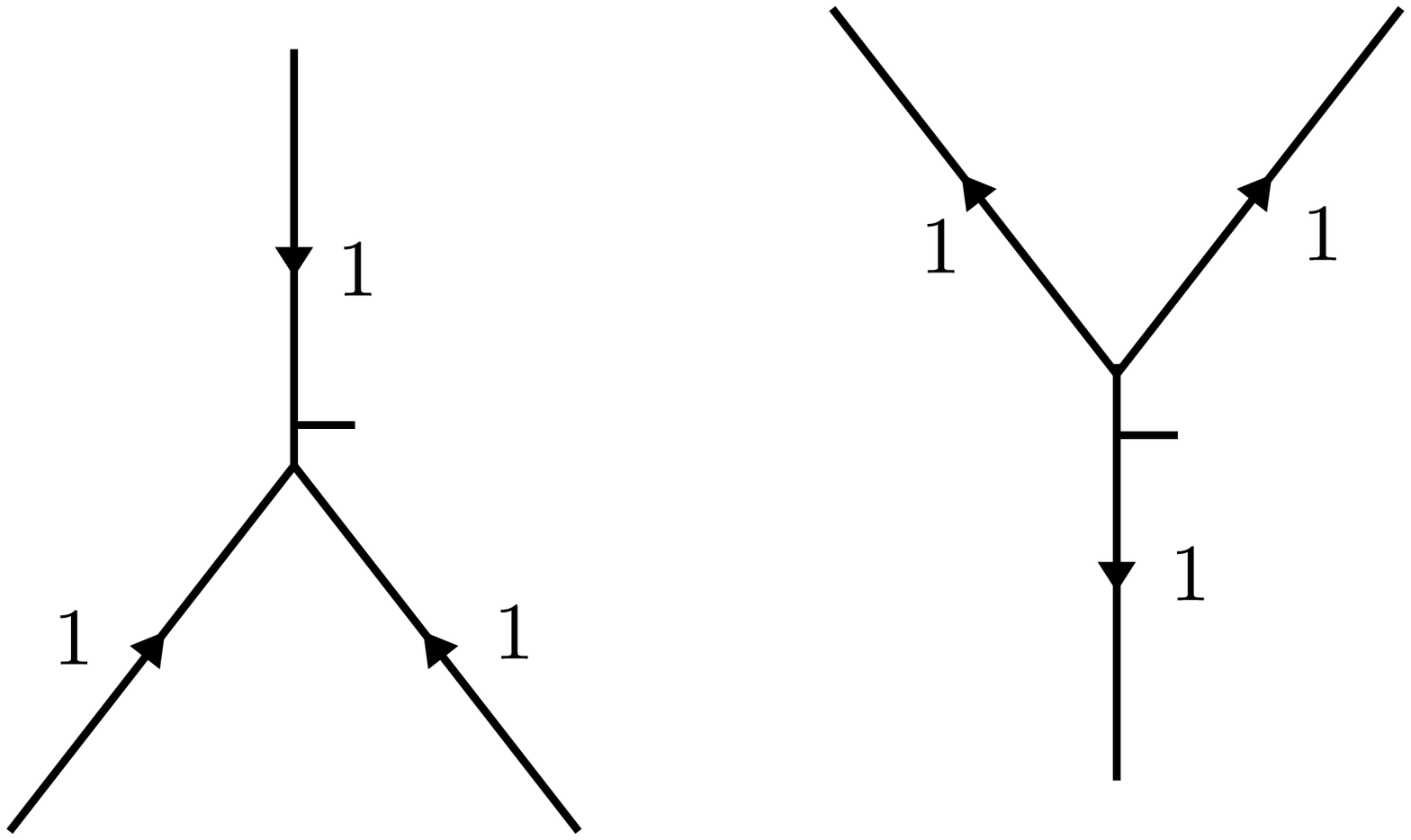}}} \label{two_tagged_vertices}
\end{align}
We thus obtain a new web $c'$, in which any vertex has the form \eqref{two_tagged_vertices}.
Moreover, two equivalent simple webs differ by transition cycles of edges carrying labels $1$ and $2$ in an alternating way, when they are not simply loops.
In either case, it is not difficult to see that one can get the same web after the procedure given above. Thus, we get the same web $c'$ starting from any
other web $\tilde{c}$ equivalent to $c$ (i.e., $\tilde{c} \sim c$). The image $h([c])$ is then defined to be the Kuperberg web obtained from $c'$ by removing tags,
as well as the labels and orientations of edges labelled by $2$. In other words, only oriented edges labelled by $1$ are kept, so that the two vertices of
\eqref{two_tagged_vertices} become respectively the sink and source vertex of the Kuperberg web model.

To construct the inverse of $h$ we first recall some results from the classical graph theoretical topic of graph coverings.
A (vertex) cycle cover of a graph $G$ is a set of cycles which are subgraphs of $G$ and contain all vertices of $G$. 
If the cycles of the cover have no vertices in common, the cover is called a {\em disjoint cycle cover}. A review of graph
covering problems can be found in \cite{Harary}. We shall need the simple result that any Kuperberg web $c'$ (being
a planar bipartite cubic graph) admits a disjoint cycle cover. Choosing an arbitrary disjoint cycle cover, we change the orientation of edges composing all of its  cycles in an alternating way. The
edges which have their orientation changed in this process are then labelled by~$2$, while all other edges of $c'$ are labelled by~$1$. It is easy to see that there is then strict flow conservation at the vertices, which are all of the type $1+1=2$. This argument makes crucial use of the bipartiteness of the graph. In the end one obtains a $U_q(\mathfrak{sl}_3)$ web~$c$ indeed.
Notice also that a different choice of cycle cover 
(in case the cycle cover is not unique) will lead to a different web $\tilde{c}$,
which will however be equivalent to the former ($\tilde{c} \sim c$), since they differ by a set of CPT transformations.
This concludes the construction of $h$ and 
it shows a bijection between the two configuration sets.

A class $[c]$ of simple $U_q(\mathfrak{sl}_3)$ webs and its image under the bijection $h([c])$, a Kuperberg web,
have the same non-local weight, $w_s([c]) = w_{\rm K}(h([c]))$, as can be seen by comparing the respective reduction rules,
\eqref{spiderrules} and \eqref{3rules}. For the local parts of the weights to be equal as well, one has to set
\begin{subequations}
\begin{eqnarray}
 y &=& y_{1,1;2} \,, \\
 z &=& y_{2;1,1} \,.
\end{eqnarray}
\end{subequations}
With this choice, we have therefore demonstrated that $Z_K$ in \eqref{Z_K} and $Z$ in \eqref{Z_web} are equal.

\subsection{Equivalence between the $U_q(\mathfrak{sl}_2)$ web model and Nienhuis' loop model}

The $U_q(\mathfrak{sl}_2)$ web model is equivalent to the well-known Nienhuis O($N$) model of dilute loops~\cite{Nienhuis} on the hexagonal lattice $\mathbb{H}$.
Indeed, observe that for $n=2$, there cannot be any vertices in the webs, since when all labels are $1$, no orientation of edges is compatible with the flow 
conservation around a vertex. Webs therefore consist only of loops, and the reduction rules \eqref{spiderrules} weight them by $N = [2]_q$, regardless of their 
orientation. By removing the labels and orientations of loops, any representative of a configuration $[c]$ of the $U_q(\mathfrak{sl}_2)$ web model produces the
same collection of unoriented loops on $\mathbb{H}$, hence a configuration of the loop model. Moreover, the bond weight $x_1$ is identified with the
monomer fugacity in the loop model.

\section{Relation with $\mathbb{Z}_n$ spin interfaces}
\label{sec:spin}
As we have seen in Section \ref{sec:Z3-interfaces}, Kuperberg webs describe, at a specific point in their parameter spaces, interfaces of $\mathbb{Z}_3$ spin models.
This generalises a well-known correspondence between the $O(N)$ loop model with $N=1$ (i.e., $N = [2]_q$ with $q=e^{i \frac{\pi}{3}}$) and Ising spin interfaces \cite{Nienhuis}.

In this section, we shall provide a further generalisation to all $n \ge 2$. We show that, by an appropriate tuning of the vertices fugacities of the \U\ web models
on $\mathbb{H}$, at the special point
\begin{equation}
 q=e^{i\frac{\pi}{n+1}} \label{special-point}
\end{equation}
the partition functions equal, up to an overall factor, those of some nearest-neighbour interaction $\mathbb{Z}_n$-symmetric spin models
on the dual triangular lattice $\mathbb{T}$. Those chiral clock models are very general and contain as special cases enhanced-symmetry
spin models, such as the clock models with symmetry group $\mathcal{D}_n$, the chiral Potts models with symmetry group $\mathcal{A}_n$,
and the usual Potts models with symmetry group $\mathcal{S}_n$.

Under the mappings that we shall exhibit, the equivalence classes of webs
that define the configurations of the \U\ web models are identified with the {\em interfaces of spin clusters}.
It thus transpires that \U\ web models provide a lattice regularisation of the local%
\footnote{The reformulaton of the \U\ web models as {\em local} vertex models will form the object of a forthcoming paper \cite{loc}.}
quantum field theory describing such non-local objects.

\subsection{Simplification at the special point}

In this subsection we show that, at the special point \eqref{special-point}, there exists a particular choice of the vertex weights, $y_{k,l;k+l}$ and $y_{k+l;k,l}$,
which leads to a trivial value of the combination of the vertex and non-local weights:
\begin{align}
    w'([c]) := \left( \prod_{\substack{k,l\in \llbracket 1,n-1 \rrbracket \\ k+l\in \llbracket 1,n-1 \rrbracket}} y_{k,l;k+l}^{N_{k,l;k+l}} \right)
    \left( \prod_{\substack{k,l\in \llbracket 1,n-1 \rrbracket \\ k+l\in \llbracket 1,n-1 \rrbracket}} y_{k+l;k,l}^{N_{k+l;k,l}} \right) w_s(c)=1 \,,\label{simp}
\end{align}
for any configuration $[c]$. The total weight \eqref{total_weight} of a configuration is then 
\begin{align}
    w([c])=\prod_{k\in \llbracket -n+1,n-1 \rrbracket} x_k^{N_k} \,,
\end{align}
where as before we have the normalisation constraint $x_0 = 1$.

\medskip

To show
\eqref{simp}, observe that one can compute $w'([c])$ in a similar way as $w_s([c])$, but with certain deformed relations,
in the same spirit as for the Kuperberg webs. Following \eqref{dressed_vertices_K}, we first incorporate the vertex weights into the
vertex diagrams, defining the ``dressed'' vertices
\begin{subequations}

\begin{align}
    \vcenter{\hbox{\includegraphics[scale=0.2]{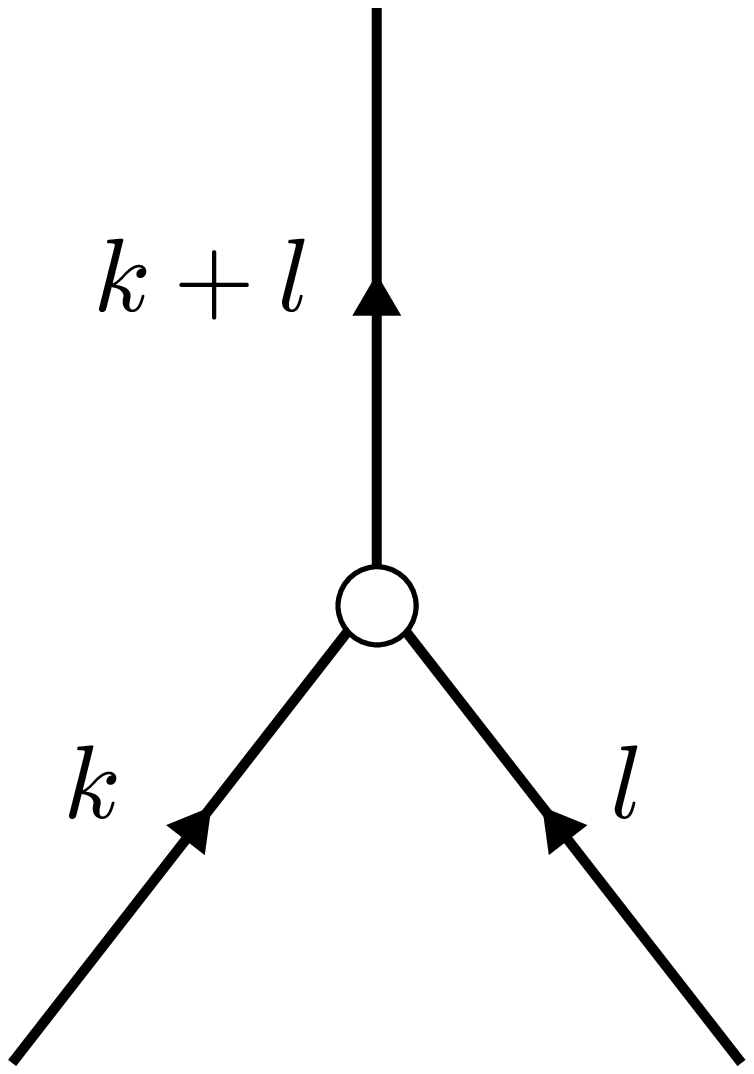}}}&= y_{k,l;k+l}\vcenter{\hbox{\includegraphics[scale=0.2]{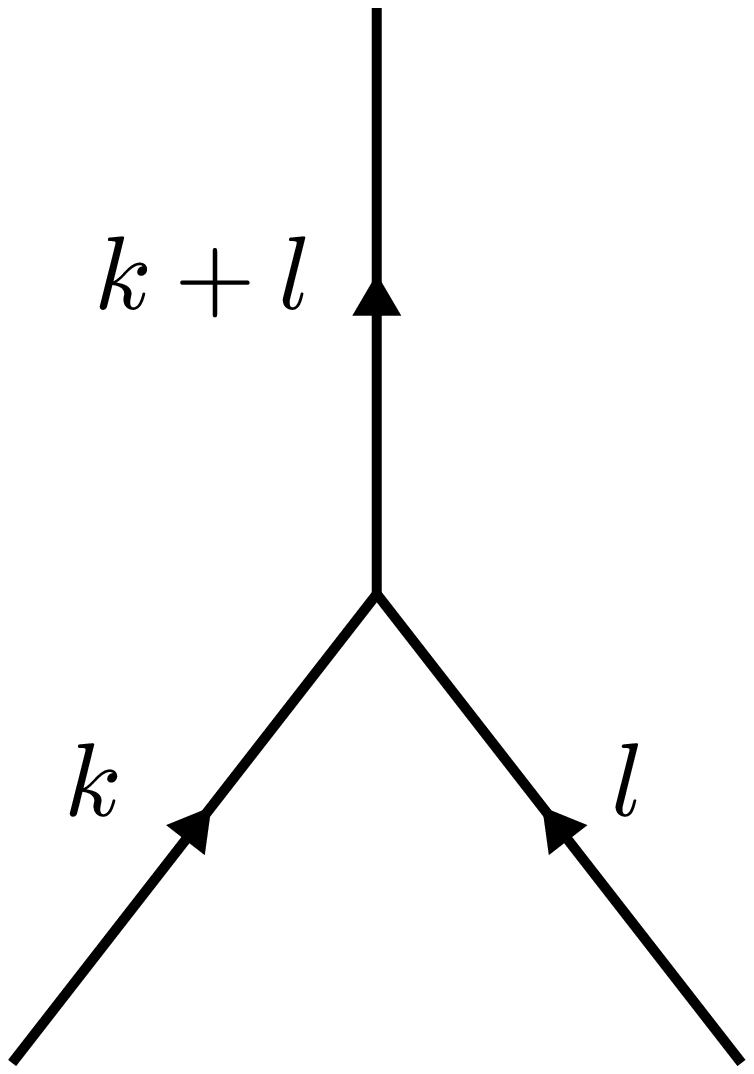}}}
\end{align}
and
\begin{align}
    \vcenter{\hbox{\includegraphics[scale=0.2]{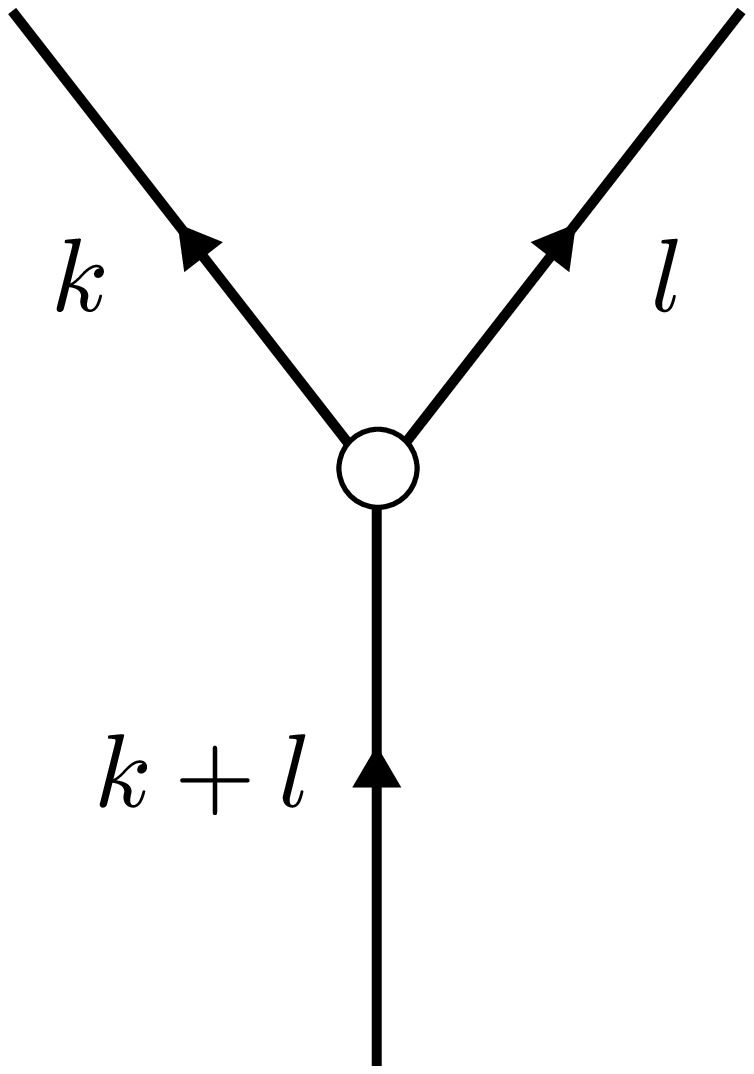}}}&= y_{k+l;k,l}\vcenter{\hbox{\includegraphics[scale=0.2]{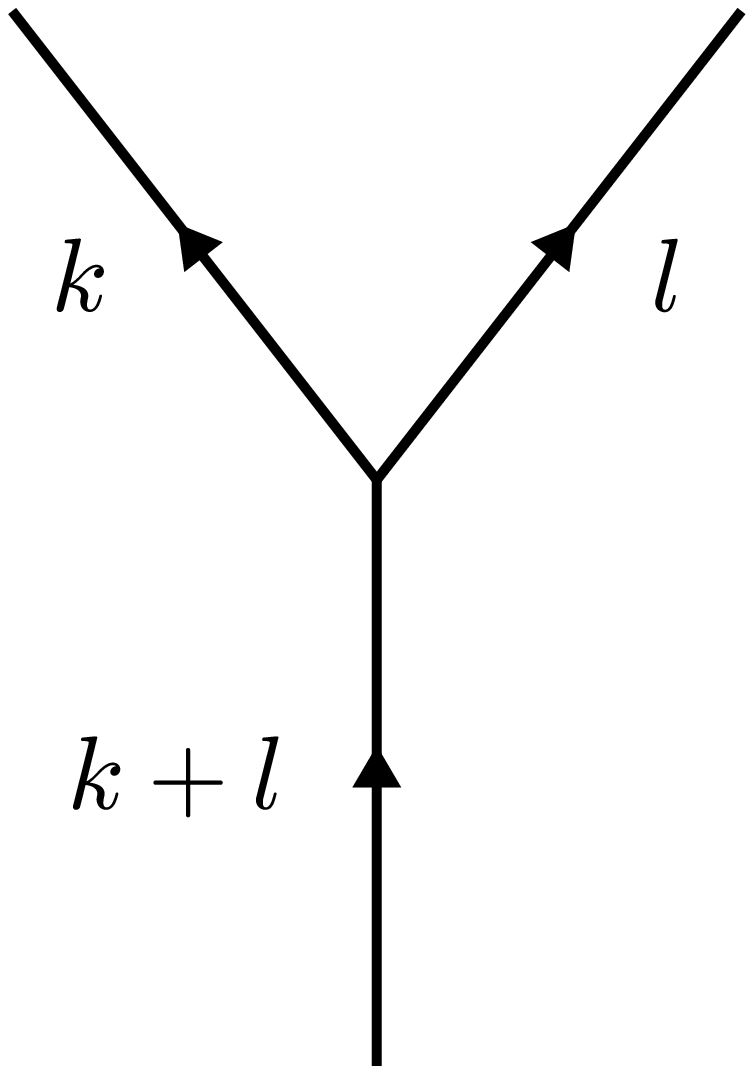}}}
\end{align}
\label{dressed_vertices}
\end{subequations}
It will turn out useful to allow the edges in \eqref{dressed_vertices} to carry the label $0$ or $n$, even when the original web $c$ does not.
We then augment the range of indices for the $y$-coefficients, such that $y_{k,l;k+l}=y_{k+l;k,l}=1$ whenever one of their arguments is $0$ or $n$.
Note that this modification changes nothing for closed simple webs. 

\medskip 

We now introduce a technical tool that we use only  for the purpose of
showing~\eqref{simp}. First, remark that in relations \eqref{eq1}-\eqref{eq9} we could have allowed any edge to take the label $0$ or $n$, with the additional convention that webs having a label outside of the interval $\llbracket0,n\rrbracket$ are zero.
Notice that, in particular, with the prescription \eqref{prescription} we would then retrieve the tag relations \eqref{eq6} and~\eqref{eq7} as special cases of \eqref{eq3} and
\eqref{eq4}
 at $l=n$ and $m=n$, respectively. However, if we do not use the prescription \eqref{prescription}, i.e, if we keep edges labelled by $n$ and only delete edges labelled by~$0$
 then  \eqref{eq1}-\eqref{eq9} give the relations  of so-called MOY graphs~\cite{MOY,Wu}.
In this scheme, the graphs are still closed, trivalent and oriented, and their
edges carry labels in $\llbracket 1, n\rrbracket$ such that the flow is conserved at vertices. From now on, we follow this convention about labels, and
we stress that MOY graphs do not use tags. This feature will be important below for the calculation of statistical weights at the special point~\eqref{special-point}.

In the MOY graphs, some of the seemingly lost tag rules are taken care of by the use of edges labelled by $n$. In particular, the  double tag rule~\eqref{eq6} is in fact the same as a label $n$ edge attached to the label $k$ edge in the digon shape; then the moving tag rule~\eqref{eq7} is the same as the associativity rule~\eqref{eq4} with a label $n$ edge. Let us explain the difference between our set of relations for MOY graphs and the one exposed in \cite{Wu}. The MOY graph relations (1)-(4) in \cite[Thm.\,2.3]{Wu} are exactly \eqref{eq1}-\eqref{eq4}.
Furthermore, there are several MOY graph rules from \cite{Wu} that are not  in the list  \eqref{eq1}-\eqref{eq9}, the equations (5)-(7) from~\cite[Thm.\,2.3]{Wu}, but it turns out that they are consequence of our relations~\eqref{eq1}-\eqref{eq9}. 
Indeed, the relation (5) from~\cite[Thm.\,2.3]{Wu} follows from (7):
we first use the digon rule~\eqref{eq3}  with the left edge labelled~$n$ to replace the leftmost bottom leg in (7) by the digon,  and then use the associativity rule~\eqref{eq4} several times to make the CPT transformation of the edges so the graph takes the form of~(5). Moreover, it is clear that the relation  (6) from~\cite[Thm.\,2.3]{Wu}  follows directly from (7). Finally, the relation (7)  follows by a repeated application of \eqref{eq2}, \eqref{eq4} and \eqref{eq9} as described briefly below~\eqref{prescription}. We therefore see that all the MOY graph rules follow from~\eqref{eq1}-\eqref{eq9} with the above convention on labels.

A closed simple web is a special case of a MOY graph in which no edge has label $n$.
As stated above, the first five relations \eqref{eq1}--\eqref{eq9} suffice to compute the weight of MOY graphs \cite[Thm.\,2.4]{Wu}, i.e.,
to reduce the graph to the empty one
multiplied by a number. 
The weight of a closed simple web is then the same as if it were regarded as a MOY graph. This means, in particular, that \eqref{eq5} is not
needed for evaluating the weight of a closed simple web---a fact that we shall use later. 

\medskip

As we have augmented the range of $y$-coefficients, \eqref{dressed_vertices} defines ``dressed" MOY graphs.
Assuming all the vertex weights $y$ to be non-zero, this leads to the following modification of their relations~\eqref{eq1}-\eqref{eq4}
where all the labels are allowed to take the value $n$:
\begin{subequations}\label{eq:rel-dressed-n}
\begin{align}
    \vcenter{\hbox{\includegraphics[scale=0.2]{e}}}&\;=\;\qbinom{n}{k}_q\\[5pt]
    y_{k+l;k,l}^{-1}y_{k,l;k+l}^{-1}\;\vcenter{\hbox{\includegraphics[scale=0.2]{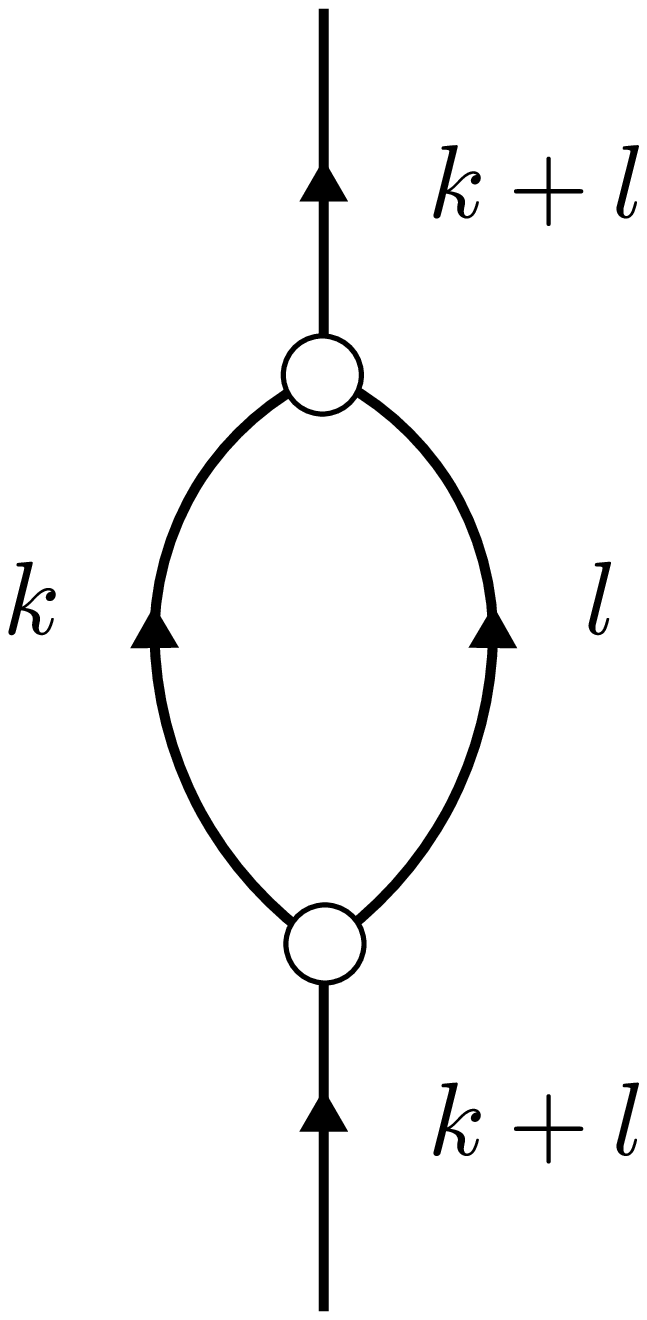}}}&\;=\;\qbinom{k+l}{k}_q\vcenter{\hbox{\includegraphics[scale=0.2]{i}}}\\[5pt]
    y_{k+l;k,l}^{-1}y_{k,l;k+l}^{-1}\;\vcenter{\hbox{\includegraphics[scale=0.2]{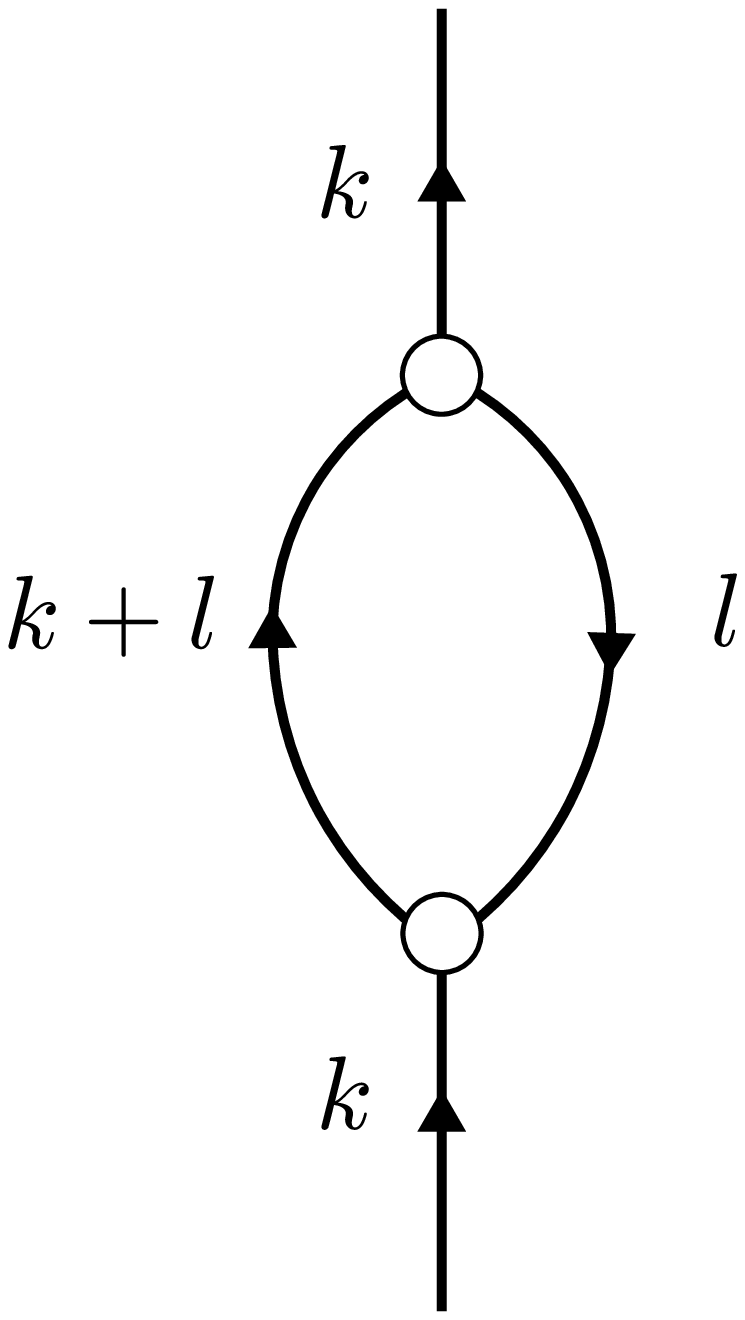}}}&\;=\;\qbinom{n-k}{l}_q\vcenter{\hbox{\includegraphics[scale=0.2]{h}}}\\[6pt]
    y_{k,l;k+l}^{-1}y_{k+l,m;k+l+m}^{-1}\;\vcenter{\hbox{\includegraphics[scale=0.2]{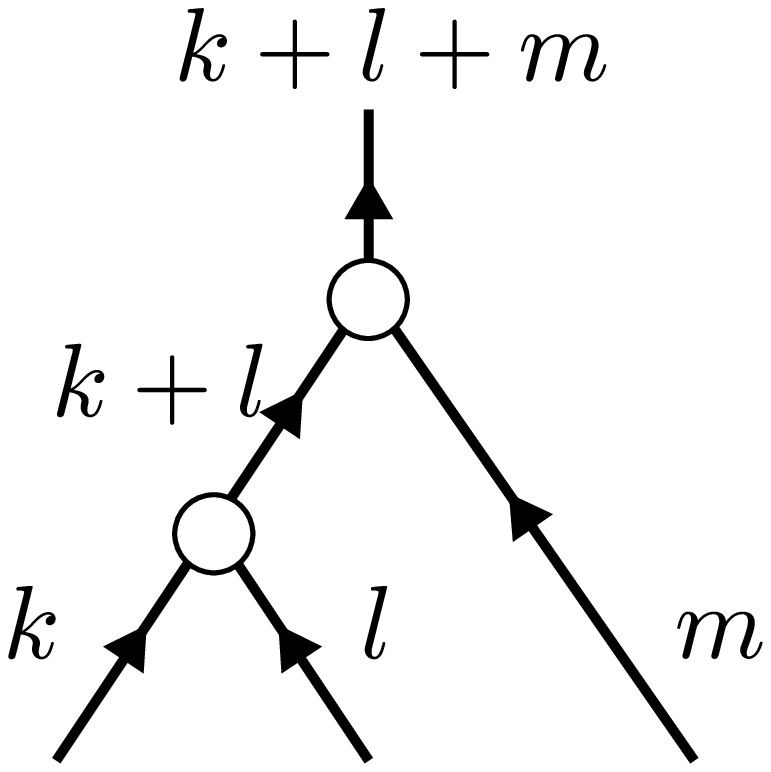}}}&\;=\;y_{l,m;l+m}^{-1}y_{k,l+m;k+l+m}^{-1}\;\vcenter{\hbox{\includegraphics[scale=0.2]{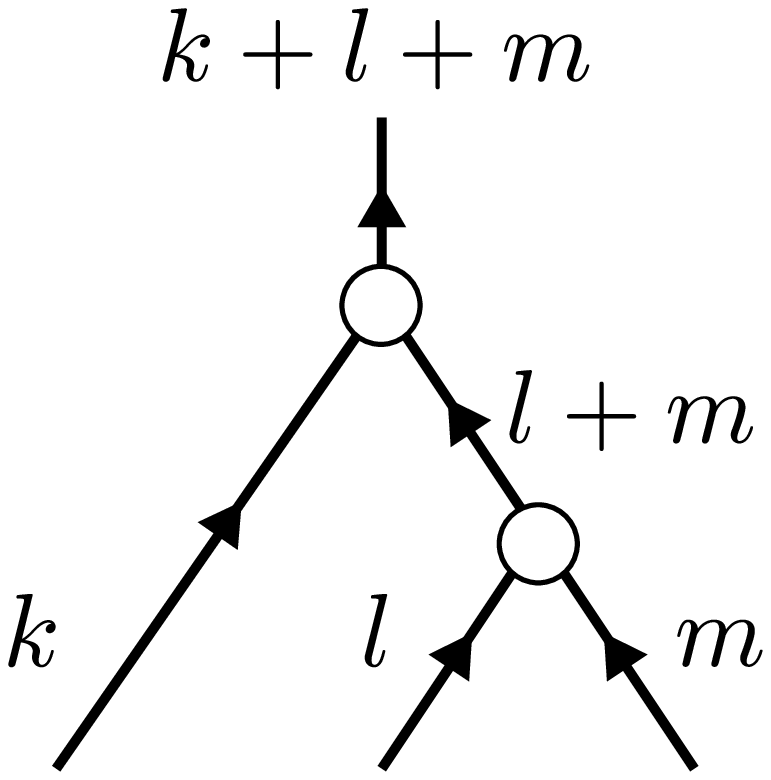}}}\\[7pt]
    y_{k;k-1,1}^{-1}y_{k-1,1;k}^{-1}y_{1,l;l+1}^{-1}y_{l+1;l,1}^{-1}\;\vcenter{\hbox{\includegraphics[scale=0.2]{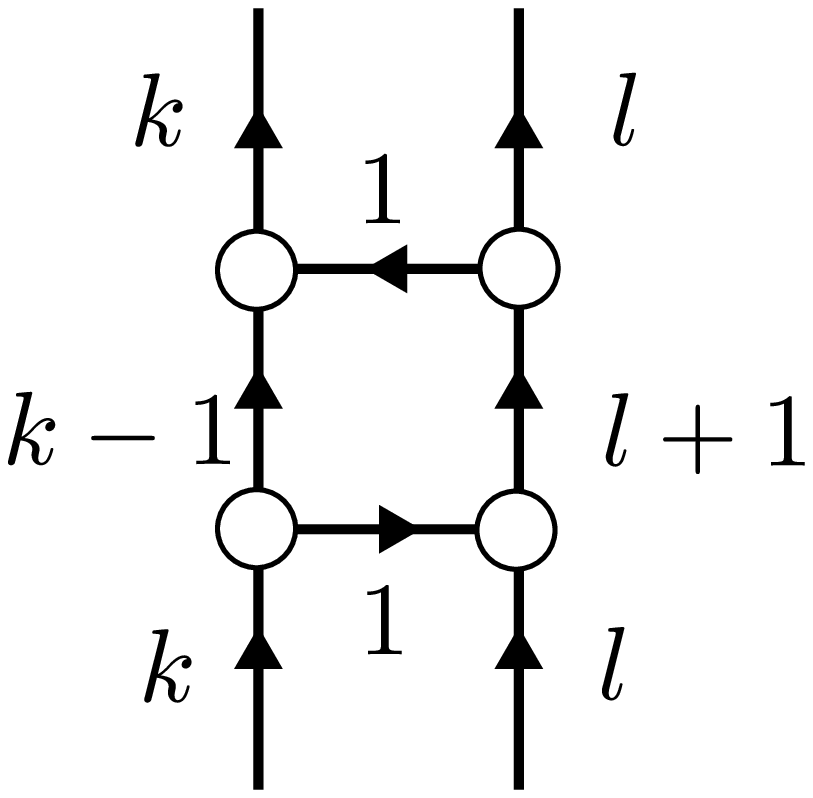}}}&\;=\;y_{k,1;k+1}^{-1}y_{k+1;k,1}^{-1}y_{l;l-1,1}^{-1}y_{1,l-1;l}^{-1}\;\vcenter{\hbox{\includegraphics[scale=0.2]{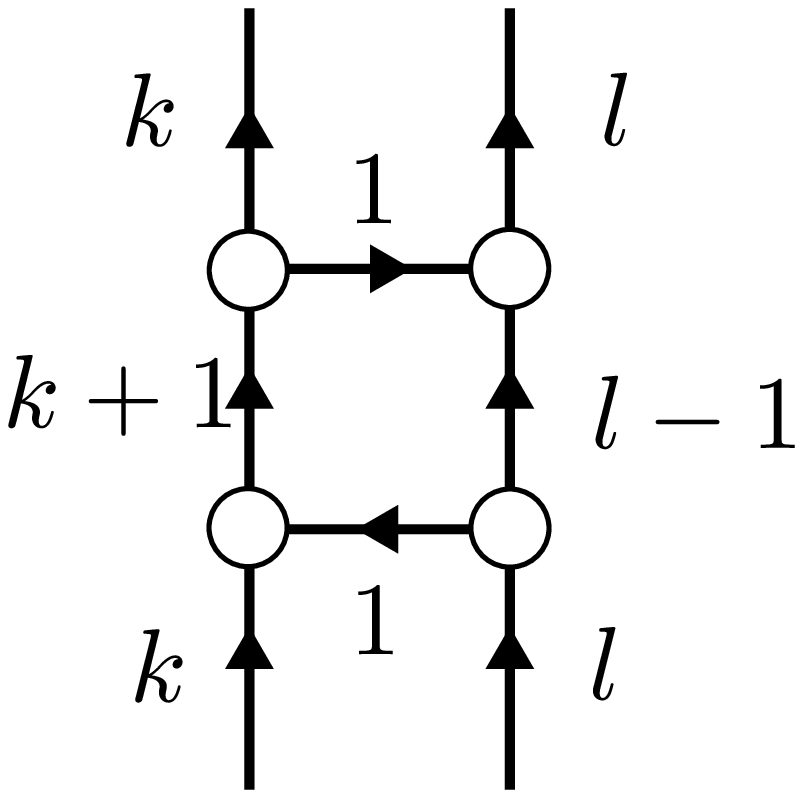}}}+[k-l]_q\;\vcenter{\hbox{\includegraphics[scale=0.2]{u}}} \label{lastreldressed}
\end{align}
\end{subequations}
together with their mirrored and arrow-reversed versions, where the vertex weights $y$ have to be changed accordingly with the change of flows.

Similarly to the case of the dressed webs in~\eqref{eq:deformed-rules-Kup}, these deformed  relations are well defined:
an evaluation with the rules~\eqref{eq:rel-dressed-n} is equivalent to the two-step process: first count all the vertex weights, then remove the dressing on the vertices and evaluate with the standard MOY graph rules, or with the rules~\eqref{eq1}-\eqref{eq9} using our convention on labels. This process  does not obviously depend on the way of reduction of the webs. In particular these dressed rules allow one to evaluate any dressed simple web to a number.

We now take for the vertex weights the particular choice
\begin{equation} \label{special-y}
  y_{k+l;k,l}=y_{k,l;k+l}=\qbinom{k+l}{k}_q^{-\frac{1}{2}} \,.
\end{equation}
At the special point \eqref{special-point} we have the identities
\begin{subequations}
\begin{align}
\qbinom{n}{k}_q&=1 \,, \\
\qbinom{k+l}{l}_q&=\qbinom{n-k}{l}_q \,.
\end{align}
\end{subequations}
They imply that $y_{k,l;k+l}$ and $y_{k+l;k,l}$ indeed equal $1$ when one of the labels is $0$ or $n$, as desired. Moreover, the
constraints~\eqref{constraints} are satisfied.
We finally notice that all the vertex weights in~\eqref{special-y} are well defined at the special point~\eqref{special-point}, because the label $k+l$ never equals $n+1$ (otherwise $[k+l]_q=0$) and therefore the $q$-binomials never vanish.

With \eqref{special-point} and \eqref{special-y} the relations with the dressed vertices become
\begin{subequations}
\begin{align}
    \vcenter{\hbox{\includegraphics[scale=0.2]{e}}}&\;=\;1\\[5pt]
    \vcenter{\hbox{\includegraphics[scale=0.2]{diagrams/fnew.eps}}}&\;=\;\vcenter{\hbox{\includegraphics[scale=0.2]{i}}}\\[5pt]
    \vcenter{\hbox{\includegraphics[scale=0.2]{diagrams/gnew.eps}}}&\;=\;\vcenter{\hbox{\includegraphics[scale=0.2]{h}}}\\[5pt]
    \vcenter{\hbox{\includegraphics[scale=0.2]{diagrams/jnew.eps}}}&\;=\;\vcenter{\hbox{\includegraphics[scale=0.2]{diagrams/knew.eps}}}\\[5pt]
    [k]_q[l+1]_q\;\vcenter{\hbox{\includegraphics[scale=0.2]{diagrams/snew.eps}}}&\;=\;[k+1]_q[l]_q\;\vcenter{\hbox{\includegraphics[scale=0.2]{diagrams/tnew.eps}}}+[k-l]_q\;\vcenter{\hbox{\includegraphics[scale=0.2]{u}}}
\end{align}
\end{subequations}
As in the case of the Kuperberg webs, for any of these relations, the sum of prefactors for the graphs on the left-hand side is equal to the sum of prefactors for
the graphs on the right-hand side. Then, by the same argument
as given below~\eqref{reducexample},
 the weight of any ``dressed" MOY graph is $1$ and in particular, the partial weight $w'([c]) = 1$ for any configuration $[c]$. 
This therefore shows~\eqref{simp}. 

Remark that we could have written the modified relations for webs instead of MOY graphs in a similar way
(without rescaling tags, as those do not account for vertex fugacities). Then, the relation~$\eqref{eq5}$ does not, in general, possess the crucial property that the sum of prefactors on the left-hand side equals the sum of prefactors on the right hand-side. Although when $n$ is odd, it actually does, we needed to use the equivalence with MOY graphs to
show
\eqref{simp} for any $n$. The point is that the identification with MOY graphs 
ensures the existence of a reduction process of a closed simple web that 
does not use $\eqref{eq5}$.

The partition function of the web model with \eqref{special-y}, at the special point \eqref{special-point}, thus reads simply
\begin{align}
    Z=\sum_{[c]\in \mathcal{C}} \prod_{k\in \llbracket -n+1,n-1 \rrbracket} x_k^{N_k} \,. \label{part}
\end{align}

\subsection{Low-temperature expansion}

Consider now a spin model on the triangular lattice $\mathbb{T}$, dual to the hexagonal lattice $\mathbb{H}$ where the web model is defined.
To each node $i$, attach a spin variable $\sigma_i\in \llbracket 0,n-1 \rrbracket$. To each link $(ij)$, with $i$ to the left of $j$ when viewed along the
chosen fixed direction, attach a local Boltzmann weight $x_{\sigma_i-\sigma_j}$, subject to the normalisation $x_0 = 1$.
In general $x_{\sigma_i-\sigma_j}\neq x_{\sigma_j-\sigma_i}$, so the model possesses a chirality.
The corresponding partition function reads
\begin{align}
    Z_{\rm spin}=\sum_{\sigma_i} \prod_{\langle ij \rangle} x_{\sigma_i-\sigma_j} \,, \label{ZspinLT}
\end{align}
where $\langle ij \rangle$ denotes the set of links.

The weights define a mapping $x: \llbracket -n+1,n-1 \rrbracket \rightarrow \mathbb{C}$, but we now impose
$x_{-k}=x_{n-k}$ for any $k \in \llbracket 1,n-1\rrbracket$, so that we have effectively $x: \llbracket 0,n-1 \rrbracket \rightarrow \mathbb{C}$
with $x_0 = 1$. The group $\mathbb{Z}_n$ acts by cyclically permuting the values $\llbracket 0,n-1 \rrbracket$ of all spins, and this action
leaves \eqref{ZspinLT} invariant as it only depends on the differences $\sigma_i-\sigma_j$.

Let $\mathcal{T}$ be the set of nodes of the triangular lattice $\mathbb{T}$. Denote by $\mathcal{S}$ the space of equivalence classes
of the mappings $\sigma : \mathcal{T}\rightarrow \llbracket 0,n-1 \rrbracket$ with respect to a global shift. We write $[s] \in \mathcal{S}$ for
such a class; there are exactly $n$ representatives for each class.
The partition function is then rewritten
\begin{align}
    Z_{\rm spin}=n\sum_{[s] \in \mathcal{S}} w_{\rm spin}(s) \,,
\end{align}
where the weights
\begin{equation}
 w_{\rm spin}(s)=\prod_{\langle ij \rangle} x_{s_i-s_j} = x_1^{N_1} x_2^{N_2} \cdots x_n^{N_n}
\end{equation}
generalise those of \eqref{Z_spin}.

Observe that the weight of a configuration $[s]$ is concentrated on links of $\mathbb{H}$ separating pairs of nodes $(ij) \in \mathbb{T}$
such that $\sigma_i\neq \sigma_j$. Now, take a representative $\sigma$ of $[s]$ and build a graph on $\mathbb{H}$ in the following way.
Give to each bond that separates different spins, $\sigma_i \neq \sigma_j$, the label $|\sigma_i-\sigma_j|$ and an orientation upward 
(resp.\ downward) if $\sigma_i<\sigma_j$ (resp.\ $\sigma_i > \sigma_j)$. Such a graph is closed and has flow conservation at vertices;
it is thus a simple web. Moreover, different representatives of $[s]$ give equivalent webs with respect to the equivalence relation
of Section~\ref{sec:Cautis}.

\begin{figure}
\begin{center}
    \includegraphics[scale=0.3]{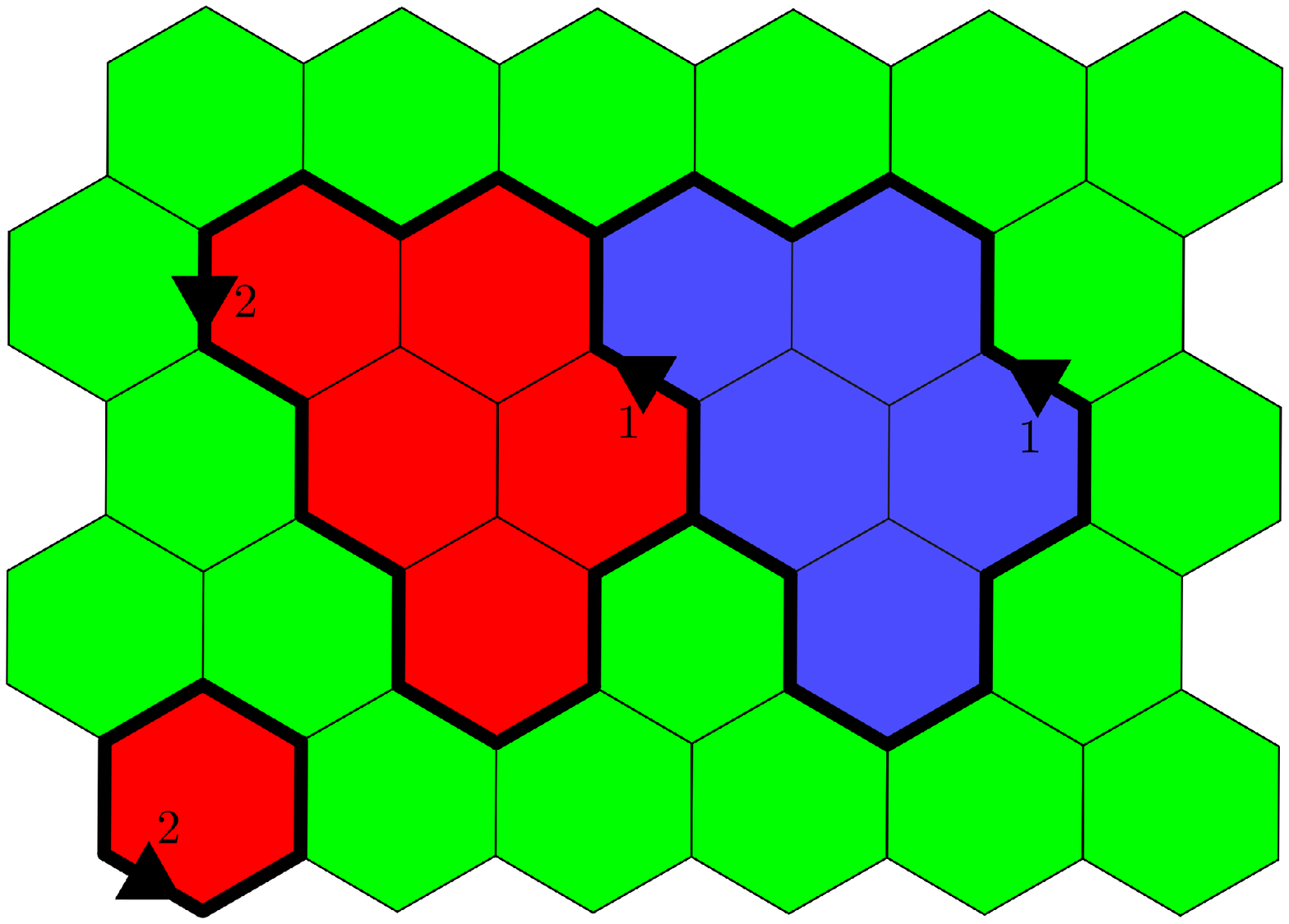} \quad \includegraphics[scale=0.3]{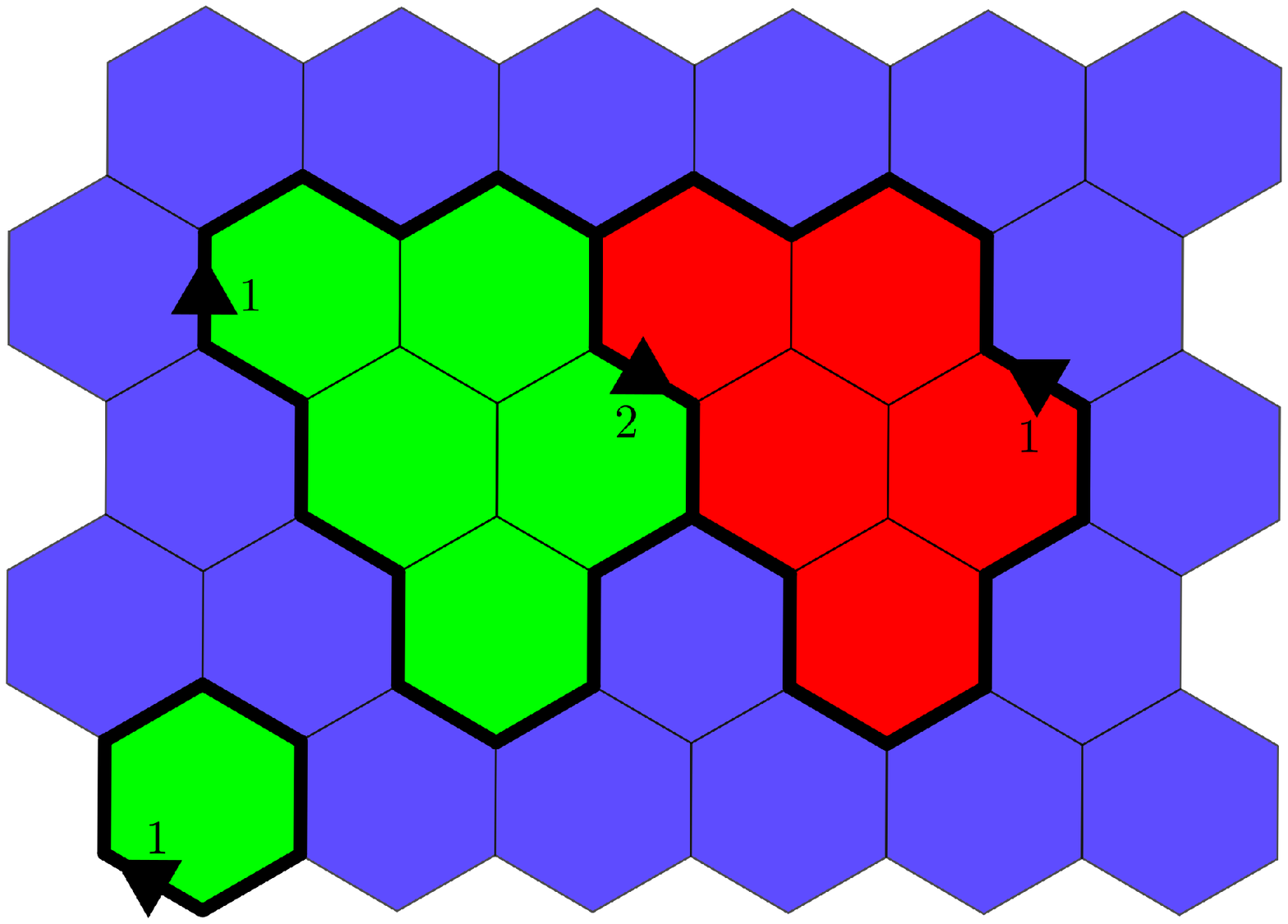}
\end{center}
    \caption{Two spin configurations in the $\mathbb{Z}_n$ model with $n=3$, with the colours $\{{\rm red, blue, green}\}$ representing the spin values
    $\mathbb{Z}_3 := \{0,1,2\}$. These configurations are related by a global $\mathbb{Z}_3$ shift and hence correspond to equivalent webs.}
\label{spinconf}
\end{figure}

This construction thus provides a well-defined mapping $f:\mathcal{S}\rightarrow \mathcal{C}$, where we
recall that $\mathcal{C}$ denotes the set of configurations of the web model. It is illustrated in Figure \ref{spinconf}.
We can remark that whereas any class in $\mathcal{S}$ has always $n$ representatives, the same is not true for classes
in $\mathcal{C}$. It is clear that $w_{\rm spin}(s)=w(f([s]))$ at the special point \eqref{special-point} introduced above.

Thus, to prove that the spin and web models are equivalent, one has to show that $f$ is a bijection. To do so, we now construct
its inverse. Consider a web $c$, a representative of $[c]$, on $\mathbb{H}$. Build a spin configuration on the dual lattice $\mathbb{T}$
in the following way. Colour one face of the web by $0$, i.e., all nodes of $\mathbb{T}$ inside this face are taken to have spin $\sigma_i = 0$.
Then the colours of the other faces are fixed by using the prescription that when an edge labelled by $k$ flowing to the left (resp.\ right) is crossed, we add $k$
(resp.\ $-k$) to the preceding colour and take the result modulo $n$. The resulting colouring is independant of the chosen representative of $[c]$.
Taking the class $[s]$ of the spin configuration built this way, one finally obtains a map $g:\mathcal{C}\rightarrow \mathcal{S}$.
Clearly, $f\circ g=Id$ and $g\circ f=Id$. We then have the desired result
\begin{align}
    Z_{\rm spin} = n Z \,,
\end{align}
where $Z$ is the partition function \eqref{part} of the web model.
The latter can therefore be identified as the low-temperature expansion of the spin model.

\section{Open boundaries and defects}
\label{sec:open}

Physical observables in the \U\ web models can be obtained by taking the ratio of modified partition functions---in which the configurations of webs have been
constrained, or had their weighting modified, or been subjected to the introduction of defects---with respect to the unmodified partition
function. While these possibilities have been thoroughly investigated in the O($N$) loop model, appearing here as the special case $n=2$, the
situation for $n>2$ still forms a pristine working ground.

As a first example, we discuss in this section the definition of the web models with open boundary condition,
in the presence of defects on the boundary. The latter take the form of {\em defect vertices}, which are $1$-valent nodes on the boundary.
We formulate everything in terms of \U\ webs, from which the special case of Kuperberg webs ($n=3)$ can be deduced straightforwardly.

\subsection{The modified partition function}

\begin{figure}
\begin{center}
    \includegraphics[scale=0.3]{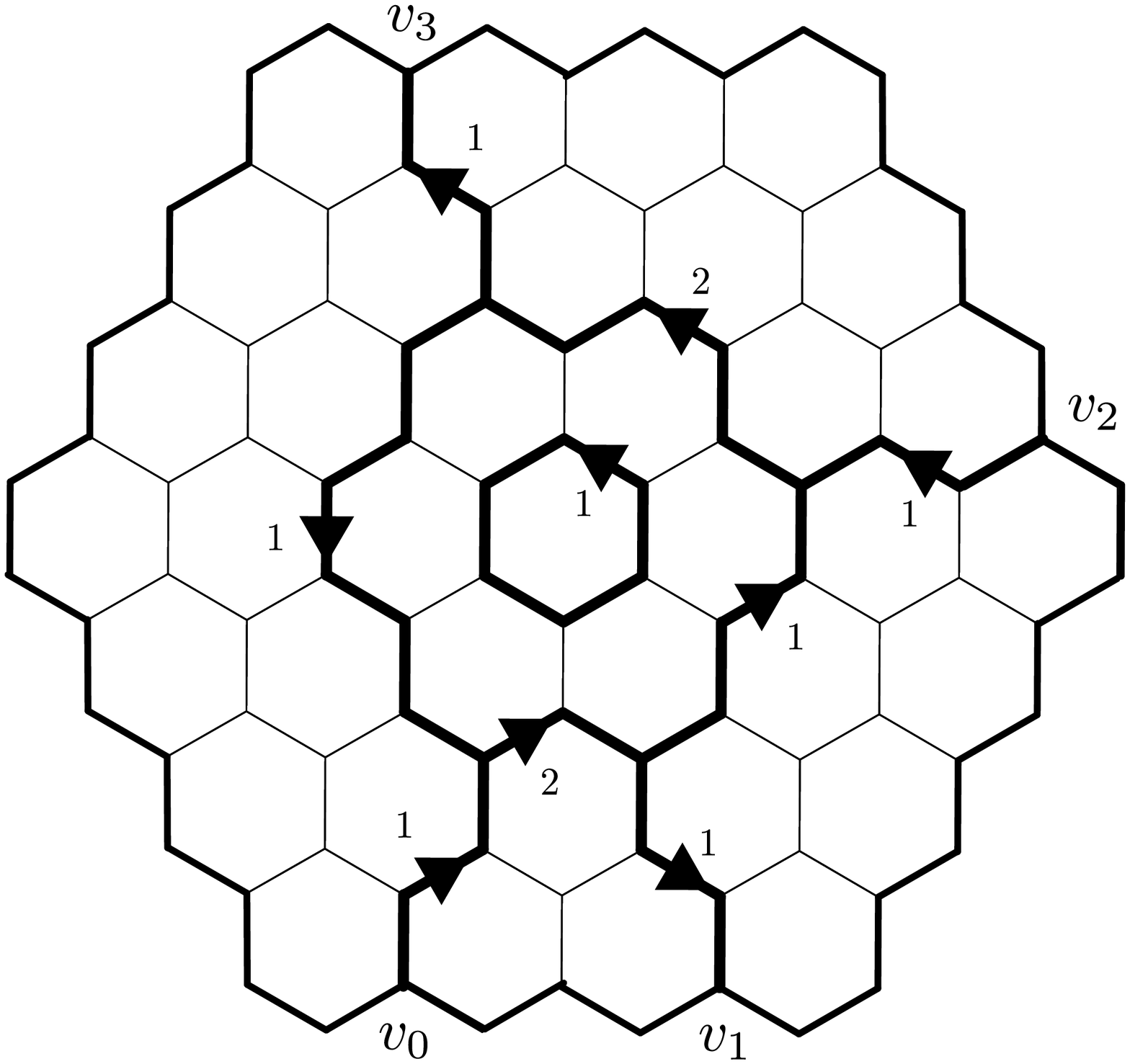} \quad \includegraphics[scale=0.3]{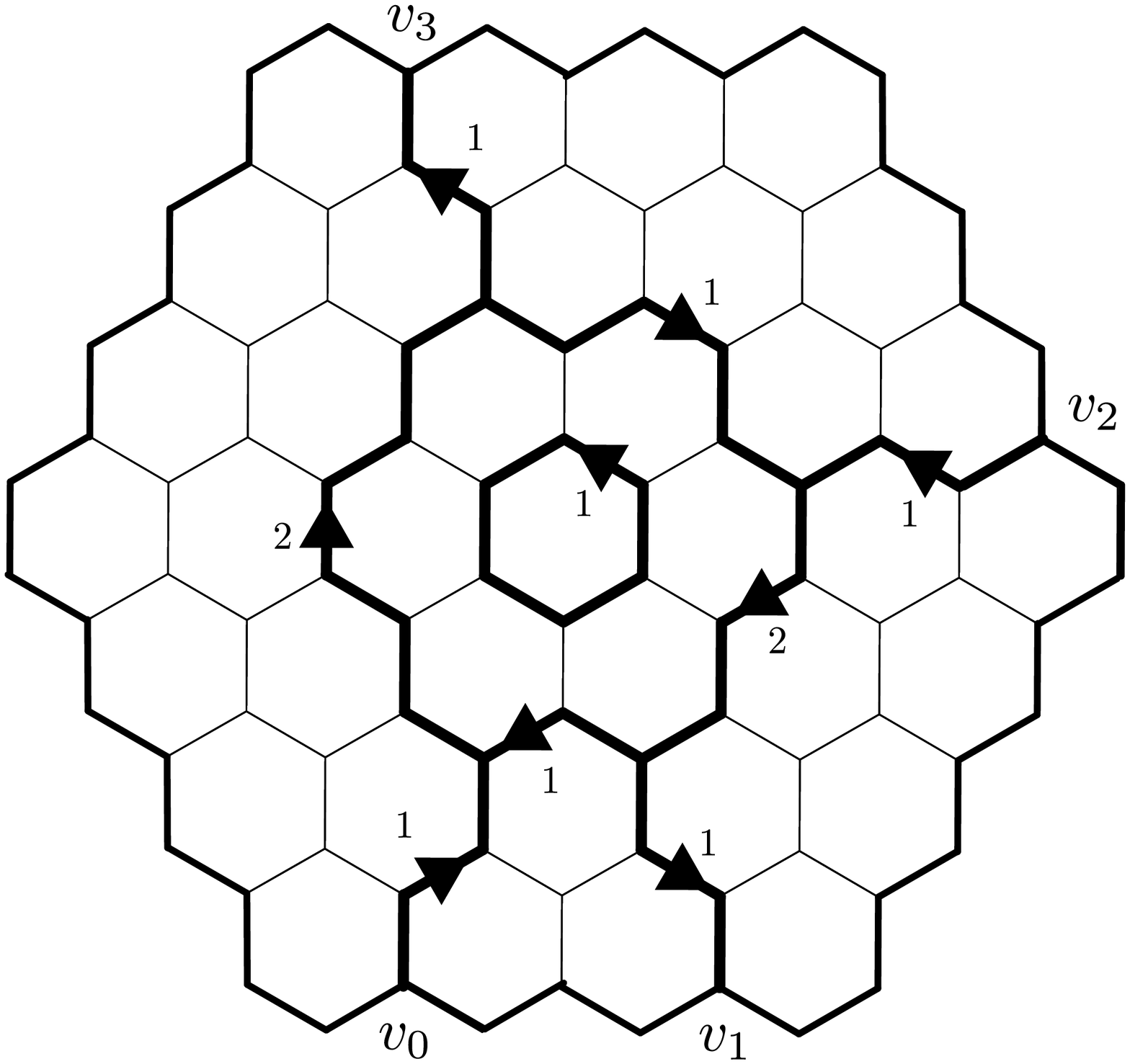}
        \end{center}
    \caption{Two equivalent open simple webs (shown here for the case $n=3$), with boundary condition $\{(v_0,1),(v_1,-1),(v_2,1),(v_3,-1)\}$.}
\label{openwebconf}
\end{figure}

Consider a simply connected domain $\mathbb{D}$ of the hexagonal lattice, again such that one third of its links are vertical (see Figure \ref{openwebconf}).
Given a subset of $I$ boundary nodes, $\{v_i \,|\, i\in \llbracket 0,I-1 \rrbracket\}$, cyclically ordered anticlockwise, we associate an integer
$k_i\in \pm \llbracket 1,n-1 \rrbracket$ to each node $v_i$. On the domain~$\mathbb{D}$, we consider simple webs that have, for each
$i \in \llbracket 0,I-1 \rrbracket$, 
one of their edges incident to $v_i$ and labelled by $|k_i|$ such that it is oriented towards the exterior (resp.\ interior) of $\mathbb{D}$ if $k_i<0$ (resp.\ $k_i>0$).

The abstract webs without tags satisfying these conditions are called \textit{open simple webs} with boundary condition
$\mathcal{B}=\{(v_i,k_i) \,|\, i\in\llbracket 0,I-1 \rrbracket\}$. The edges of the web that end at the boundary will be called
\textit{free edges}. Each node on the boundary of $\mathbb{D}$ incident on a free edge $v_i$ is called a \textit{defect node}
of charge $k_i$. Remark that the strict flow conservation at vertices of the open simple web implies the global
defect charge neutrality:
\begin{align}
    \sum_{i\in\llbracket 0,I-1 \rrbracket} \!\!\!\!\! k_i = 0 \,.
\end{align}

We will now take as configurations the equivalence classes of such embedded open simple webs,
with respect to the equivalence relation generated by CPT transformations that leave their free edges
\textit{unchanged} (see Figure \ref{openwebconf}). Note that, in principle, the argument given in Section~\ref{sec:def-of-models}
would allow the CPT transformations that connect two webs to be given not only by transition cycles
(as in the closed case), but also by transition paths going from one defect node to another. But the
requirement that free edges remain unchanged amounts to allowing only the CPT transformations of transition cycles.
 We will denote the corresponding configuration space of equivalence classes by~$\mathcal{C}_{\mathcal{B}}$.

To any equivalence class of embedded open simple webs, $[c]\in \mathcal{C}_\mathcal{B}$, we can specify fugacities for bonds and vertices, as in the closed
boundary case. In the open case, it is possible to allow for the bond fugacities to be different, depending on whether the
bond is situated on the boundary of $\mathbb{D}$ or in its interior. In other words, the web tension $t_i$ on the boundary
can be taken different from the bulk tension $x_i$. For the loop model ($n=2$) it is known that taking $t_i > x_i$ can drive
the model to the so-called special surface transition, whose critical behaviour is different from that of the ordinary transition
at $t_i = x_i$ (see \cite{DubailSp1,DubailSp2} and references therein). For $n > 2$ one could similarly envisage the existence of a special surface transition; it is conceivable that to reach it
one would also have to modify the vertex weights $y$ at the boundary, but we do not need such a modification in the present paper.

We now turn to the non-local weights. In \cite{Cautis_2014}, it is shown that abstract open webs, modulo the relations \eqref{spiderrules},
form a finite-dimensional vector space $V_{\mathcal{B}}$. On the representation theoretical side, $V_{\mathcal{B}}$ is isomorphic to the space of \U-invariants in the tensor product of representations:
\begin{align}
 V_{\mathcal{B}} \;\cong\;   \text{Inv}\left(V_{|k_0|}^{\text{sign}(k_0)}\otimes V_{|k_1|}^{\text{sign}(k_1)}\otimes \cdots \otimes V_{|k_{I-1}|}^{\text{sign}(k_{I-1})}\right) \,,
\end{align}
where $V_i$ denotes the $i$th fundamental reprensetation of \U, and we have used the notation $V_i^1=V_i$ for that representation itself,
and $V_i^{-1}=V_i^*$ for its dual.
To the abstract open web $c$, we can thus attach a vector $\ket{c} \in V_\mathcal{B}$ in this space. Because of the argument of Appendix \ref{sec:app-equiv-weights},
this vector does not depend on the representative and we may write $\ket{[c]}$ instead, with $[c] \in \mathcal{C}_\mathcal{B}$.
We thus obtain a ``partition vector"
\begin{align}
    \ket{Z}=\sum_{[c]\in \mathcal{C}_{\mathcal{B}}} w_{\text{loc}}([c]) \ket{[c]} \,,
\end{align}
where the local weight is
\begin{align}
    w_{\text{loc}}([c])= \left(\prod_{k\in \llbracket -n+1,n-1 \rrbracket} t_k^{M_k}\right)\left(\prod_{k\in \llbracket -n+1,n-1 \rrbracket} x_k^{N_k}\right) \left(\prod_{\substack{k,l\in \llbracket 1,n-1 \rrbracket \\ k+l\in \llbracket 1,n-1 \rrbracket}} y_{k,l;k+l}^{N_{k,l;k+l}} \right) \left( \prod_{\substack{k,l\in \llbracket 1,n-1 \rrbracket \\ k+l\in \llbracket 1,n-1 \rrbracket}} y_{k+l;k,l}^{N_{k+l;k,l}} \right) \,,
\end{align}
where $M_k$ is the number of boundary bonds covered by an edge labelled by $k$, and the other occurrence numbers are as in \eqref{total_weight}.

In order to obtain a partition function for the open case, one must choose a linear form $F$ on~$V_{\mathcal{B}}$:
\begin{align}
    Z_F=F\big(\ket{Z}\big) =\sum_{[c]\in \mathcal{C}} \Big( w_{\text{loc}}([c])\ F\big(\ket{[c]}\big) \Big) \,.
\end{align}
Remark that when the boundary condition is trivial (no free edges), we may interpret the partition function of Section \ref{sec:Cautis} in this picture,
if we simply forget the presence of the boundary. In that case the webs are closed, so every vector $\ket{[c]}$ can be expressed as a multiple of the
empty web,
\begin{align}
    \ket{[c]}=w_s([c])\ket{\emptyset } \,,
\end{align}
and by choosing $F$ such that $F(\ket{\emptyset })=1$ we get the desired result. Remark that, in the same spirit, it is possible to relax
the condition of simple connectedness to define a partition function on any connected domain when the boundary condition is trivial. In particular,
we could obtain the web model on the annulus in this way. However, the spiders from~\cite{Cautis_2014} do not lead to a definition of non-trivial defects attached to the boundary of the annulus.\footnote{For such a definition, one would need to introduce the periodic (or affine) version of the spiders.}

When the boundary condition $\mathcal{B}$ is non-trivial, we can pick a basis of $V_{\mathcal{B}}$ that is \textit{minimal} in the sense that the total number of vertices and tags of the webs forming the basis is minimal (note that in general tags may be needed to form a basis). This means that we would reduce $\ket{[c]}$ as much as possible. When ${\rm dim} \, V_{\mathcal{B}} = 1$  the weight of the web must be the component (up to a scalar) of the unique basis vector,
so there is no arbitrariness involved in choosing the weight (up to a scalar). This is the case when the boundary condition is trivial. This situation also occurs, for instance, when a single defect is propagating from one point of the boundary
to another, a setup reminiscent of chordal SLE \cite{SLE}. 

When ${\rm dim} \, V_{\mathcal{B}} \ge 2$ there is, in general, no obvious natural choice of $F$.
However, we have a very natural  linear form $F$ if the set of  boundary conditions splits into two subsets dual to each other, namely, to a set of labels $S$ and the set $S^*$ of the corresponding dual labels (in particular this requires the number of points $I$ to be even). For such boundary conditions,  $F$ can be taken as the quantum trace of \U. More precisely, there is an isomorphism  between the space $V_{\mathcal{B}}$, i.e., invariants from 
$$
\mathrm{Inv}\left(\bigotimes_{i\in S}V_i \otimes \bigotimes_{i\in S}V_i^*\right),
$$
 and 
the space of morphisms of the object $\{(i,\epsilon_i)\}_{i\in S}$ to itself  in the category of spider webs of Cautis {\em et al.}~\cite{Cautis_2014} (here, $\epsilon_i$ is a sign).  In simpler terms, these morphisms are nothing but linear operators on $\otimes_{i\in S}V_i$ intertwining the  \U \  action. Having such an operator,  one can evaluate on it the \U \ invariant trace---the so-called quantum trace. It is the usual trace function but with an inserted pivotal element, which is a certain product of Cartan generators of \U.
This procedure is the direct analogue of the Markov trace used in the context of loop models or Temperley-Lieb algebra.

More physically, the above procedure with the quantum trace corresponds to deforming the domain $\mathcal{D}$ so that
$S$ and $S^*$ reside at opposite ends of a strip, and then gluing together the two ends so as to form a web on an annulus, and finally evaluating the weight of the corresponding planar graph.
We note however that in this way we obtain the web in a different geometry (an annulus instead of a disc) and only in one specific sector---the defects propagate around the annulus but never end on the boundary.

\medskip
We will show in our next paper\cite{loc} that the \U\ web model at the special point \eqref{special-point} is equivalent to the $\mathbb{Z}_n$ spin model on \textsl{the same} hexagonal lattice, via a high-temperature expansion. In this framework, the free end of a defect will correspond to a spin operator inserted into a correlation function. A single defect propagating from one point of the boundary to another thus measures a boundary spin-spin correlator. We can therefore expect an interesting surface critical behaviour for the \U\ web model, also at other points of the parameter space than \eqref{special-point}.

\subsection{Relation with spin interfaces}

We can formulate an equivalence between a web model on $\mathbb{D}$ and the $\mathbb{Z}_n$ spin model on the dual triangular lattice
(without the point at infinity) with spins fixed on the boundary. First, one fixes the value of the boundary spins---i.e., those residing on the outermost
layer of hexagons on $\mathbb{D}$---once and for all. This boundary condition will be denoted $\Gamma$, and we write $\mathcal{S}_{\Gamma}$
for the space of spin configurations satisfying $\Gamma$. Second, we set the boundary bond fugacities to $t_i=0$ for all $i$. This implies that the
webs can only touch the boundary by means of free ends incident on the defect vertices.

\begin{figure}
\begin{center}
    \includegraphics[scale=0.3]{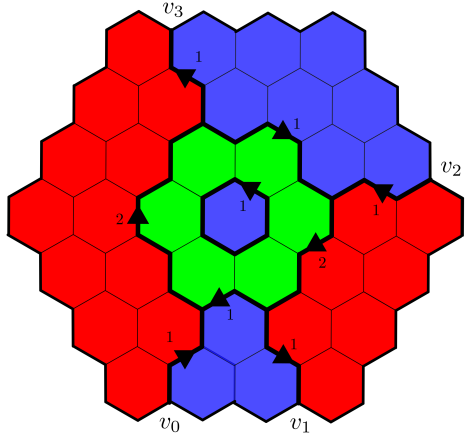} 
        \end{center}
    \caption{A spin configuration of the $\mathbb{Z}_3$ spin model, with the colours $\{{\rm red, blue, green}\}$ representing the spin values
    $\mathbb{Z}_3 := \{0,1,2\}$, in the open case.}
\label{openspinconf}
\end{figure}

In the spirit of Section \ref{sec:spin}, for each pair of neighbouring, unequal boundary spins, $\sigma_i\neq \sigma_j$, where 
$\sigma_j$ is next to $\sigma_i$ in an anticlockwise manner, one assigns the label (defect charge) $k_i = \sigma_j-\sigma_i$
to the free end in-between the pair of spins. This fixes the boundary condition $\mathcal{B}$ for the web model (see Figure \ref{openspinconf}).
We say that the boundary of the web model is \textit{compatible} with the boundary condition~$\Gamma$ of the spin model.

Given compatible boundary conditions, $\Gamma$ and $\mathcal{B}$, an equivalence class of open simple webs with boundary condition
$\mathcal{B}$ and a (unique) spin configuration satisfying $\Gamma$ are in bijection, thanks to mappings defined as in Section \ref{sec:spin}.
Take a configuration $\sigma\in \mathcal{S}_{\Gamma}$ and assign to each bond that separates different spins, $\sigma_i \neq \sigma_j$,
the label $|\sigma_i-\sigma_j|$ and an orientation upward (resp.\ downward) if $\sigma_i<\sigma_j$ (resp.\ $\sigma_i > \sigma_j)$.
This produces a simple web with the boundary condition $\mathcal{B}$. By taking its equivalence class, we obtain a mapping
$f_{\text{o}} : \mathcal{S}_{\Gamma} \rightarrow \mathcal{C}_{\mathcal{B}}$.

Conversely, given a web configuration $[c] \in  \mathcal{C}_{\mathcal{B}}$, any representative $c$ of $[c]$ defines a spin configuration
in $\mathcal{S}_{\Gamma}$ by the following construction. The colours of spins on the boundary are fixed by $\Gamma$. Then the colours
of the other faces of $\mathbb{D}$ are fixed by the prescription that when an edge labelled by $k$ flowing to the left (resp.\ right) is crossed, we add $k$
(resp.\ $-k$) to the preceding colour and take the result modulo $n$. The resulting colouring is independant of the chosen representative of $[c]$. This defines $g_{\text{o}}$ : $\mathcal{C}_{\mathcal{B}} \rightarrow \mathcal{S}_{\Gamma}$. 

The two maps, $f_{\text{o}}$ and $g_{\text{o}}$, are clearly inverses of each other. The equivalence between models is then established
by choosing a specific linear form $F$ for the web model, such that partition functions of the web and spin models agree.

\smallskip

For a web model boundary condition $\mathcal{B}$ compatible with a spin boundary condition $\Gamma$, there exists a minimal basis
for $V_\mathcal{B}$ consisting of open simple webs. Indeed, for any minimal basis of~$V_\mathcal{B}$, each basis vector is, up to a sign,
a simple web. To see this, note we can apply the above mappings also to an {\em abstract} web $w$ satisfying boundary condition $\mathcal{B}$.
In this construction the spins are defined on the graph dual to the web $w$ (rather than on the faces of $\mathbb{D}$), still with the point at infinity
being removed. Applying $g_{\text{o}}$ to $w$ yields a well-defined colouring of these spins, even when the web contain tags.
Composing finally with $f_{\text{o}}$ gives a simple web that satisfies the boundary condition~$\mathcal{B}$. Moreover, this web differs from
the original basis web only by a possible set of tags and CPT transformed edges. Contracting the tags, one finally establishes the result.

For instance, in the $n=2$ case, a web model boundary condition compatible with a spin boundary condition has alternating defect charges
along the boundary, $k_{2i}=\pm 1$ and $k_{2i+1}=\mp 1$. A minimal basis is then given by planar perfect matchings of the defect nodes,
i.e., a Temperley-Lieb diagram.

\smallskip

Now, rescale the vector represented by a given configuration $[c] \in \mathcal{C}_\mathcal{B}$, 
\begin{align}
    \left(\prod_{\substack{k,l\in \llbracket 1,n-1 \rrbracket \\ k+l\in \llbracket 1,n-1 \rrbracket}} y_{k,l;k+l}^{N_{k,l;k+l}} \right) \left( \prod_{\substack{k,l\in \llbracket 1,n-1 \rrbracket \\ k+l\in \llbracket 1,n-1 \rrbracket}} y_{k+l;k,l}^{N_{k+l;k,l}} \right) \ket{[c]} = \ket{[c]}_d \,,
\end{align}
such that $\ket{[c]}_d$ is the vector associated to the web $c$ after dressing---i.e., the same web but with the dressed vertices of \eqref{dressed_vertices}.
The partition vector then reads
\begin{align}
        \ket{Z}=\sum_{[c]\in \mathcal{C}_{\mathcal{B}} } \left(\prod_{k\in \llbracket -n+1,n-1 \rrbracket} x_k^{N_k}\right) \ket{[c]}_d \,.
\end{align}

We would like to find a linear form $F$ such that 
\begin{align}
    F\big(\ket{[c]}_d\big) = 1
\end{align} 
for all $\ket{[c]}_d$.
To achieve this, we again go to the special point \eqref{special-point} and tune the vertex fugacities to \eqref{special-y}. We then pick a minimal
basis $\mathcal{A}$ of open simple webs. In such a basis, we expect the components $w_i$ of
\begin{align}
    \ket{[c]}_d=\sum_{\ket{v_i}\in \mathcal{A}} w_i\ket{v_i}_d
\end{align}
to satisfy
\begin{align}
\label{cond2}
    \sum_i w_i =1
\end{align}
by an argument similar to the closed case. Yet, in the case of open webs, one cannot proceed as in the proof of \eqref{simp},
using the identification with MOY graphs to avoid the use of relation \eqref{eq5}. When $n$ is odd, no sign appears in \eqref{eq5},
and the result \eqref{cond2} follows nonetheless. We believe that \eqref{cond2} also holds true for even $n$, but an additional argument
would be needed to complete the proof.

In any case, assuming \eqref{cond2}, one can choose the linear form $F$ to be simply the sum of components in the above basis of dressed webs.
One then recovers the partition function of spin models,
\begin{align}
    Z_F=\sum_{\sigma \in \mathcal{S}_{\Gamma}} \left(\prod_{k\in \llbracket -n+1,n-1 \rrbracket} x_k^{N_k}\right) \,.
\end{align}

\section{Conclusion}
\label{sec:conclusion}

In this paper we have introduced statistical models based on \U\ webs. They are geometrical models providing higher-rank generalisations of the
well-known $O(N)$ loop model \cite{Nienhuis}, which is retrieved for $n=2$. The most striking difference between the former and the $n>2$
models is the fact, that in the latter case branchings are allowed. The spiders \cite{Cautis_2014,Kuperberg_1996} allowed for the definition of the models 
on a genus $0$ surface in the closed boundary case, or on a simply connected domain with boundary defects in the open boundary case.
We have shown that, in a similar way as for the loop model, there exists a
special point \eqref{special-point} at which the webs are in correspondence with interfaces of $\mathbb{Z}_n$ spin models. 

We therefore believe that the web models at the special point provide an interesting framework for understanding the critical properties of
the $\mathbb{Z}_n$ spin interfaces, in the same way that the loop model was a useful tool for understanding domain walls in the Ising model (including
in the limit of infinite temperature, which is formally equivalent to site percolation). 

Moreover, we expect that the special points are part of one (or several) critical submanifolds of the parameter space. To be able to explore
the critical behaviour of such geometrical objects for a wider range of parameter values---in particular for generic $q$ on the unit circle---it
is crucial to have a local formulation of their Boltzmann weights. Such a local formulation of the weights will be given in a forthcoming paper \cite{loc}.
The analysis of the phase space of such a local model, and of its corresponding critical behaviour, then becomes amenable to the use of
transfer matrix techniques. Indeed, it is not clear to us whether the non-local reduction rules, \eqref{3rules} or \eqref{spiderrules}, can be handled
in a transfer matrix formalism. A local reformulation is also a prerequisite for the use of powerful quantum field theory techniques, in particular
those of CFT. 

Another outstanding issue is whether the continuum limit of web models at criticality can be described by Coulomb Gas methods---historically a
strong tool for dealing with the $n=2$ case of loop models. The Coulomb Gas has also been succesful in describing the fully-packed loop model on the hexagonal lattice with a special non contractible loop weight, chosen so as to make the continuum limit exhibit $\mathcal{W}_3$ symmetry \cite{FPL}. This fully-packed loop model should be a particular instance of our $U_q(\mathfrak{sl}_3)$ web model.
Within the Coulomb Gas approach, the goal would be to find analytical expressions for (geometrical) critical exponents,
both in the open and closed boundary cases. This is an objective that we plan to pursue in the near future.

We also believe it is worth investigating the possible description of the growth of a web by a stochastic process, in the same spirit that
Schramm-Loewner evolution \cite{SLE} provides a growth process related to the loop models. As $\mathbb{Z}_n$ spin models at the Fateev-Zamolodchikov
point are believed to be lattice regularisations of parafermions, we can further ask whether web models and their branchings will provide a geometrical
insight into $\mathcal{W}_n$ extended symmetry within CFT, a question that, so far, has been pursued only in the simple-curve context of SLE \cite{SLELie,santa}. Note also that, in order to define defects in the bulk, both in the closed and open boundary cases, one would need an affine version of the diagrammatic categories given by spiders.

Finally, it is  clear that a local formulation of the web model will provide physical motivation for the study of diagrammatic algebras based
on the spiders of \cite{Kuperberg_1996} and \cite{Cautis_2014}, generalising the Temperley-Lieb algebra and its variations. In addition, as webs
are related to \U\ intertwiners in the representation theoretical context, it is natural to ask whether there are integrable points for the web models.

\subsection*{Acknowledgments}

We thank H.\ Saleur and T.\ Dupic for interesting discussions. This work was supported by the European Research Council through the advanced grant NuCFT.
The work of AMG was supported by the CNRS, and partially by the ANR grant JCJC ANR-18-CE40-0001 and the RSF Grant No.\ 20-61-46005. 
AMG is also grateful to IPHT Saclay and ENS Paris for their kind hospitality in 2019 and 2020.
\appendix

\section{Proof of equation \eqref{appAeq}}
\label{sec:app-equiv-weights}
Here we show that if two simple webs $c$ and $c'$ are equivalent---that is, related to each other by CPT transformations---then $w_s(c)=w_s(c')$, or $\ket{c}=\ket{c'}$ in the open boundary case. In fact, we will show that one can go from the web $c$ to $c'$ by using the rules \eqref{spiderrules}. Hence, they get weighted the same way in the closed case, or represent the same vector in the open case. To prove this, we focus on one transition cycle of edges of $c$ that are CPT transformed in $c'$. When this cycle has only one edge, the proof is straightforward. Suppose it has more than one edge, as depicted in the following figure:
\begin{center}
    \includegraphics[scale=0.3]{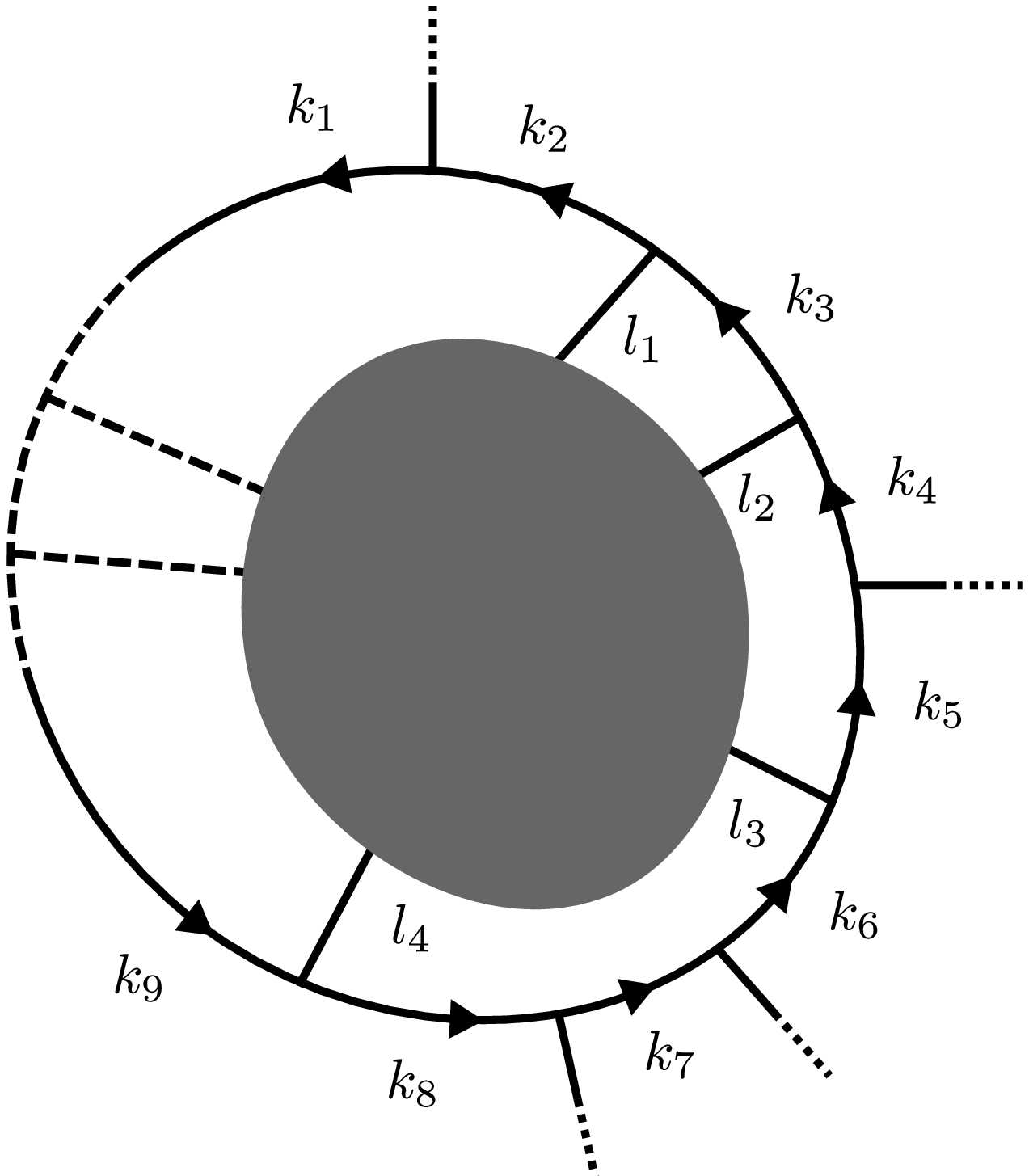}
\end{center}{}
The orientation of the edges connected to the cycle is immaterial. In order to CPT transform the whole cycle, one can introduce pairs of opposite tags,
by applying \eqref{eq6} on each edge of the cycle, obtaining:
\begin{center}
    \includegraphics[scale=0.3]{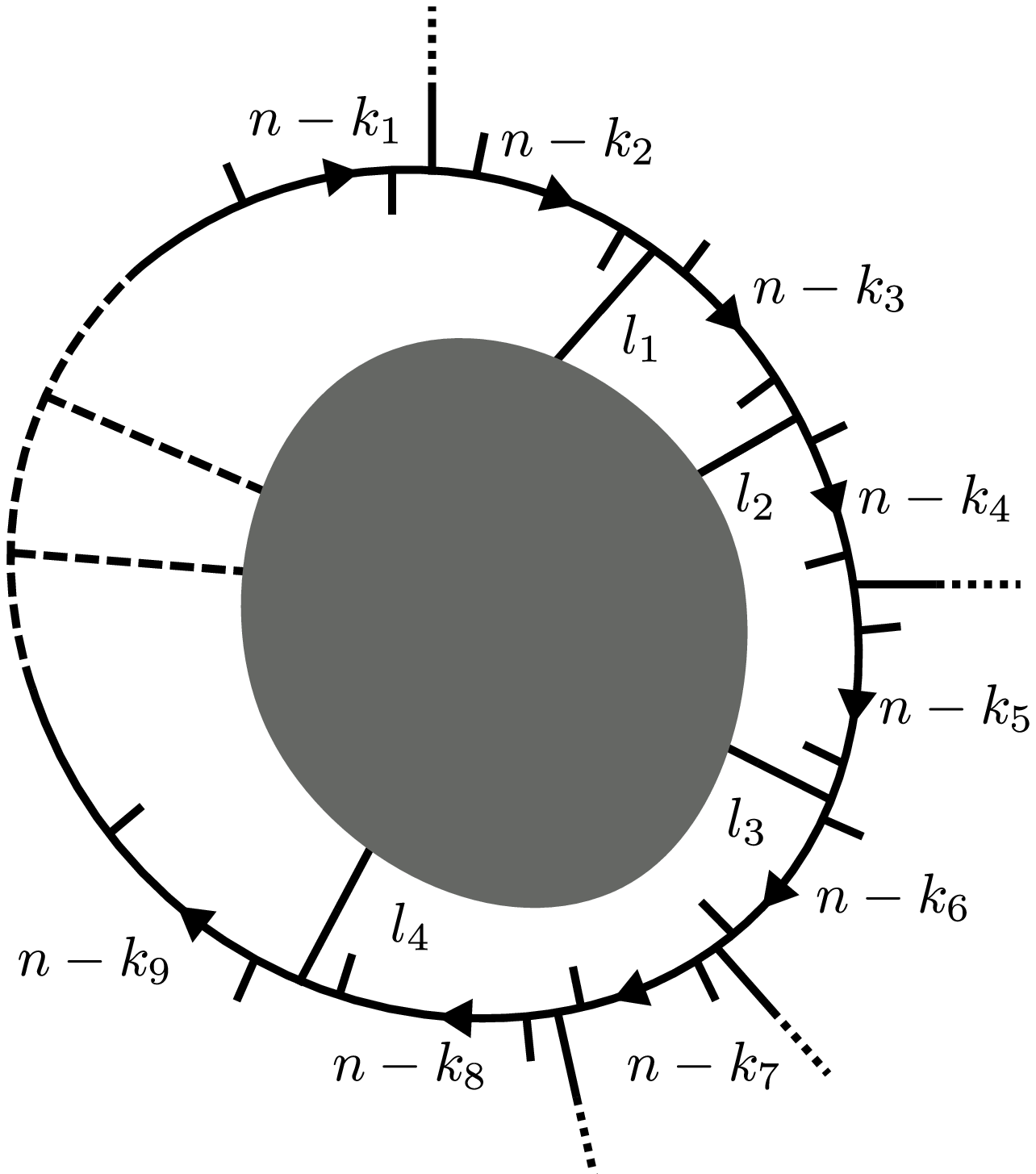}
\end{center}{}
Now thanks to \eqref{eq7}, \eqref{eq8} and \eqref{eq5}, one can move each tag in front of the arrow of the transformed edges through the vertex in front of it,
and contract it with the following tag, at the price of a possible sign. When $n$ is odd, there is no sign. When $n$ is even, the sign occurs for crossing edges
connected to the interior, grey part of the diagram, and is equal to $(-1)^{l_i}$ whatever the orientation of the edge carrying the label $l_i$. Doing this contraction
for each edge of the cycle thus creates a total factor of $(-1)^{\sum_i l_i}$. Now, due to the flow conservation in the gray part of the diagram, there exists
a set of signs $\epsilon_i=\pm 1$, such that $\sum_i \epsilon_i l_i=0$. This implies that $(-1)^{\sum_i l_i}=1$, so finally no non-trivial factor appears
when transforming the cycle. Repeating the procedure for each transition cycle then leads to the result.


\begin{thebibliography}{99}

\bibitem{DS-NPB87} B.\ Duplantier and H.\ Saleur, {\em Exact critical properties of two-dimensional dense self-avoiding walks}, Nucl.\ Phys.\ B {\bf 290}, 291--326 (1987).
\bibitem{KH95} J.\ Kondev and C.L.\ Henley, {\em Geometrical exponents of contour loops on random Gaussian surfaces}, Phys.\ Rev.\ Lett.\ {\bf 74}, 4580 (1995).
\bibitem{GamsaCardy07} A.\ Gamsa and J.\ Cardy, {\em Schramm-Loewner evolution in the three-state Potts model---a numerical study},
 J.\ Stat.\ Mech.:\ Theor.\ Exp.\ P08020 (2007); {\tt arXiv:0705.1510}.
\bibitem{GLR99} I.A.\ Gruzberg, A.W.W.\ Ludwig and N.\ Read, {\em Exact exponents for the spin quantum Hall transition}, Phys.\ Rev.\ Lett.\ {\bf 82}, 4524 (1999); {\tt arXiv:cond-mat/9902063}.
\bibitem{Nienhuis} B.\ Nienhuis, {\em Exact critical point and critical exponents of O(n) models in two dimensions}, Phys.\ Rev.\ Lett.\ {\bf 49}, 1062 (1982).
\bibitem{Baxter86} R.J.\ Baxter, {\em $q$-colourings of the triangular lattice}, J.\ Phys.\ A:\ Math.\ Gen.\ {\bf 19}, 2821 (1986).
\bibitem{WNS92} S.O.\ Warnaar, B.\ Nienhuis and K.A.\ Seaton, {\em New construction of solvable lattice models including an Ising model in a field}, Phys.\ Rev.\ Lett.\ {\bf 69}, 710 (1992).
\bibitem{IntReview} J.L.\ Jacobsen, {\em Integrability in statistical physics and quantum spin chains}.
 In P.\ Dorey, G.\ Korchemsky, N.\ Nekrasov, V.\ Schomerus and D.\ Serban (eds.), 
 {\em Les Houches Summer School session CVI. Integrability: From statistical systems to gauge theory}
 (Oxford University Press, 2019).
\bibitem{Kauf87}  L.H.\ Kauffman, {\em State models and the Jones polynomial}, Topology, {\bf 26}, 395--407 (1987).
\bibitem{AlgReview} J.J.\ Graham and G.I.\ Lehrer, {\em Cellular algebras and diagram algebras in representation theory}.
 In T.\ Shoji, M.\ Kashiwara, N.\ Kawanaka, G.\ Lusztig and K.\ Shinoda (eds.),
 {\em Representation theory of algebraic groups and quantum groups},  Adv.\ Stud.\ Pure Math.\, 141--173
 (Mathematical Society of Japan, Tokyo, 2004).
\bibitem{Wang} Zh.\ Wang, {\em Topological Quantum Computation} (AMS, Providence, RI, 2010).
\bibitem{GL1} J.J.\ Graham and G.I.\ Lehrer, {\em The representation theory of affine Temperley-Lieb algebras}, Enseign.\ Math.\ {\bf 44}, 173 (1998).
\bibitem{GS16} A.M. Gainutdinov, H. Saleur, {\em Fusion and braiding in finite and affine Temperley-Lieb categories}, preprint Hamburger Beitr\"age zur Mathematik {\bf 596}; {\tt arXiv:1606.04530}.
\bibitem{LoopReview} J.L.\ Jacobsen, {\em Conformal field theory applied to loop models}.
 In A.J.\ Guttmann (ed.), {\em Polygons, polyominoes and polycubes},
 Lecture Notes in Physics {\bf 775}, 347--424 (2009).
\bibitem{CFTbook} P.\ Di Francesco, D.\ S\'en\'echal and P.\ Mathieu, {\em Conformal field theory} (Springer-Verlag, New York, 1997).
\bibitem{CLE} S.\ Sheffield, {\em Exploration trees and conformal loop ensembles}, Duke Math.\ J.\ {\bf 147}, 79--129 (2009); {\tt arXiv:math/0609167}.
\bibitem{SLE} J.\ Cardy, {\em SLE for theoretical physicists}, Ann.\ Phys.\ {\bf 318}, 81--118 (2005); {\tt arXiv:cond-mat/0503313}.
\bibitem{Dubail10} J.\ Dubail, J.L.\ Jacobsen and H.\ Saleur, {\em Critical exponents of domain walls in the two-dimensional Potts model},
J.\ Phys.\ A:\ Math.\ Theor.\ {\bf 43}, 482002 (2010); {\tt arXiv:1008.1216}.
\bibitem{Dubail10b} J.\ Dubail, J.L.\ Jacobsen and H.\ Saleur, {\em Bulk and boundary critical behaviour of thin and thick domain walls in the two-dimensional Potts model}, J.\ Stat.\ Mech.:\ Theor.\ Exp.\ P12026 (2010); {\tt arXiv:1010.1700}.
\bibitem{PiccoSantachiara11} M.\ Picco and R.\ Santachiara, {\em Critical interfaces and duality in the Ashkin-Teller model},
Phys.\ Rev.\ E {\bf 83}, 061124 (2011); {\tt arXiv:1011.1159}.
\bibitem{Kitaev03} A.Y.\ Kitaeev, {\em Fault-tolerant quantum computation by anyons}, Annals of Phys.\ {\bf 303}, 2--30 (2003); {\tt arXiv:quant-phy/9707021}.
\bibitem{LevinWen05} M.A.\ Levin and X.-G.\ Wen, {\em String-net condensation: A physical mechanism for topological phases}, Phys.\ Rev.\ B {\bf 71}, 045110 (2005); {\tt arXiv:cond-mat/0404617}.
\bibitem{Fendley08} P.\ Fendley, {\em Topological order from quantum loops and nets}, Annals of Phys. {\bf 323}, 3113--3136 (2008); {\tt arXiv:0804.0625}.


\bibitem{kuperberg1991quantum} G.\ Kuperberg, {\em The quantum $G_2$ link invariant}, Int.\ J.\ Math.\ {\bf 5}, 61--85 (1994); {\tt arXiv:math/9201302}.
\bibitem{Kuperberg_1996} G.\ Kuperberg, {\em Spiders for rank-2 Lie algebras}, Comm.\ Math.\ Phys. {\bf 180}, 109--151 (1996); {\tt arXiv:q-alg/9712003}.
\bibitem{MOY} H.\ Murakami, T.\ Ohtsuki and  S.\ Yamada, {\em HOMFLY polynomial via an invariant of colored
plane graphs}, Enseign.\ Math.\ {\bf 44}, 325--360 (1998).
\bibitem{kim2003graphical} D.\ Kim, {\em Graphical calculus on representations of quantum Lie algebras}, J.\ Knot Theor.\ Ramifications {\bf 15}, 453--469 (2006); {\tt arXiv:math/0310143}.
\bibitem{JK05} M.-J.\ Jeong and D.\ Kim, {\em The quantum $sl(n,\mathbb{C})$ representation theory and its applications},
J.\ Korean Math.\ Soc.\ {\bf 49}, 993--1015 (2012); {\tt arXiv:math/0506403}.
\bibitem{Mor07} S.\ Morrison, {\em A diagrammatic category for the representation theory of $U_q (sl_n)$}, PhD thesis (University
of California, Berkeley, 2007); {\tt arXiv:0704.1503}.
\bibitem{Wu} H.\ Wu, {\em A colored $sl(N)$-homology for links in $S^3$}, (2009); {\tt arXiv:0907.0695}.
\bibitem{Cautis_2014} S.\ Cautis, J.\ Kamnitzer and S.\ Morrison, {\em Webs and quantum skew Howe duality}, Mathematische Annalen {\bf 360}, 351--390 (2014); {\tt arXiv:1210.6437}.

 \bibitem{TL71} H.N.V.\ Temperley and E.T.\ Lieb, {\em Relation between the `percolation' and `colouring' problem and other graph-theoretical problems associated with planar lattices: some exact results for the `percolation' problem}, Proc.\ Roy.\ Soc.\ London A {\bf 322}, 251 (1971).
 
\bibitem{loc} A.\ Lafay, A.M.\ Gainutdinov and J.L.\ Jacobsen, {\em Local vertex-model formulation of web models}, in preparation.

\bibitem{BirmanWenzl} J.S.\ Birman and H.\ Wenzl, {\em Braids, link polynomials and a new algebra}, Trans.\ Am.\ Math.\ Soc.\ {\bf 313}, 249--273 (1989).
\bibitem{Murakami} J.\ Murakami {\em The Kauffman polynomial of links and representation theory}, Osaka J.\ Math.\ {\bf 24}, 745 (1987).
\bibitem{Grimm} U.\ Grimm, {\em Dilute algebras and solvable lattice models}. In M.-L.\ Ge and F.Y.\ Wu (eds), {\em Statistical models, Yang-Baxter equation and related topics---Proceedings of the satellite meeting of Statphys 19}, pp.\ 110--117 (World Scientific, Singapore 1996).
\bibitem{GrimmPearce} U.\ Grimm and P.A.\ Pearce, {\em Multi-colour braid-monoid algebras}, J.\ Phys.\ A:\ Math.\ Gen.\ {\bf 26}, 7435 (1993); {\tt arXiv:hep-th/9303161}.
\bibitem{GrimmMartin} U.\ Grimm and P.P.\ Martin, {\em The bubble algebra: structure of a two-colour Temperley-Lieb algebra}, J.\ Phys.\ A:\ Math.\ Gen.\ {\bf 36}, 10551 (2003); {\tt arXiv:math-ph/0307017}.
\bibitem{KdGN96} J.\ Kondev, J.\ de Gier and B.\ Nienhuis, {\em Operator spectrum and exact exponents of the fully packed loop model}, J.\ Phys.\ A:\ Math.\ Gen.\ {\bf 29}, 6489 (1996); {\tt arXiv:cond-mat/9603170}.
\bibitem{JK98} J.L.\ Jacobsen and J.\ Kondev, {\em Field theory of compact polymers on the square lattice}, Nucl.\ Phys.\ B {\bf 532}, 635--688 (1988); {\tt arXiv:cond-mat/9804048}.
\bibitem{KJ98} J.\ Kondev and J.L.\ Jacobsen, {\em Conformational entropy of compact polymers}, Phys.\ Rev.\ Lett.\ {\bf 81}, 2922--2925 (1998); {\tt arXiv:cond-mat/9805178}.
\bibitem{Salas} J.L.\ Jacobsen, J.\ Salas and C.R.\ Scullard, {\em Phase diagram of the triangular-lattice Potts antiferromagnet}, J.\ Phys.\ A:\ Math.\ Theor.\ {\bf 50}, 345002 (2017); {\tt arXiv:1702.02006}.
\bibitem{Harary} F.\ Harary, {\em Covering and packing in graphs, I.}, Ann.\ New York Acad.\  Sci.\ {\bf 175}, 198 (1970).
\bibitem{DubailSp1} J.\ Dubail, J.L.\ Jacobsen and H.\ Saleur, {\em Conformal boundary conditions in the critical O($n$) model and dilute loop models}, Nucl.\ Phys.\ B {\bf 827}, 457--502 (2010); {\tt arXiv:0905.1382}.
\bibitem{DubailSp2} J.\ Dubail, J.L.\ Jacobsen and H.\ Saleur, {\em Exact solution of the anisotropic special transition in the O($n$) model in 2D}, Phys.\ Rev.\ Lett.\ {\bf 103}, 145701 (2009); {\tt arXiv:0909.2949}.
\bibitem{santa} R.\ Santachiara,  {\em SLE in self-dual critical $Z(N)$ spin systems : CFT predictions}, Nucl. Phys. B {\bf 793}, 396-424 (2008); {\tt arXiv:0705.2749}.
\bibitem{SLELie} E.\ Bettelheim, I.A.\ Gruzberg, A. W. W.\ Ludwig and P.\ Wiegmann, {\em Stochastic Loewner evolution for conformal field theories with Lie group symmetries}, Phys.\ Rev.\ Lett.\ {\bf 95}, 251601 (2005); {\tt arXiv:hep-th/0503013}.
\bibitem{FPL}  T.\ Dupic, B.\ Estienne and Y.\ Ikhlef, {\em The fully packed loop model as a non-rational $\mathcal{W}_3$ conformal field theory}, J.\ Phys.\ A:\ Math.\ Theor.\ {\bf 49}, 505202 (2016); {\tt arXiv:1606.05376}.

\end{thebibliography}
\end{document}